\documentclass[12pt,preprint]{aastex}
\usepackage{url}
\usepackage[usenames]{color}
\usepackage{amsmath}
\newcommand{\logg} {\log \textsl{\textrm{g}}}
\newcommand{\qh} {q({\rm H})}
\newcommand{\qhe} {q({\rm He})}
\newcommand{\logqh} {\log q({\rm H})}

\newcommand{\Te} {T_{\rm eff}}
\newcommand{\mh} {M_{\rm H}}
\newcommand{\mhe} {M_{\rm He}}

\newcommand{\msun} {$M_\odot$}

\newcommand\lta{\lower 0.5ex\hbox{$\buildrel < \over \sim\ $}} 

\newcommand{\ha} {$\rm{H}{\alpha}$}
\newcommand{\hb} {$\rm{H}{\beta}$}

\newcommand{\lsun} {$L_{\odot}$}
\newcommand{\logh} {\log \rm{H/He}}

\begin{document}

\title{On the Spectral Evolution of Helium-Atmosphere White Dwarfs
  Showing Traces of Hydrogen}

\author{B. Rolland, P. Bergeron \& G. Fontaine}
\affil{D\'epartement de Physique, Universit\'e de Montr\'eal,
  C.P.~6128, Succ.~Centre-Ville, Montr\'eal, Qu\'ebec H3C 3J7, Canada}
\email{rolland@astro.umontreal.ca, bergeron@astro.umontreal.ca, fontaine@astro.umontreal.ca}

\begin{abstract}
We present a detailed spectroscopic analysis of 115 helium-line (DB)
and 28 cool, He-rich hydrogen-line (DA) white dwarfs based on
atmosphere fits to optical spectroscopy and photometry. We find that
63\% of our DB population show hydrogen lines, making them DBA stars.
We also demonstrate the persistence of pure DB white dwarfs with no
detectable hydrogen feature at low effective temperatures. Using
state-of-the art envelope models, we next compute the total quantity
of hydrogen, $\mh$, that is contained in the outer convection zone as
a function of effective temperature and atmospheric H/He ratio. We
find that some $(\Te,\mh)$ pairs cannot physically exist as a
homogeneously mixed structure; such combination can only occur as
stratified objects of the DA spectral type. On that basis, we show
that the values of $\mh$ inferred for the bulk of the DBA stars are
too large and incompatible with the convective dilution scenario. We
also present evidence that the hydrogen abundances measured in DBA and
cool, helium-rich white dwarfs cannot be globally accounted for by any
kind of accretion mechanism onto a pure DB star. We suggest that cool,
He-rich DA white dwarfs are most likely created by the convective
mixing of a DA star with a thin hydrogen envelope; they are not cooled
down DBA's. We finally explore several scenarios that could account
for the presence of hydrogen in DBA stars.
\end{abstract}


\section{INTRODUCTION}

The extreme chemical purity of white dwarf atmospheres can be
attributed to the intense gravitational field present at the surface
of these stars, causing all the heavy elements to sink rapidly out of
sight \citep{schatzman45}. This gravitational settling process should
thus produce white dwarf atmospheres that are completely dominated by
hydrogen --- or DA stars. However, it is well known that a significant
fraction of the white dwarf population is hydrogen-deficient --- e.g.,
PG 1159, DO, DB, DQ, DZ, and some DC stars \citep{wesemael93} --- and
a very small fraction even have carbon-dominated atmospheres
\citep{dufourNature07}. More importantly, the relative number of white
dwarfs of a given spectral type varies considerably as a function of
effective temperature, indicating that there exist several physical
mechanisms that compete with gravitational settling to alter the
chemical composition of the outer layers of white dwarfs as they
evolve along the cooling sequence. Such physical mechanisms include
convective mixing, convective dredge-up from the core, accretion from
the interstellar medium or circumstellar material, radiative
levitation, and stellar winds. Understanding the so-called {\it
  spectral evolution of white dwarf stars} has always remained a
fundamental topic of research in the white dwarf field, in particular
with the large number of new discoveries in the Sloan Digital Sky
Survey (SDSS; see, e.g., \citealt{SDSS_DR7}).

Probably the most significant evidence for the spectral evolution of
white dwarfs, discussed at length in \citet{FW87}, is the existence of
a ``DB-gap'', a range in effective temperature between
$\Te\sim30,000$~K and 45,000~K where only DA stars are found, while
helium-atmosphere white dwarfs exist both above (the DO stars) and
below (the DB stars) the gap.  One model proposed by Fontaine \&
Wesemael to account for this gap starts with hot white dwarf
progenitors with hydrogen-deficient atmospheres (PG 1159 or DO stars)
containing only minute amounts of hydrogen thoroughly diluted within
the stellar envelope. As these stars cool off, hydrogen would
gradually float up to the surface, thus building an atmosphere
enriched with hydrogen. The fact that all white dwarfs turn into DA
stars by the time they reach $\Te\sim45,000$~K imposes a lower limit
on the total amount of hydrogen present in the hot progenitors, of the
order of $\mh\sim 10^{-16}$ \msun.  Hybrid white dwarfs with
thinner hydrogen layers floating in diffusive equilibrium on top of
the helium envelope would appear as DAO stars, bearing the signature
of chemically stratified atmospheres (see \citealt{manseau16} and
references therein).

Below the red edge of the gap ($\Te\lesssim 30,000$~K), the
reappearance in large numbers of helium-atmosphere white dwarfs --- DB
stars in this case --- has been interpreted in terms of the dilution
of a thin, superficial hydrogen {\it radiative} layer ($\mh\sim
10^{-15}$ \msun) by the underlying and more massive convective helium
envelope \citep{FW87}. In this paper, we refer to this mechanism as
the {\it convective dilution scenario} (see also \citealt{MV91}). Even
though the large number of white dwarfs discovered in the SDSS has
unveiled the existence of many hot DB stars in the gap, the fraction
of DB white dwarfs within the gap remains significantly lower than
that found at lower temperatures, and we are thus dealing with a DB
deficiency rather than a true gap. Nevertheless, the float-up model and
convective dilution scenario discussed above must occur for a
significant fraction of white dwarfs, perhaps of the order of
$\sim$20\% \citep{bergeron11}.

Another important signature of the spectral evolution of white dwarfs,
also discussed in \citet{FW87}, is the spectacular increase in the ratio
of non-DA to DA stars below $\Te\sim 10,000$~K, which jumps from a
value around 25\% above this temperature to a value near unity below it. This
sudden increase in the number of non-DA stars (i.e., DQ, DZ, DC) in this
temperature range has been interpreted as the result of the mixing of
the superficial {\it convective} hydrogen layer with the deeper and
much more massive convective helium envelope
\citep{koester76,vauclair77,dantona79}. We will refer to this
mechanism as the {\it convective mixing scenario}, as opposed to the
{\it convective dilution scenario} discussed above, in which the hydrogen
superficial layer is purely radiative.

One way to further our understanding of the spectral evolution of
helium-atmosphere white dwarfs below $\Te\sim30,000$~K is to determine
the hydrogen abundance in these stars, often present as a trace
element, and to study the hydrogen abundance pattern as a function of
effective temperature. Indeed, a large fraction of DB white dwarfs
shows traces of hydrogen --- the DBA stars --- if observed at
sufficiently high signal-to-noise ratio (S/N). \citet{KK15} even
suggested, based on their analysis of the DB stars in the DR10 and
DR12 of the SDSS, that perhaps {\it all} DB white dwarfs would show
hydrogen if the resolution and S/N were high enough. The origin of
hydrogen in DBA stars has remained a mystery, and the subject of
controversy as well. While it seems reasonable to assume that the
presence of hydrogen in these stars has a residual origin --- the
leftovers from the convective dilution scenario discussed above ---
the total mass of hydrogen inferred in those stars, which is
homogeneously mixed in the convective stellar envelope, lies in the
range $\mh=10^{-13}$ to $10^{-10}\ M_\odot$
\citep{Voss07,bergeron11,KK15}. The problem with these estimates is
that DA progenitors with such thick hydrogen layers would easily
survive the convective dilution process, and thus never turn into DB
stars in the first place \citep{MV91}. The most common way around this
problem is to assume that the DA progenitors have hydrogen layers thin
enough (of the order of $\mh\sim10^{-15}\ M_\odot$) to allow the
DA-to-DB transition below $\Te\sim30,000$~K, and that significant
amounts of hydrogen are then accreted onto the DB star from external
sources such as the interstellar medium, disrupted asteroids, small
planets, and even comets \citep{MV91,bergeron11,KK15,fusillo17}. These
accretion scenarios can easily account for the observed hydrogen
abundances in DBA stars, assuming reasonable accretion rates.

Also of key interest is the presence of hydrogen in much cooler
($\Te\lesssim10,000$~K) helium-atmosphere white dwarfs, the prototypes
of which are the DZA stars L745-46A and Ross 640 (see, e.g., Figure 14
of \citealt{Giamm12}), which show broad and shallow H$\alpha$
absorption features resulting from van der Waals broadening in a
helium-dominated atmosphere. Traces of hydrogen have now been
detected in many DZ stars from the SDSS \citep{dufour07}. The origin
of hydrogen in these objects, whether it has a residual origin ---
cooled off DBA stars or convectively mixed DA stars --- or has been
accreted from external bodies, remains an open question.

In this paper, we revisit the problem of the spectral evolution of
helium-atmosphere white dwarfs below $\Te\sim30,000$~K, by studying
the hydrogen abundance pattern in these stars as a function of
effective temperature. We first present in Section \ref{sec:abun} a
detailed model atmosphere analysis of relatively bright DB and DBA
white dwarfs, as well as cool He-rich DA stars drawn from the SDSS.
In Section \ref{sec:models}, we describe our stellar envelope models
with stratified and homogeneous chemical compositions appropriate for
these stars, which are then used in Section \ref{sec:evol} to explore
and test various scenarios that could account for the observed hydrogen
abundance pattern, and discuss possible evolutionary channels that
could produce DB, DBA, and cool He-rich DA and DZA stars. Our
conclusions follow in Section \ref{sec:concl}.

\section{HYDROGEN ABUNDANCE PATTERN IN HELIUM-ATMOSPHERE WHITE DWARFS}\label{sec:abun}

\subsection{Hydrogen in DBA stars}\label{sec:dba}

\subsubsection{Spectroscopic Observations}\label{sec:specobs}

\begin{figure}[bp]
\centering
\includegraphics[width=0.8\linewidth]{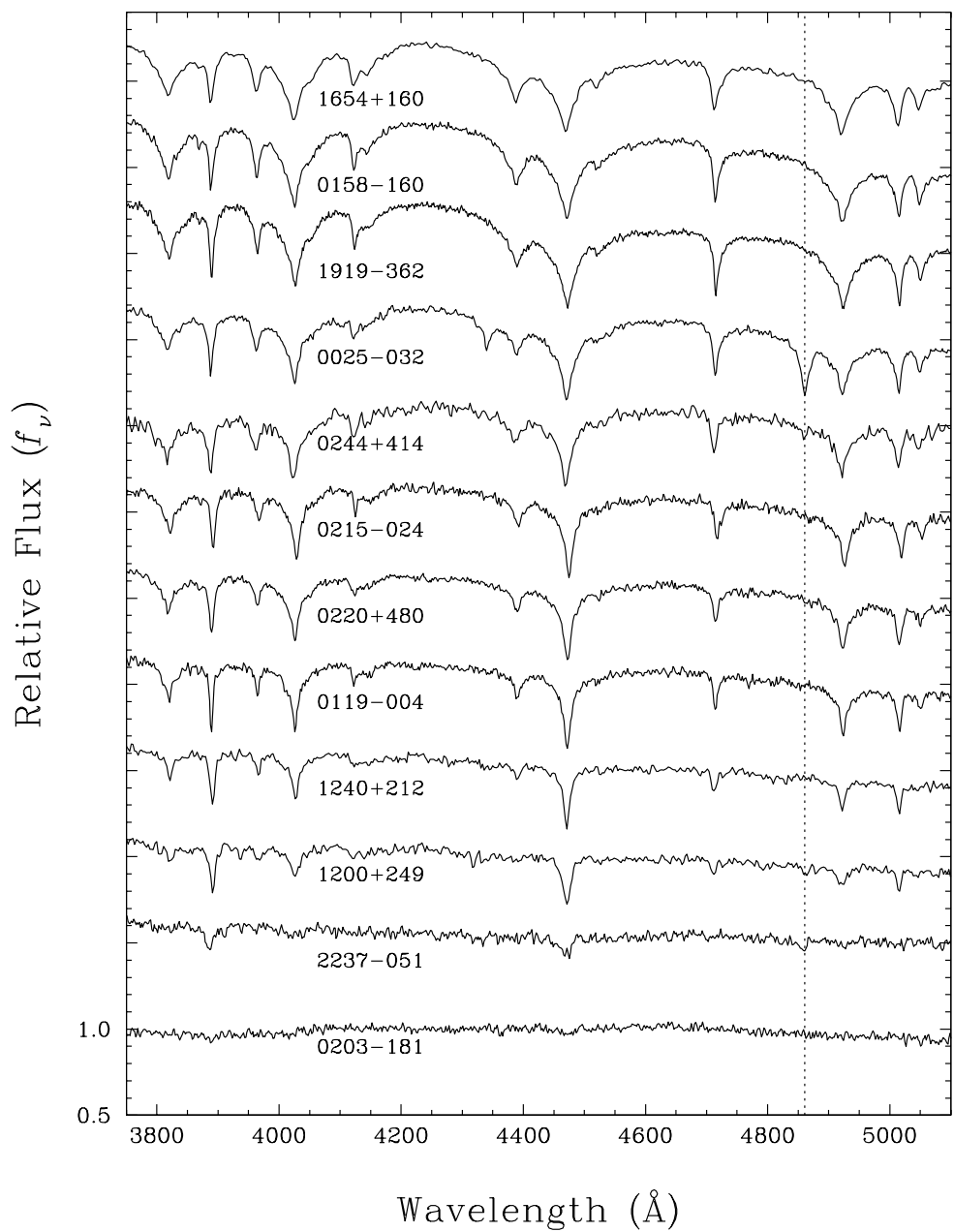}
\caption{Optical (blue) spectra for the 11 new DB stars included in
  our extended sample, as well as a new spectrum for PG 1654$+$160;
  the remaining spectra used in our analysis are displayed in Figure 5
  of \citet{bergeron11}.  The spectra are normalized at 4500 \AA~and
  shifted vertically from each other by a factor of 0.5 for
  clarity. The effective temperature decreases from top to bottom. The
  location of \hb\ is shown by a dotted line.\label{spec_DBA_blue}}
\end{figure}

Our sample of bright, helium-line DB and DBA stars is based on an
extension of the 108 DB white dwarfs analyzed in detail in
\citet{bergeron11}. In particular, high S/N spectra of 6 additional DB
stars, selected from the electronic version of the Catalogue of
Spectroscopically Identified White
Dwarfs\footnote{http://www.astronomy.villanova.edu/WDCatalog/index.html}
\citep[][hereafter WD Catalog]{mccook99}, have been secured with the
Steward Observatory 2.3 m Bok Telescope equipped with the Boller \&
Chivens spectrograph. The 4$\farcs$5 slit together with the 600 line
mm$^{-1}$ grating blazed at 3568 \AA~in first order provides a
spectral coverage from about 3500 to 5250 \AA~at a resolution of
$\sim$6 \AA~FWHM (see also \citealt{bergeron15}). Also included are
1919$-$362\footnote{We omit in the remainder of this paper the WD
  prefix for conciseness.} from \citet{sub17}, as well as the 4 new DB
stars discovered by \citet[][see their Figure 15]{LBL15} in the course
of their spectroscopic survey of the SUPERBLINK proper motion
catalog. We also secured a new optical spectrum for PG 1654$+$160
using the same setup. These additional optical spectra are displayed
in Figure \ref{spec_DBA_blue} in order of decreasing effective
temperature; the other blue spectra in our sample have already been
displayed in Figure 5 of \citet{bergeron11}. Note the particular
strength of the hydrogen lines in PB 8252 (0025$-$032).

\begin{figure}[bp]
\centering
\includegraphics[width=0.8\linewidth]{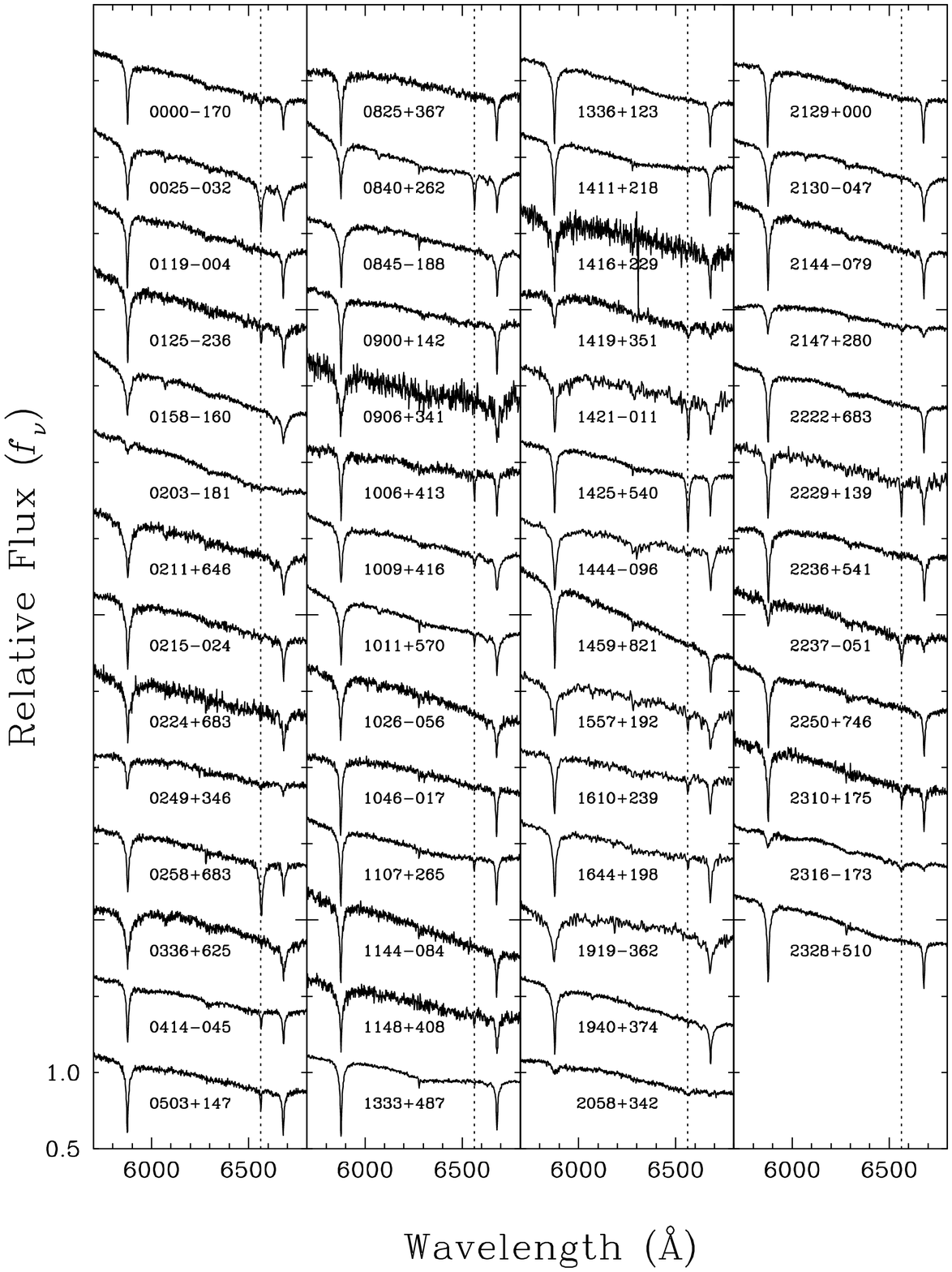}
\caption{New spectra near the \ha\ region for 54 DB and DBA stars in
  our extended sample, together with the DB star from
  \citet[][1919$-$362]{sub17}.  The spectra are shown in order of
  right ascension, normalized at 6200 \AA, and shifted vertically from
  each other by a factor of 0.5 for clarity. The location of \ha\ is
  shown by a dotted line. \label{spec_DBA_red}}
\end{figure}

Because we are mostly interested here in studying the hydrogen
abundance pattern in helium-atmosphere white dwarfs, we also improved
the sample of DB stars from \citet{bergeron11} by acquiring 54 high
signal-to-noise \ha\ spectra missing from our original data set (see
also \citealt{bergeron15}). These spectra have been obtained with the
NOAO Mayall 4-m telescope; the adopted configuration allows a spectral
coverage of $\lambda\lambda$3800--6700, at an intermediate resolution
of $\sim$6 \AA\ FWHM. These spectra are displayed in Figure
\ref{spec_DBA_red}. Note how the strength of \ha\ varies considerably
from object to object, and how it is particularly strong for
PB 8252 (0025$-$032) and Lan 143 (0258$+$683). Hydrogen is now detected in 63\%
of the DB stars in our sample, a value somewhat lower than the
estimated 75\% fraction of DBA white dwarfs obtained by \cite{KK15}
using their best spectra. Four objects in our sample still lack
\ha\ spectroscopic data --- L715$-$34 (0308$-$565), BPM 17731
(0418$-$539), L151-81A (1454$-$630.1), and GD 27 (0220$+$480) --- and
these are excluded from our analysis for homogeneity purposes. Our
final sample thus includes 115 DB white dwarfs, among which 73 are DBA
stars.

A significant advantage of our extended sample is its homogeneity,
both in terms of wavelength coverage and S/N. Even though its size is
modest in comparison to the SDSS sample \citep{KK15}, the quality of
our spectra is superior in terms of S/N. Indeed, because the exposure
time of a given SDSS spectrum is constant for all targets on a given
plate, the corresponding S/N is necessarily a function of the
magnitude of the star, resulting in typical sensitivity between
${\rm S/N}\sim5$ and 20. In comparison, the majority of our spectra
have S/N well above 50, with an average around 80.

\subsubsection{Model Atmosphere Analysis}

\begin{figure}[bp]
\centering
\includegraphics[width=0.8\linewidth]{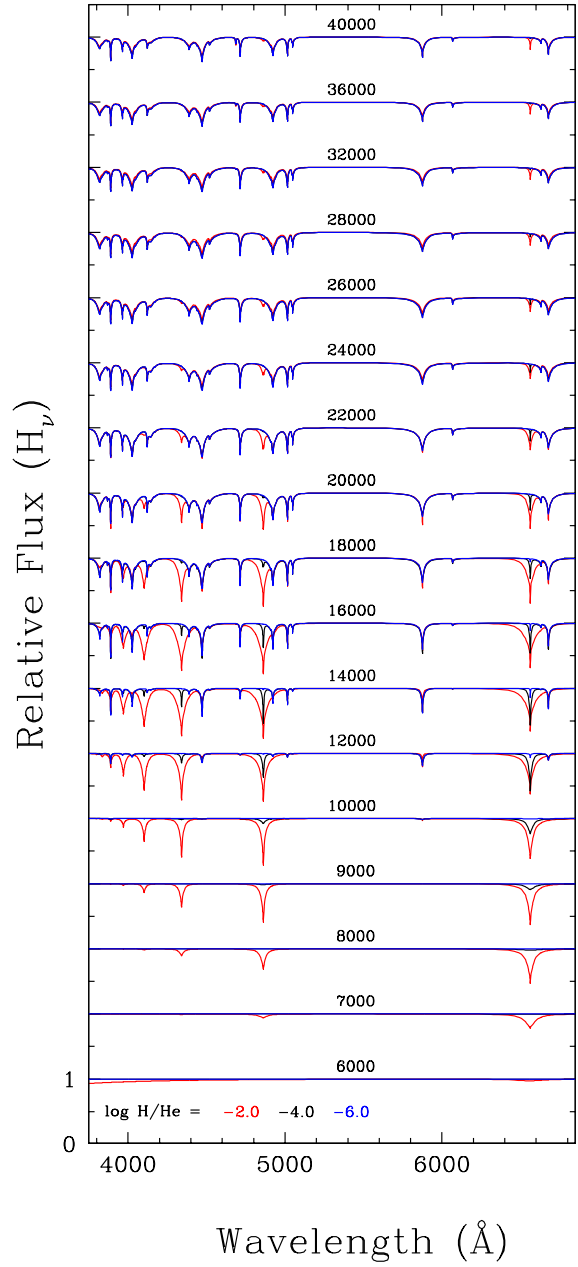}
\caption{Synthetic spectra of homogeneous H/He models at $\logg=8$
  for various effective temperatures and hydrogen abundances of
  $\rm{H/He}=10^{-6}$ (blue), $10^{-4}$ (black), and $10^{-2}$ (red). All
  spectra are normalized to a continuum set to unity and offset from
  each other by a factor of 1.0 for clarity. \label{spec_DBA_synth}}
\end{figure}

Our model atmospheres and synthetic spectra are identical to those
described at length in \cite{bergeron11} and references therein.
These models are built from the LTE model atmosphere code described in
\cite{TB09} and references therein, in which the improved Stark
profiles of neutral helium of \cite{BWB97} have been
incorporated. These detailed profiles of more than 20 neutral helium
lines take into account the transition from the impact to the
quasistatic regime for electrons, the transition from quadratic to
linear Stark broadening, as well as forbidden components.  We also
include in the cooler models van der Waals broadening following the
treatment of \cite{DvR76}.  For the treatment of the hydrogen lines,
we rely on the improved calculations for Stark broadening of
\cite{TB09}, as well as resonance broadening and nonresonant
broadening for the Balmer lines in the cooler models, as also
described in Tremblay \& Bergeron. Convective energy transport is
treated within the ML2 version of the mixing-length theory with a
value of $\alpha=\ell/H$ --- the ratio of the mixing length to the
pressure scale height --- set to 1.25, following the prescription of
\citet{bergeron11}. All hydrogen and helium level populations are
computed using the occupation probability formalism of \cite{HM88},
which is also included in the calculations of the corresponding
bound-bound, bound-free, and pseudo-continuum opacities. Our model
grid has been extended to the regime of cool, He-rich DA stars, and
now covers a range of effective temperature between $\Te=40,000$~K and
6000~K by steps of 1000~K, while $\logg$ ranges from 7.0 to 9.0 by
steps of 0.5 dex. In addition to pure helium models, we also
calculated models with $\logh=-7.0$ to $-1.5$ by steps of
0.5. Illustrative spectra are displayed in Figure 1 of
\citet{bergeron11} for various values of effective temperatures,
surface gravities, and convective efficiencies, while we show in
Figure \ref{spec_DBA_synth} synthetic spectra at $\logg=8$ for various
effective temperatures --- including our extension to low temperatures
--- and hydrogen-to-helium abundance ratios. Note that for the largest
hydrogen abundance shown in this plot ($\rm{H/He}=10^{-2}$), the star
would appear as a normal DB white dwarf at high effective
temperatures, but as a pure DA star below $\Te\sim 12,000$~K, when the
helium lines vanish.

Because of the complexity and degeneracy of the atmospheric parameter
space for DB stars, the values derived from various studies for
$\Te$, $\logg$, and H/He can differ significantly, as
illustrated, for instance, by \citet[][see their Figures 3 and
  4]{Voss07} who compare the results of the SPY survey with those of
independent studies (\citealt{Beauchamp99}; \citealt{FKCRW00};
\citealt{CKHK06}). One major reason for these discrepancies is the
lack of sensitivity of the neutral helium lines to effective
temperature in hotter DB stars. Indeed, for a given set of H/He, $\logg$, and
$\alpha$, the equivalent width of He~\textsc{i} $\lambda$4471 reaches
a plateau between 20,000~K and 30,000~K (see Figure 2 of
\citealt{bergeron11}); a similar behavior can be observed in our model
spectra displayed in Figure \ref{spec_DBA_synth}. As a result, two solutions
exist for a given DB star, one on each side of the maximum
strength of He~\textsc{i} $\lambda$4471. As discussed by
\cite{bergeron11}, this degeneracy can be lifted with the use of
spectroscopic observations at \ha, which add an additional constraint
to the solution.

\begin{figure}[bp]
\centering
\includegraphics[width=0.8\linewidth]{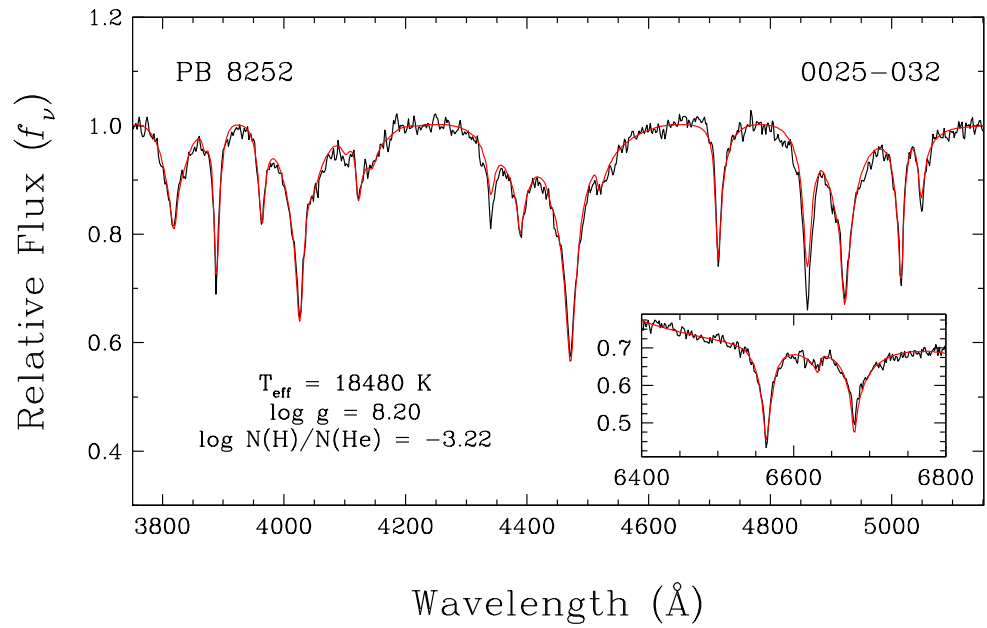}
\caption{Example of a full spectroscopic fit where the \ha\ line
  profile (inset) is used to measure, or constrain, the hydrogen
  abundance of the overall solution. \label{fits_DBA_exa}}
\end{figure}

Our fitting procedure is similar to that described at length in
\citet{bergeron11}.  Since in most cases our \ha\ spectra are
independent of our blue data, we first fit the blue spectrum with pure
helium models to obtain an estimate of $\Te$ and $\log g$. The
\ha~spectrum is then used to determine the hydrogen abundance --- or
upper limits on H/He --- at these particular values of $\Te$ and
$\logg$. The procedure is then repeated iteratively until an internal
consistency is reached. An example of our solution for PB 8252
(0025$-$032, HE 0025$-$0317) --- a new object added to the sample of
\citet{bergeron11} --- is displayed in Figure \ref{fits_DBA_exa}. The
\ha\ absorption feature shown in the inset serves as an important
constraint on the hydrogen abundance.  This fitting procedure is
reliable when the \ha~absorption line is present in the optical
spectrum. When no feature is visible, only upper limits on the
hydrogen abundance can be determined. We adopt a detection limit of
200 m\AA~for the equivalent widths of \ha\ based on the S/N of our DB
spectra. For high S/N spectra with no detectable hydrogen feature, the
value of H/He is set to the appropriate upper limit for the
corresponding temperature. In cases where the spectrum is noisier,
however, our fitting procedure may find an upper limit to the hydrogen
abundance that is larger than that inferred from this upper limit.

\subsubsection{Selected Results}\label{sec:res}

Because of the inclusion of additional objects and new
\ha\ spectroscopic data to our sample of bright DB stars, we present
here an update of the relevant results from \citet{bergeron11}. The
atmospheric parameters for all 115 DB and DBA white dwarfs in our
enlarged sample are provided in Table \ref{table_fits_DBA}; pure DB
stars in this table correspond to the objects where only an upper
limit on the hydrogen abundance is given. For completeness, we also
include the four objects listed in Section \ref{sec:specobs} that still
lack \ha\ data (these are noted in the table). For each star we give
the stellar mass ($M$) and white dwarf cooling age ($\log\tau$)
obtained from evolutionary models similar to those described in
\citet{FBB01} but with C/O cores, $\qhe\equiv \mhe/M_{\star}=10^{-2}$
and $\qh=10^{-10}$, which are representative of helium-atmosphere
white dwarfs\footnote{See
  http://www.astro.umontreal.ca/$\sim$bergeron/CoolingModels.}. The
absolute visual magnitude ($M_V$) and luminosity ($L$) are determined
using the improved calibration of \citet{HB06}, while the distance $D$
is obtained by combining $M_V$ with the magnitude $V$, also given in
the table. Since the presence or not of hydrogen features is crucial
to our understanding of the origin of DB stars, we provide as online
material our spectroscopic fits for all white dwarfs in our sample,
where the left panels show the blue portion of our spectroscopic fits,
while the right panels show the corresponding region near \ha.

\begin{figure}[bp]
\centering
\includegraphics[width=0.8\linewidth]{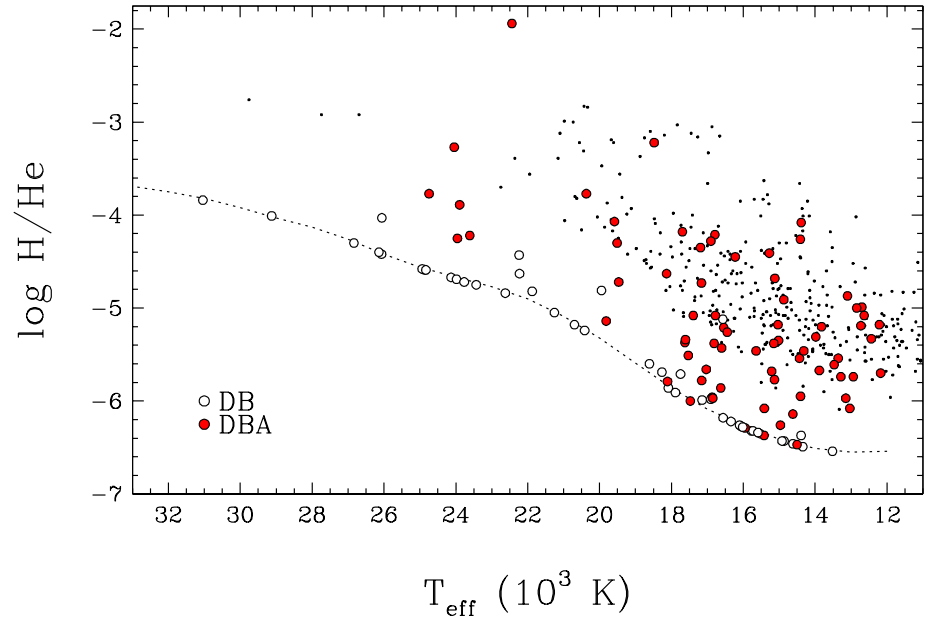}
\caption{Hydrogen-to-helium abundance ratio as a function of effective
  temperature for all DB (white symbols) and DBA (red symbols) white
  dwarfs in our sample. Limits on the hydrogen abundance set by our
  spectroscopic observations at \ha\ (200 m\AA\ equivalent width) are
  shown by the dotted line; the hydrogen abundances for DB stars
  thus represent only upper limits. Also shown as small dots are the
  results from \citet{KK15} for DBA white dwarfs (hydrogen detected) in the
  SDSS.\label{correlty}}
\end{figure}

The hydrogen abundances as a function of effective temperature for all
DB stars in our sample, but with \ha\ spectra available to us, are
displayed in Figure \ref{correlty}. Also shown are the upper limits on
the hydrogen abundance for DB stars, as determined from the absence of
\ha. In general, the pure DB stars are aligned on these observational
limits but as discussed above, some objects have noisier data and
these limits are simply not reached. The ratio of DBA stars to the
total number of white dwarfs in our sample now reaches 63\%,
significantly higher than the value of 44\% reported by
\citet{bergeron11}, thanks to our improved high S/N spectroscopic data
at \ha, which revealed the presence of hydrogen in objects where
\hb\ was spectroscopically invisible. This higher ratio now compares
favorably well with the value of 75\% reported by \citet{KK15} for the
highest S/N DB spectra in the SDSS, although this ratio drops to a
value of 32\% in their overall sample. The results of Koester \&
Kepler for the DBA white dwarfs from the SDSS --- i.e., with hydrogen
features detected --- are also reproduced in Figure
\ref{correlty}. Although both sets of hydrogen abundance
determinations overlap very nicely, the upper limits for the DB stars
in our sample are about 1 dex smaller due to the much higher S/N of
our observations compared to the SDSS spectra, as discussed above,
thus putting more severe constraints on the amount of hydrogen present
in the DB stars with no detectable \ha\ feature.

\begin{figure}[bp]
\centering
\includegraphics[width=0.8\linewidth]{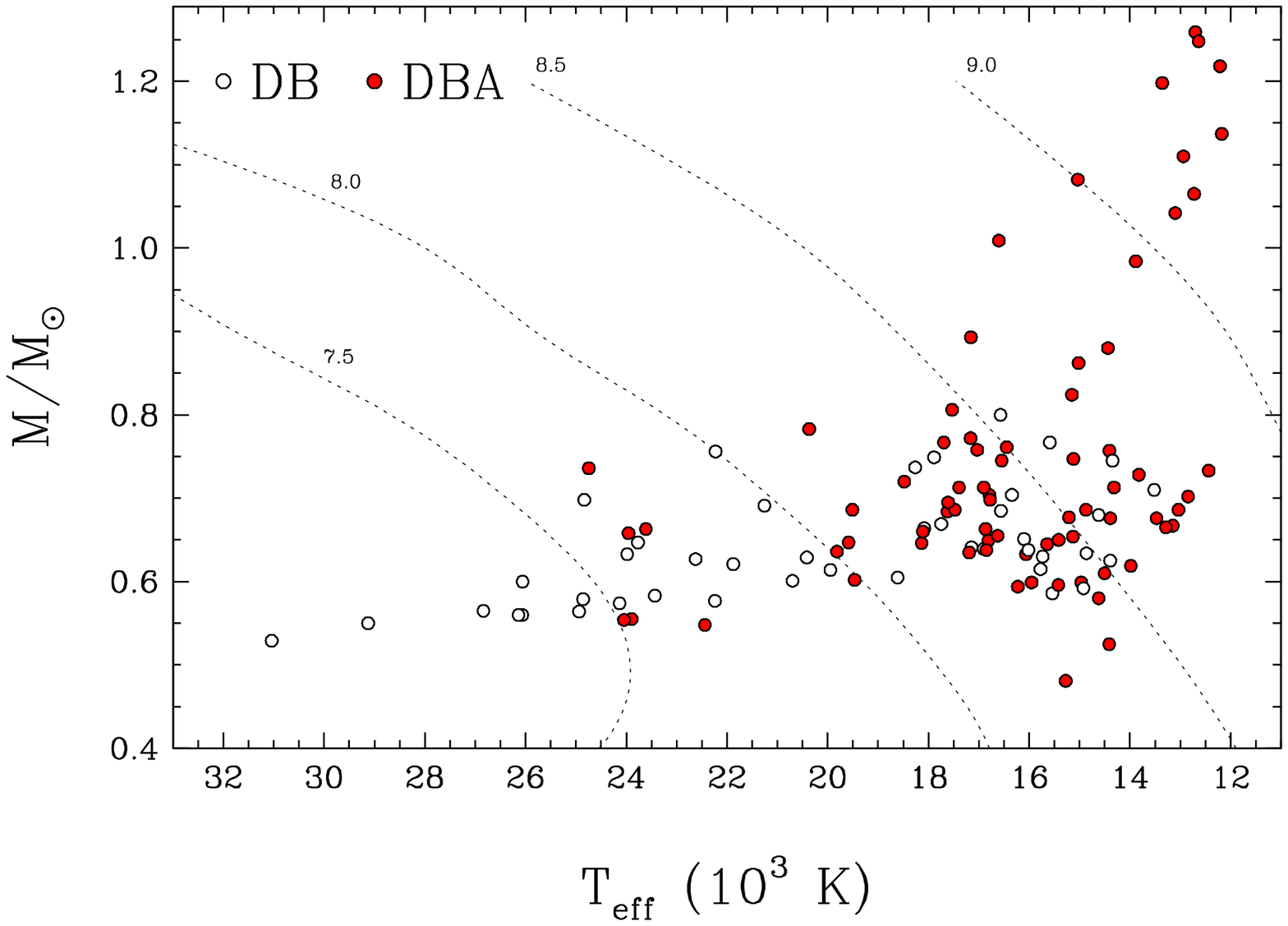}
\caption{Distribution of mass as a function of effective temperature
  for the 42 DB (white symbols) and 73 DBA (red symbols) white dwarfs
  in our sample. Also shown as dotted lines are the theoretical
  isochrones from our evolutionary models, labeled as $\log\tau$ where
  $\tau$ is the white dwarf cooling age in years.\label{correltm}}
\end{figure}

The mass distribution as a function of effective temperature is shown
in Figure \ref{correltm} for the same white dwarfs as in Figure
\ref{correlty}. These results are comparable to those displayed in
Figure 21 of \citet{bergeron11}, although all cool white dwarfs that
appeared massive in their analysis now {\it all show hydrogen
features}, while DB stars in our sample without detectable \ha\ have
normal masses. As discussed by Bergeron et al., the high masses
inferred for these cool DBA stars can probably be attributed to some
inaccurate treatment of van der Waals broadening in our models
\citep{BWBLS96}.  We can also see a definite trend for the bulk of
white dwarfs in our sample to show higher masses ($\sim$0.7 \msun) at
low effective temperatures than at the hot end of the sample
($\lesssim 0.6$ \msun).

\subsection{Hydrogen in cool, He-rich DA stars}

\subsubsection{Spectroscopic and Photometric Observations}

Because we are interested in studying the hydrogen abundance pattern
in all helium-atmosphere white dwarfs below $\Te\sim30,000$~K, we also
need to extend our search to the cool end of the white dwarf sequence
by including cool, helium-atmosphere DA and DZA stars such as L745-46A
and Ross 640 discussed in the Introduction, which are usually found in
the $\sim$8000--12,000~K temperature range. In general, the only
hydrogen line visible in their spectra is \ha, which appears very
broad and shallow as a result of van der Waals broadening. These
He-rich DA white dwarfs can be easily distinguished from the
much cooler ($\Te\lesssim6000$~K), hydrogen-atmosphere DA stars in
which the \ha\ absorption feature is much sharper (see, e.g., Figure
23 of \citealt{BRL97}). The latter are also much redder
photometrically.

With this idea in mind, we searched the Data Release 7 from the SDSS
catalog of \citet{SDSS_DR7} with $(g-r)<0.5$, $(u-g)<0.8$, and a
magnitude cut-off of $g<19.5$, and retained 28 He-rich DA white dwarf
candidates, previously identified by P.~Dufour (private
communication). This sample is not complete in any sense, but it is
considered representative of the hydrogen abundance pattern in cool,
helium-rich atmospheres. These cool, He-rich DA stars can be analyzed
using a hybrid photometric and spectroscopic technique, described in
the next section, which requires $ugriz$ photometry as well as optical
spectra in the region around \ha. The $ugriz$ photometry for our 28
SDSS white dwarfs was taken from \cite{SDSS_DR7} while the
corresponding spectra were retrieved from the SDSS database; these
cover the 4000--9200 \AA\ wavelength range, with an average
signal-to-noise ratio of $\rm{S/N}\sim15$.

\subsubsection{Model Atmosphere Analysis}

An examination of the theoretical spectra displayed in Figure
\ref{spec_DBA_synth} reveals that the number of hydrogen and helium
absorption lines that can be detected in the optical spectra of cool
($\Te\lesssim12,000$~K), helium-dominated atmospheres becomes
increasingly small. As a result, the spectroscopic technique alone
fails to yield reliable measurements of the atmospheric parameters for
such white dwarfs.  In order to overcome this problem, we developed a
hybrid approach that relies on both photometry and spectroscopy near
the \ha\ region (see also \citealt{Giamm12}). The first step is based
on the photometric technique developed by \citet{BRL97,BLR01}, where
the observed magnitudes are converted into average fluxes and compared
to the predictions of model atmosphere calculations. Briefly, every
magnitude $m_\lambda$ is transformed into an average flux
$f_\lambda^m$ using the relation

\begin{equation}
m_\lambda=-2.5\log f_\lambda^m+c_m
\end{equation}

\noindent where 

\begin{equation}
\label{eq:avgflux}
f_\lambda^m=\frac{\int_{0}^{\infty}f_\lambda S_m\left(\lambda\right)\lambda\,d\lambda}{\int_{0}^{\infty}S_m\left(\lambda\right)\lambda\,d\lambda}\ ,
\end{equation}

\noindent $f_\lambda$ is the monochromatic flux received at Earth from the star,
$c_m$ is a zero point calibration constant, and
$S_m\left(\lambda\right)$ is the transmission function of the
corresponding bandpass. The zero points and transmission functions are
taken from \citet{HB06}, where an expression similar to that above
is also provided for the SDSS $ugriz$ photometric system (AB$_{95}$).
These average observed fluxes can then be compared
with the model predictions using the relation

\begin{equation}
\label{eq:avgflux2}
f_\lambda^m=4\pi\left(R/D\right)^2H_\lambda^m
\end{equation}

\noindent where $R/D$ is the ratio of the radius of the star to its
distance from Earth, and $H_\lambda^m$ is the average model flux
obtained by substituting $f_\lambda$ in Equation
\ref{eq:avgflux} for the monochromatic Eddington fluxes $H_\lambda$,
which depend on the atmospheric parameters $\Te$, $\logg$,
and H/He.

The fitting procedure relies on the Levenberg-Marquardt nonlinear
least-square method where the $\chi^2$ value is taken as the sum over
all bandpasses of the difference between both sides of Equation
\ref{eq:avgflux2}, properly weighted by the corresponding
observational uncertainties.  Only $\Te$ and the solid angle
$\pi\left(R/D\right)^2$ are considered free parameters at a fixed
value of the hydrogen abundance.  In principle, trigonometric parallax
measurements can be used to constrain the $\logg$ value, but since no
such data is available for our sample of cool, He-rich DA white dwarfs
from the SDSS, we simply assume $\logg=8$ throughout.  When a
satisfactory fit to the energy distribution is reached at some initial
value of the hydrogen abundance, the resulting values of
$\Te$ and $\logg$ are then used to measure the
hydrogen-to-helium abundance ratio (H/He) by fitting the
\ha\ spectroscopic data using the same fitting procedure as that
described above for the DBA white dwarfs. The entire procedure is then
repeated iteratively until a consistent set of $\Te$, $\log
g$, and H/He values is reached. Because the limits on the hydrogen
abundance depend on the S/N of the observations at \ha, and that the
average S/N of our cool white dwarf sample from the SDSS is $\sim$5.4
times lower than the average of our hotter DB sample, we adopt a
detection limit of 1100 m\AA\ for the equivalent width of \ha.

\subsubsection{Selected Results}\label{sec:sr}

Our fits to the 28 cool, He-rich DA white dwarfs in the SDSS sample
are displayed in Figure \ref{fits_DCA_all}.  The main panels show the
photometric fits to the observed $ugriz$ energy distributions, while
the insets show the corresponding spectroscopic fits in the region
covering \ha~and He {\sc i} $\lambda$5876. Since it also possible that
some of these objects could be unresolved DA+DC white dwarf binaries,
we also made sure that our solutions are consistent with the observed
spectra in the blue portion of the spectrum (H$\beta$ and blueward),
within the signal-to-noise of the observations.  The atmospheric
parameters for all objects, assuming a value of $\logg=8$, are
provided in Table \ref{table_fits_DCA}, together with the same
information as in Table \ref{table_fits_DBA}, with the exception that
the absolute magnitude is given here for the $g$ filter ($M_g$). We
also added a note for possible DA+DC systems.

\begin{figure}[bp]
\centering
\includegraphics[width=0.8\linewidth]{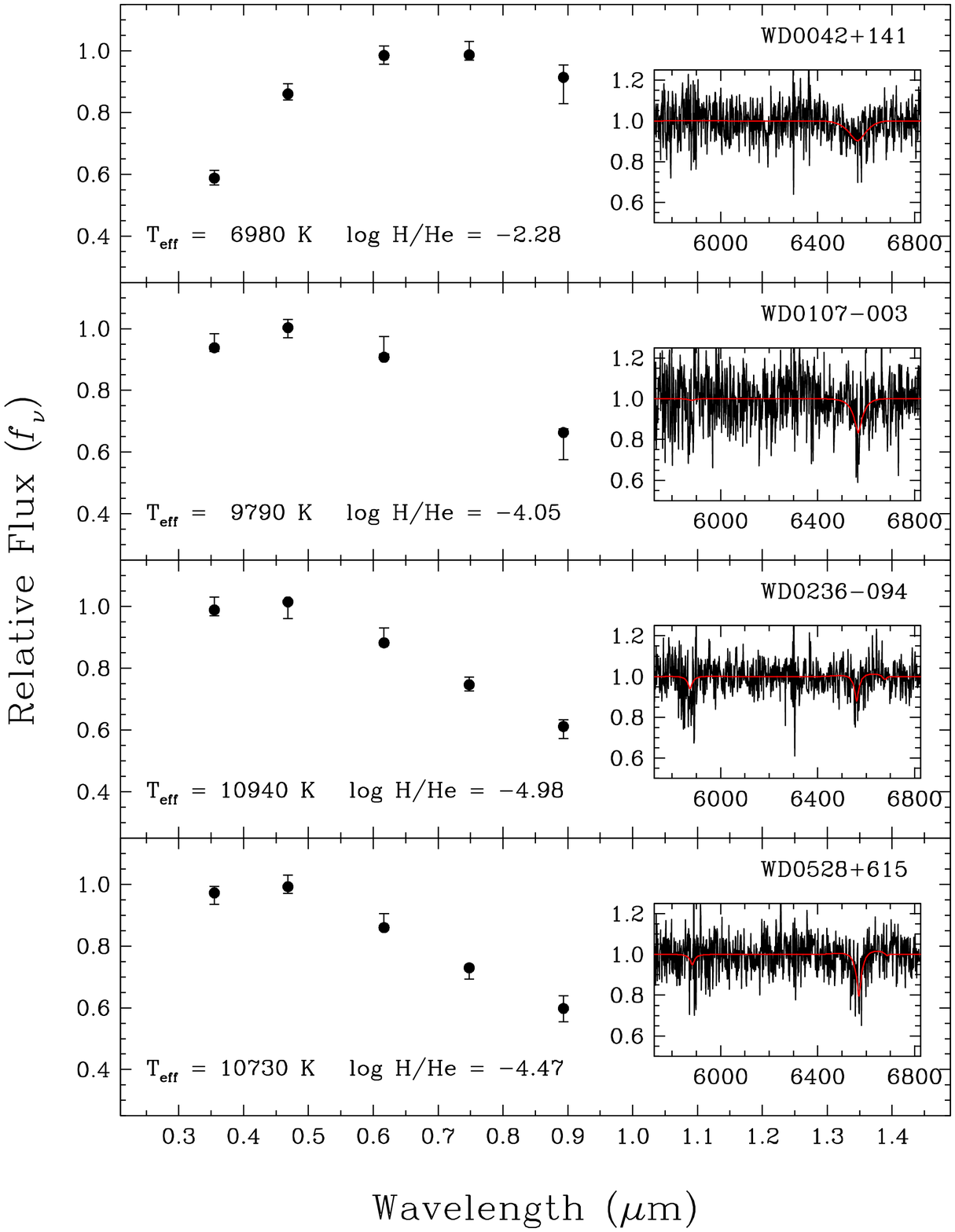}
\caption{Fits obtained from the hybrid fitting method for cool,
  He-rich DA stars with homogeneous models at $\logg=8$. The $ugriz$
  photometric observations are represented by error bars, while the
  model fluxes are shown by filled circles. The inset shows our fit to
  the spectrum near the \ha~region, normalized to a continuum set to
  unity, which is used to measure the hydrogen
  abundance.\label{fits_DCA_all}}
\end{figure}

\begin{figure}[bp]
\figurenum{7}
\centering
\includegraphics[width=0.8\linewidth]{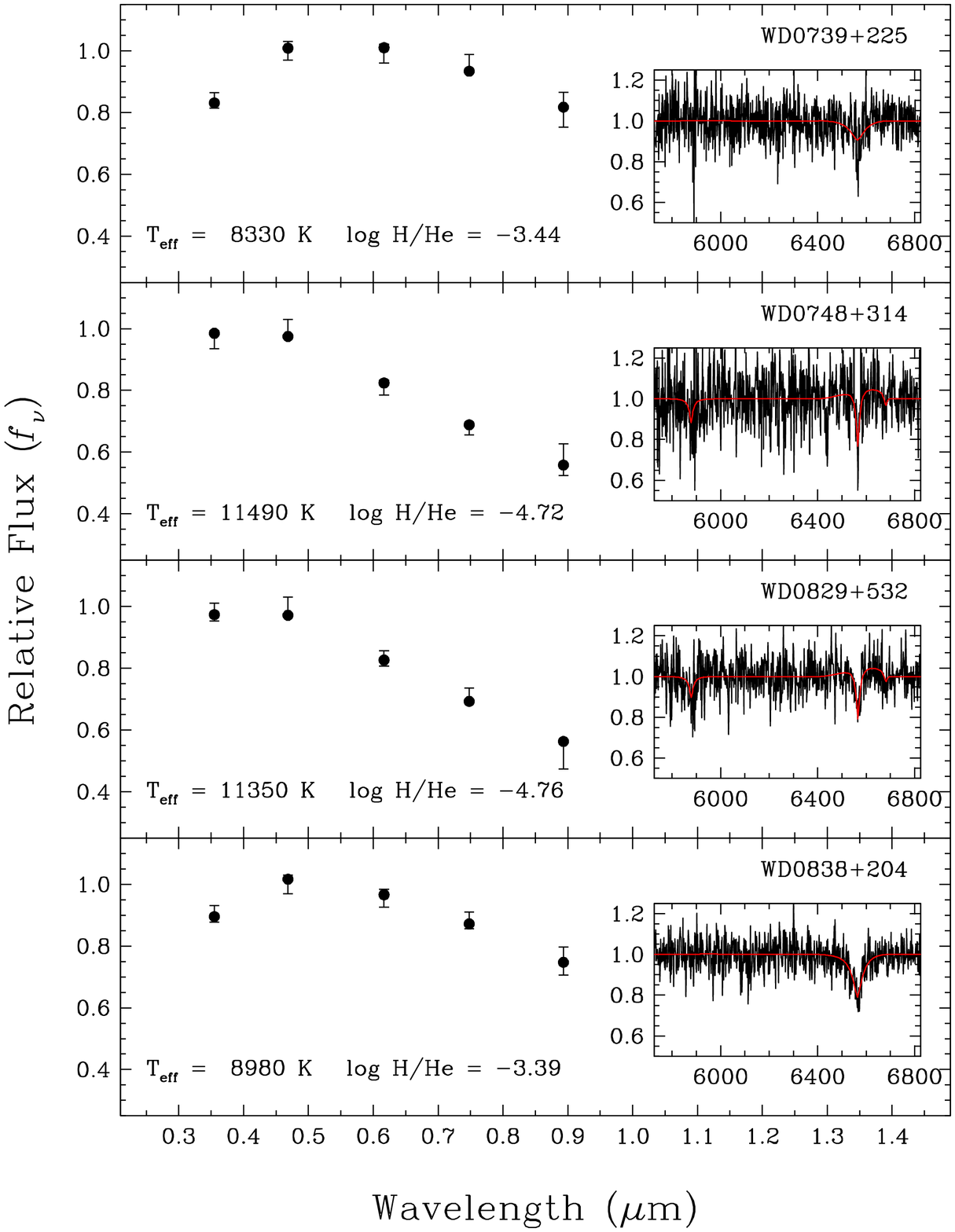}
\caption{(Continued)}
\end{figure}

\begin{figure}[bp]
\figurenum{7}
\centering
\includegraphics[width=0.8\linewidth]{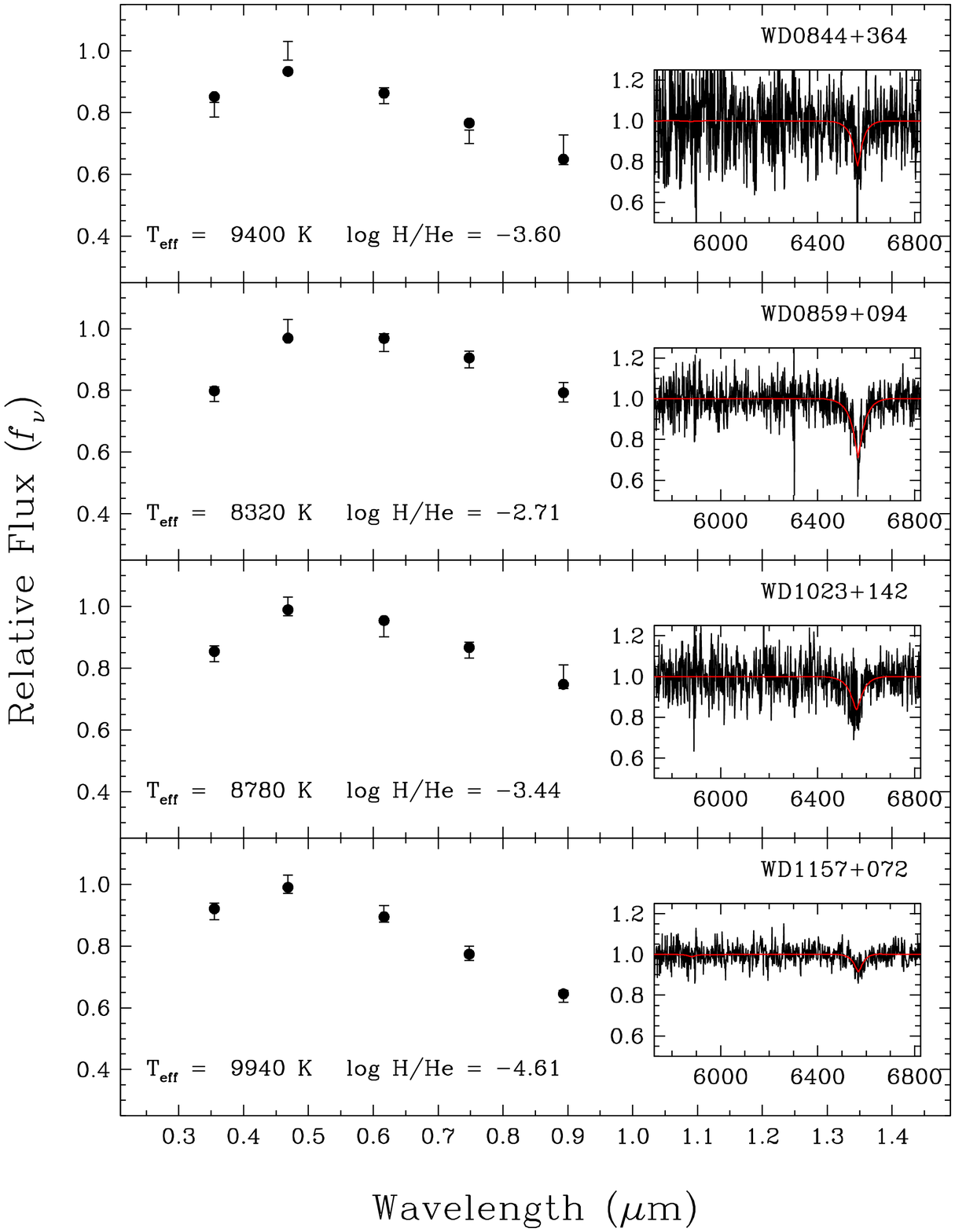}
\caption{(Continued)}
\end{figure}

\begin{figure}[bp]
\figurenum{7}
\centering
\includegraphics[width=0.8\linewidth]{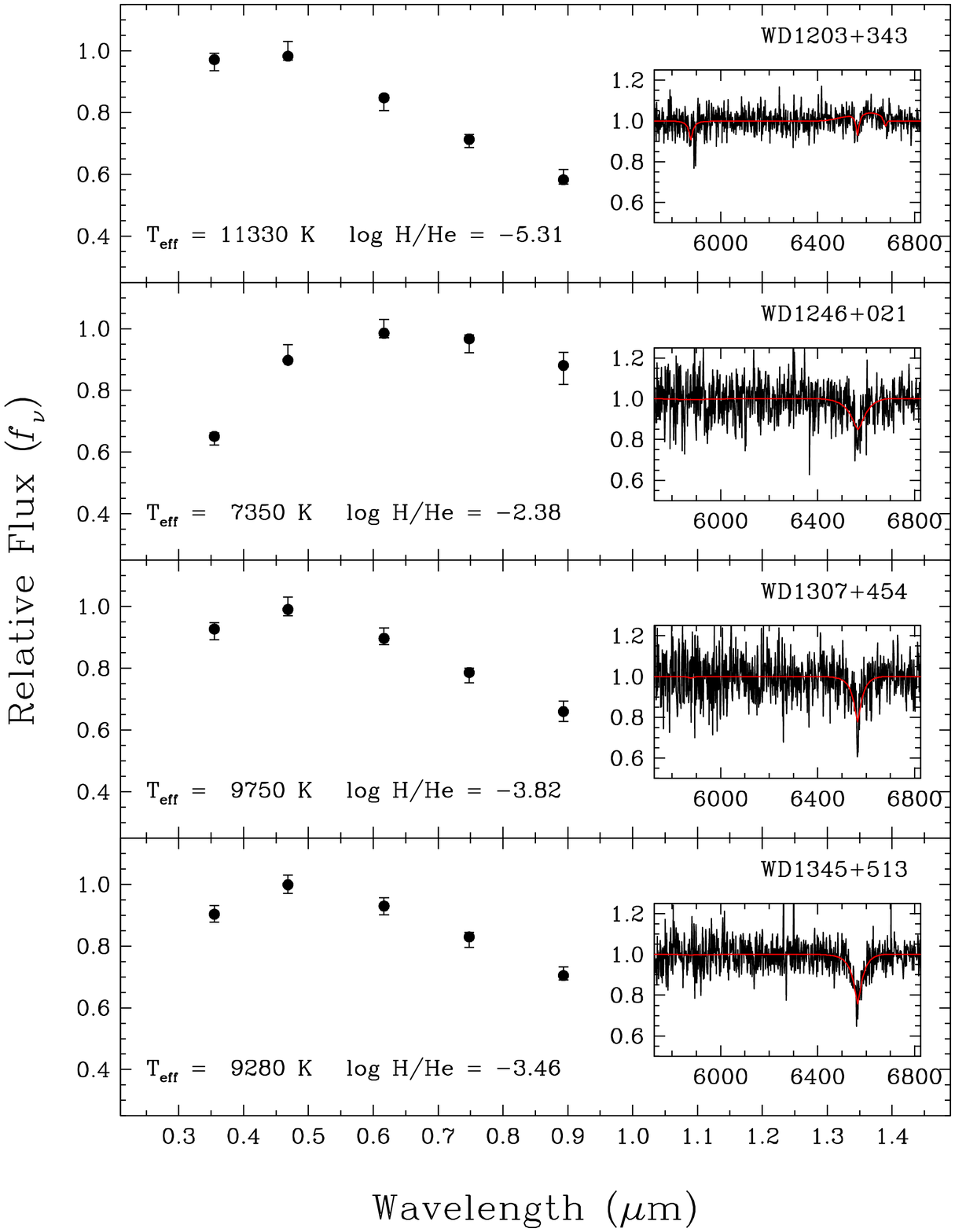}
\caption{(Continued)}
\end{figure}

\begin{figure}[bp]
\figurenum{7}
\centering
\includegraphics[width=0.8\linewidth]{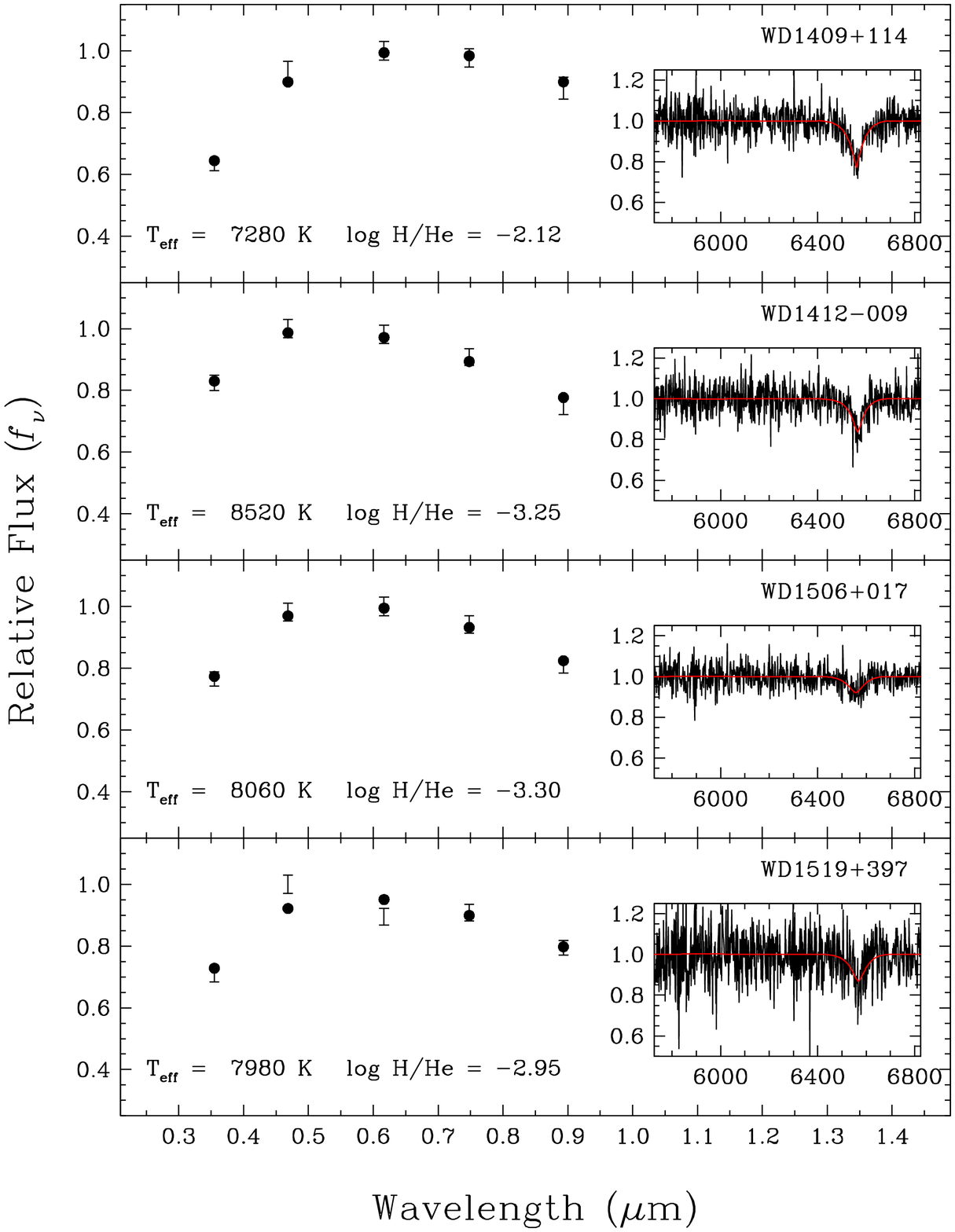}
\caption{(Continued)}
\end{figure}

\begin{figure}[bp]
\figurenum{7}
\centering
\includegraphics[width=0.8\linewidth]{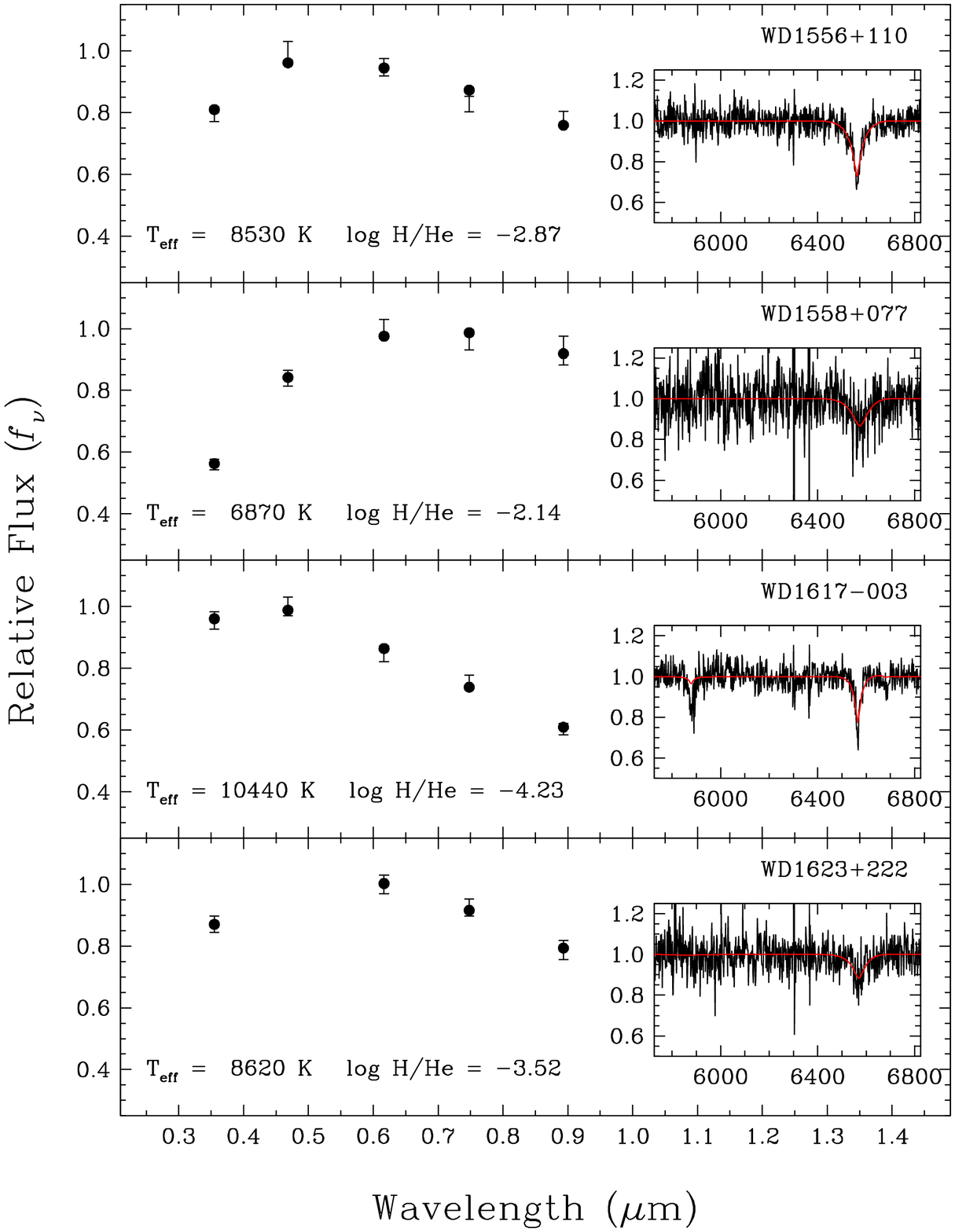}
\caption{(Continued)}
\end{figure}

\begin{figure}[bp]
\figurenum{7}
\centering
\includegraphics[width=0.8\linewidth]{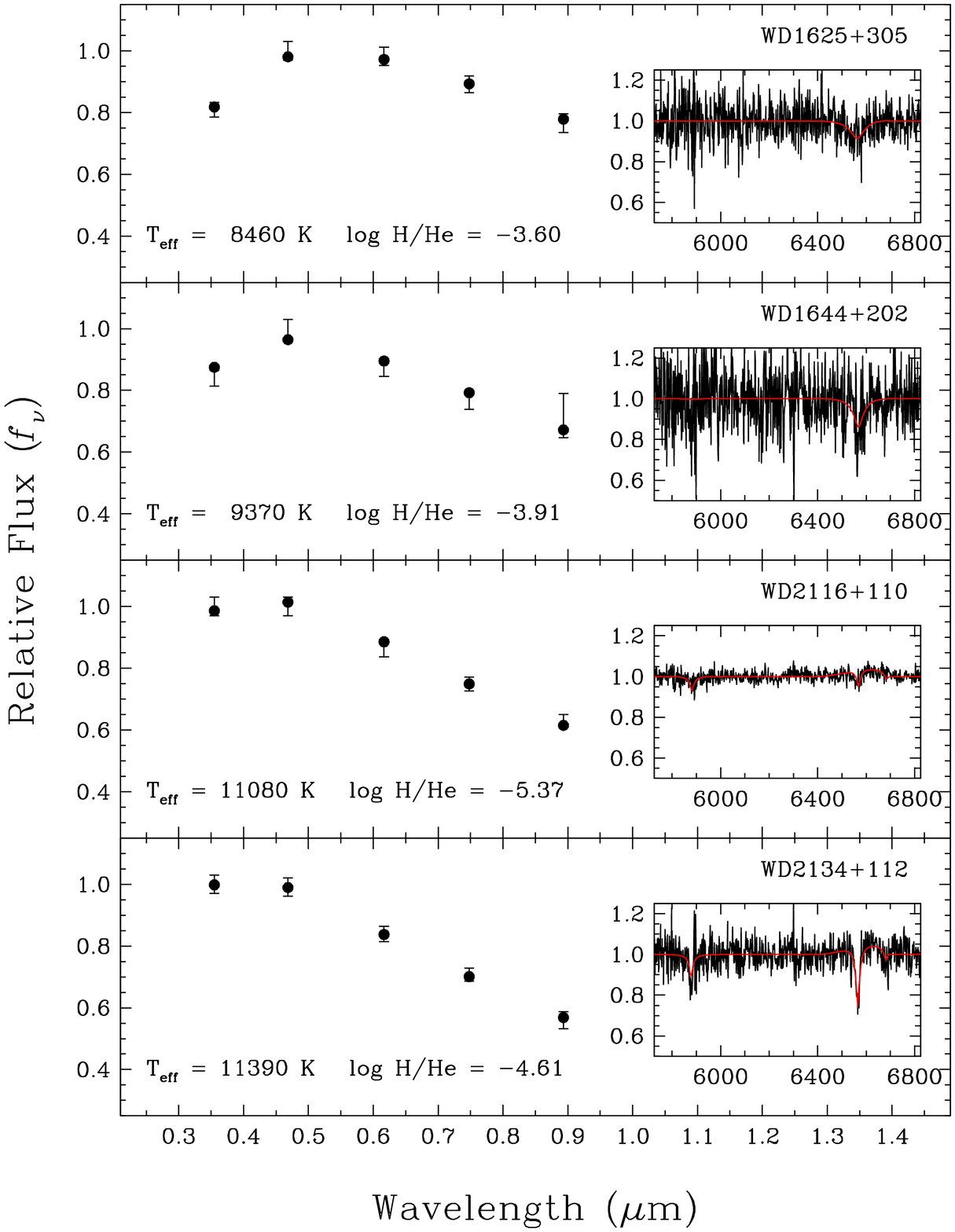}
\caption{(Continued)}
\end{figure}

\begin{figure}[bp]
\centering
\includegraphics[width=0.8\linewidth]{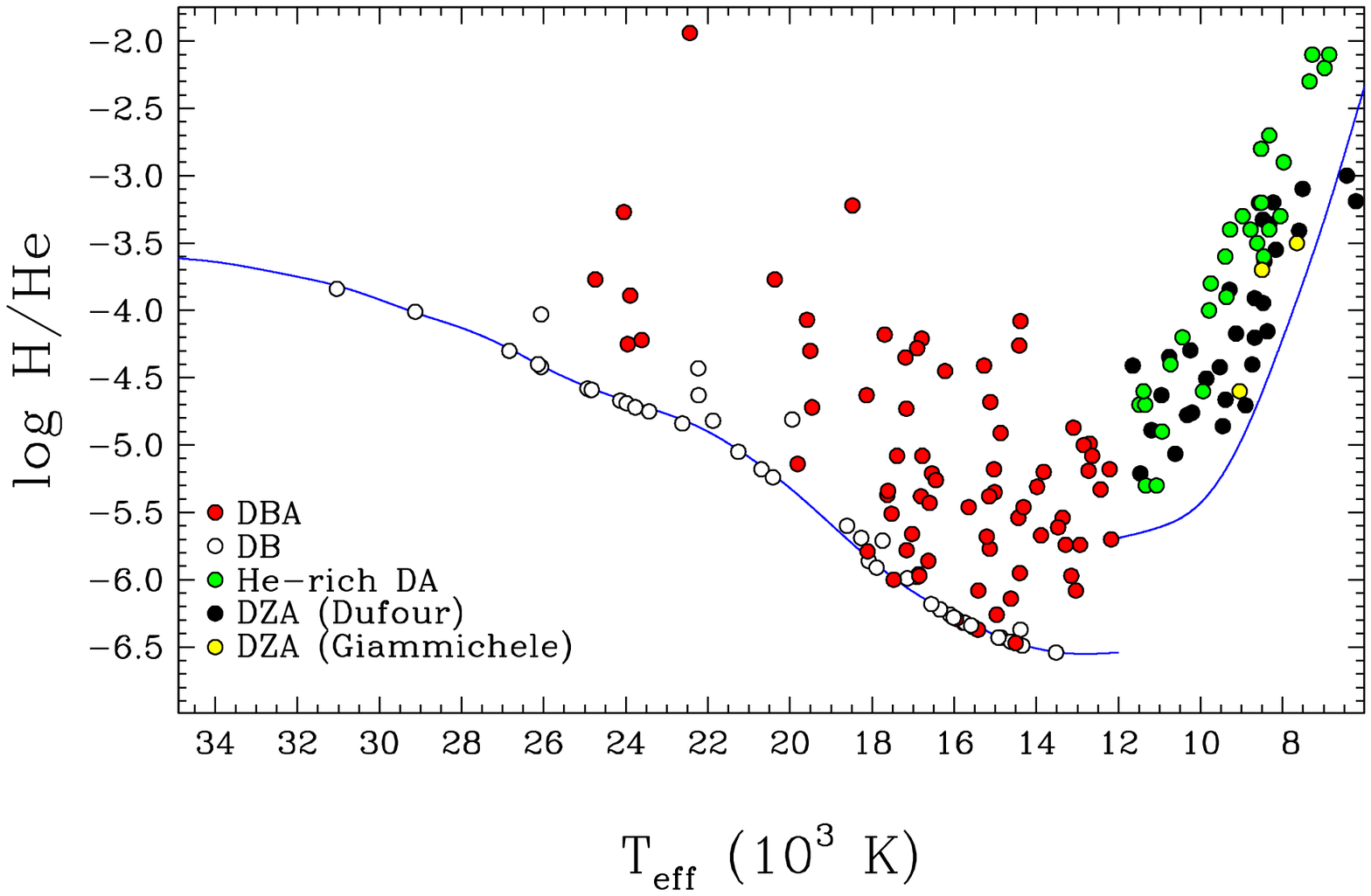}
\caption{Hydrogen-to-helium abundance ratio as a function of effective
  temperature for all DB, DBA, and cool, He-rich DA white dwarfs in
  our SDSS sample; the DZA stars from \cite{dufour07} and
  \cite{Giamm12} are also displayed. The hydrogen detection
  limits at \ha\ are indicated by blue lines for the subsamples above
  and below $\Te=12,000$~K. \label{correlty_all}}
\end{figure}

The hydrogen abundances as a function of effective temperature for all
cool, He-rich DA white dwarfs in our sample are displayed in Figure
\ref {correlty_all}, together with our spectroscopic results for the
DB and DBA stars; also shown are the limits on the hydrogen abundance
for both subsamples. For completeness, we reproduce in this figure the
results of \cite{dufour07} for the DZA white dwarfs from the SDSS, for
which the hydrogen abundances have been determined spectroscopically,
as well as the three bright DZA stars from \citet{Giamm12}. The
location of our cool, He-rich DA stars and DZA white dwarfs in this
plot indicates that these two populations are very alike. They cover
essentially the same range of hydrogen abundances, with a similar
dispersion, and most importantly, they display the same behavior with
respect to effective temperature.  Note that cool, He-rich DA/DZA
white dwarfs most certainly exist below the detection threshold at
\ha\ displayed in Figure \ref {correlty_all}, although objects with
very large hydrogen abundances have not been found in our
analysis. Since hydrogen has been detected --- or inferred --- in 27\%
of the DZ stars analyzed by \cite{dufour07}, our results suggest, as a
conservative estimate, that the cool, He-rich DA white dwarfs
represent around 25\% of the total DC population below $\sim$12,000~K,
at least in the range of temperature where \ha\ can be detected in
helium-rich atmospheres ($\Te\gtrsim 6000$~K).

The results displayed in Figure \ref {correlty_all} represent the best
picture we have so far of the hydrogen abundance pattern in
helium-atmosphere white dwarfs below $\Te\sim30,000$~K, both in terms
of the quality of the photometric and spectroscopic data currently
available to us, but also in terms of model atmospheres and fitting
techniques. Any viable model of the spectral evolution of white dwarfs
involving the convective dilution --- or mixing --- of the thin
hydrogen layer with the deeper helium envelope, or accretion from the
interstellar medium or other external bodies, must be able to account
self-consistently for the observed hydrogen abundance pattern depicted
in Figure \ref {correlty_all}.

\section{MODEL ENVELOPE STRUCTURES}\label{sec:models}

In the absence of competing mechanisms, gravitational settling would
gradually make the hydrogen present in DBA white dwarfs and cool,
He-rich DA/DZA stars float up to the surface, resulting in the
creation of a hydrogen-dominated atmosphere in a time frame much
smaller than the typical white dwarf cooling time. In the temperature
range considered here ($\Te\lesssim30,000$~K), however, convective
energy transport within the thick helium envelope is the main
mechanism competing with element diffusion, and hydrogen is thus
expected to be thoroughly mixed within the helium convection zone,
resulting in a helium-dominated atmosphere with a homogeneous H/He
abundance profile. As discussed in the Introduction, the progenitors
of some of the DB white dwarfs are probably DA stars with sufficiently
thin radiative hydrogen layers, of the order of $\mh\sim 10^{-15}$
\msun, which are transformed into helium-atmosphere white dwarfs as a
result of the {\it convective dilution} of this thin hydrogen layer
with the deeper and more massive convective helium envelope. Cool
($\Te\lesssim12,000$~K), helium-atmosphere white dwarfs that show only
traces of hydrogen (and sometimes metals), on the other hand, could be
interpreted as DBA stars that have simply cooled off, or
alternatively, they could also be interpreted as the result of
convectively mixed DA stars, when the bottom of the hydrogen
convection zone in a DA white dwarf eventually reaches the underlying
and more massive convective helium envelope, resulting in the {\it
  convective mixing} of the hydrogen and helium layers, a process
believed to occur when the hydrogen layer mass is in the range
$\mh\sim 10^{-14}-10^{-6}$ \msun.

The structure of both types of DA progenitors described above is thus
a thin hydrogen layer --- convective or not --- sitting on top of a
massive helium envelope. To describe these structures, as well as the
homogeneously mixed hydrogen and helium structures, we make use of the
latest version of the Montr\'eal white dwarf model-building code in its
envelope mode (see \citealt{brassard94} for a first description). 
This code uses the same input physics as the full evolutionary models
described at length in \citet{FBB01}, but with updates discussed briefly
in \citet{noemi16}. Models with both homogeneously mixed and
chemically stratified compositions have been computed for a large set of
effective temperatures, stellar masses, and assumed convective
efficiencies. These are described in turn.

\subsection{Homogeneously Mixed Composition Models}\label{sec:homo}

Our static, homogeneously mixed models are characterized by $q({\rm
  env}) \equiv M_{\rm env}/M_{\star}=10^{-2}$ --- which are
representative of helium-rich white dwarfs --- with a homogeneous
hydrogen and helium abundance profile from the surface to the bottom
of the stellar envelope. Note that the presence of hydrogen uniformly
distributed below the mixed H/He convection zone may not be very
realistic, but, importantly, it does not affect in any way the
location of the bottom of this convection zone in our structures. We
use these models below to compute the mass of hydrogen contained in
the convection zone only. Our homogeneous grid covers a range of
effective temperature between $\Te=60,000$~K and 30,000~K by steps of
500~K, and between $\Te=30,000$~K and 4000~K by steps of 50~K. The
hydrogen mass fraction in the envelope ranges from $\log X=-8.5$ to
$-2.5$ by steps of 0.5 dex, and from $\log X=-2.50$ to $0.35$ with a
varying mesh between 0.20 or 0.25 dex. We also explore the so-called
ML2/$\alpha=0.6$ and $\alpha=2$ parameterizations of the mixing-length
theory to treat convective energy transport. These two values bracket
the convective efficiencies mostly used in the context of white dwarf
atmospheres and envelopes (see, e.g., \citealt{TFW90,tremblay15}).

\begin{figure}[bp]
\centering
\includegraphics[width=0.8\linewidth]{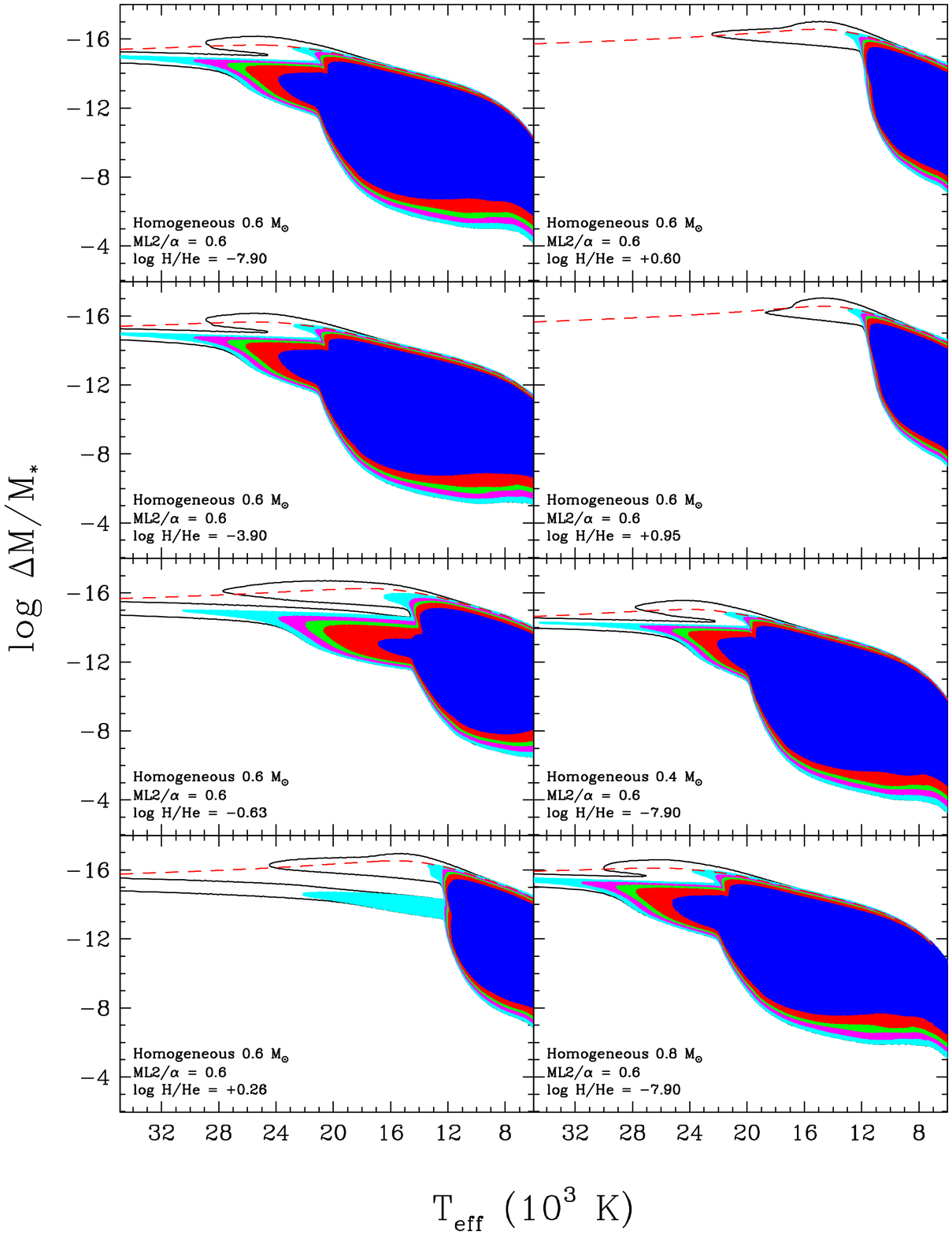}
\caption{Examples of envelope structures for white dwarf models with
  homogeneously mixed compositions as a function of effective
  temperature. The depth is expressed as the fractional mass above the
  point of interest with respect to the total mass of the star.  The
  models illustrated here are (from upper left to bottom right) for
  0.6 \msun\ with increasing hydrogen abundances, with the exception
  of the last two panels that show the results at 0.4 and 0.8
  \msun\ for an almost pure helium composition, and they all assume a
  ML2/$\alpha=0.6$ parameterization of the convective efficiency.  The
  red dashed line corresponds to the location of the photosphere, while
  the contours with various colors represent the convection zones with
  0.1, 1, 50, 75, 85, and 95\% of the total flux being transported by
  convection.
\label{envelopes_hom_ML2}}
\end{figure}

\begin{figure}[bp]
\centering
\includegraphics[width=0.8\linewidth]{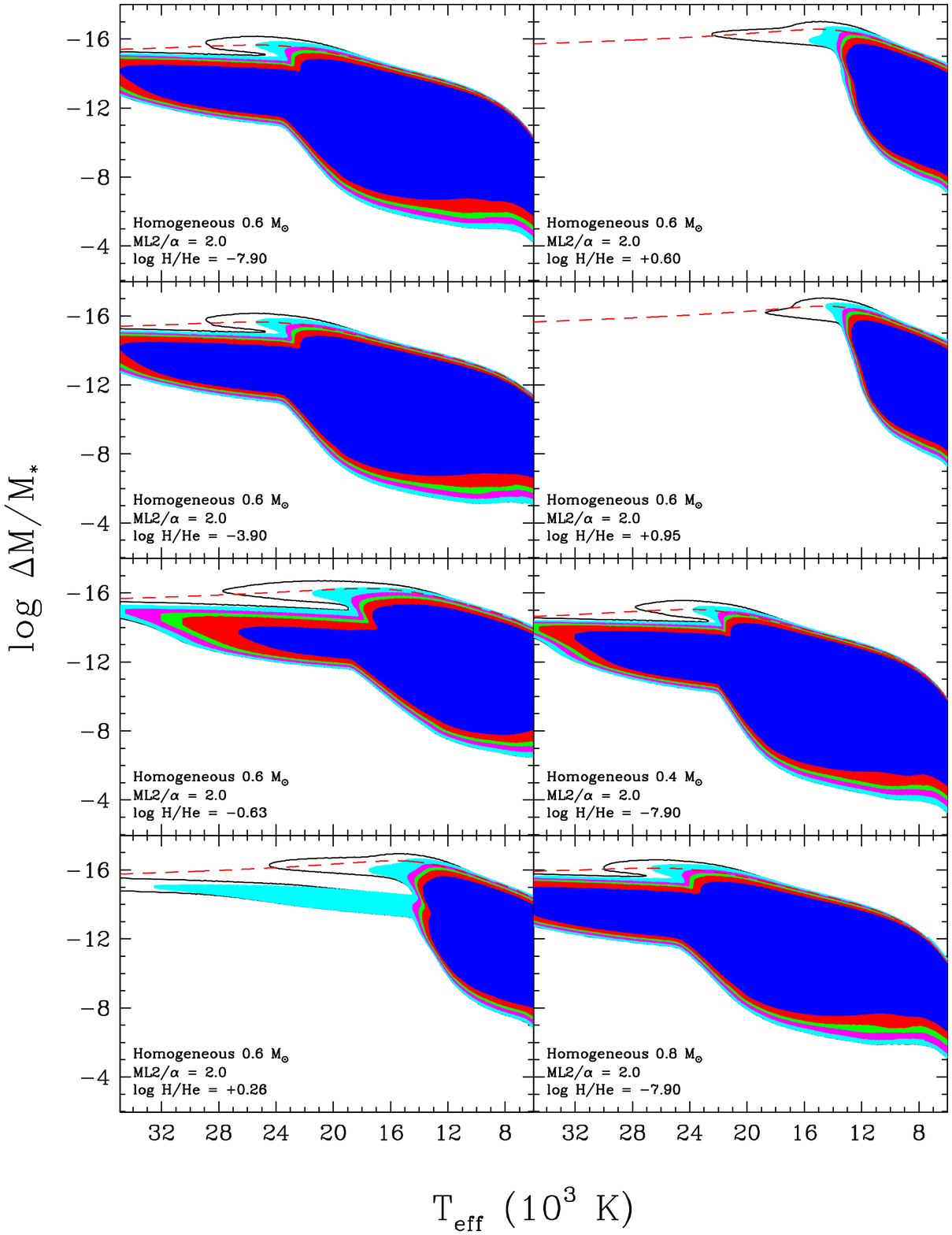}
\caption{Same as Figure \ref{envelopes_hom_ML2} but for models
  assuming a ML2/$\alpha=2$ parameterization of the convective
  efficiency.
\label{envelopes_hom_ML3}}
\end{figure}

Envelope structures for various hydrogen-to-helium abundance ratios
(H/He), stellar masses, and convective efficiencies, are displayed in
Figures \ref{envelopes_hom_ML2} and \ref{envelopes_hom_ML3} for
ML2/$\alpha=0.6$ and $\alpha=2$ models, respectively. The extent of
the convection zones (color coded with the fraction of the total flux
carried out by convection) as well as the location of the photosphere
are indicated in each panel. The smallest and largest hydrogen
abundances illustrated here correspond to almost pure helium and pure
hydrogen compositions, respectively, in terms of their structures. In
the most helium-rich models with ML2/$\alpha=0.6$, a small convection
zone is present at high temperatures, due to the partial ionization of
He~\textsc{ii}, but with only $\sim$1\% of the flux transported by
convection; the depth of this convection zone is significantly larger
with $\alpha=2$, with a much larger fraction of the total flux being
transported by convection. Below $\Te\sim28,000$~K, a second, more
superficial, convection zone appears, due this time to the partial
ionization of He~\textsc{i}. Eventually, both convection zones merge
below $\sim$25,000~K. As the hydrogen content is increased, the
temperature at which these two convection zones merge decreases, and
another convection zone develops due to the partial ionization of
hydrogen (mixed with the superficial helium convection zone), which
appears as a bump at the top of the convection zone near 16,000 K in
the models with $\logh\gtrsim+0.26$ shown in Figures
\ref{envelopes_hom_ML2} and \ref{envelopes_hom_ML3}. In addition, we
note that the increase in hydrogen content gradually delays the
development of the deep, mixed H/He convection zone.  At the largest
hydrogen abundances illustrated here, the convection zone due to the
partial ionization of He~\textsc{ii} at high temperatures is also
totally suppressed. Finally, in the range of effective temperatures
and hydrogen abundances where DBA white dwarfs are found in our sample
($\Te\sim 12,000-30,000$~K, $\logh<-4$), the structure of the helium
convection zone remains unaffected by the presence of hydrogen. In
this particular temperature range, hydrogen starts to affect the
structure of the convection zone only above $\logh\sim-3.4$.

For the cooler envelope models ($\Te\lesssim10,000$~K) displayed in
Figures \ref{envelopes_hom_ML2} and \ref{envelopes_hom_ML3} --- more
representative of the cool, He-rich DA stars analyzed above --- the
situation is somewhat different. First, we note that the top of the
convection zone, which coincides with the location of the photosphere,
becomes increasingly deeper in cooler models as a result of the
decrease in total opacity. Since neutral helium is particularly
transparent with respect to hydrogen at low temperatures ($\Te\lesssim
6000$~K), the location of the photosphere and the top of the
convection zone are orders of magnitude deeper (when expressed in
fractional mass) in the cool, hydrogen-poor models than in the
hydrogen-rich models. The effect on the location of the bottom of the
mixed H/He convection zone is also significant.

As discussed above, the behavior of the ML2/$\alpha=2$ models
displayed in Figure \ref{envelopes_hom_ML3} are qualitatively similar
to the ML2/$\alpha=0.6$ models, with the notable exception that at a
given effective temperature, the mixed H/He convection zone extends
significantly deeper in the star. This implies that for a given
hydrogen-to-helium abundance ratio observed at the photosphere, a
larger hydrogen mass will be inferred using models with more efficient
convection. Below roughly 10,000~K, however, convection becomes
adiabatic and both sets of envelope structures are identical at the
bottom of the convection zone. We also note that the assumed
convective efficiency has a negligible effect at the surface. Finally,
an examination of our almost pure helium models at 0.4 and 0.8
\msun\ reveals that the He~\textsc{ii} convection zone starts to
plunge into the star at higher temperatures in more massive white
dwarfs --- $\Te\sim30,000$~K at 0.8 \msun\ compared to $\sim$25,000~K
at 0.4 \msun\ --- but the convection zone in the more massive models
does not extend as deep below $\sim$20,000~K. Since these differences
remain small, we find it reasonable to assume only 0.6 \msun\ models
in our discussion of the various evolutionary scenarios described
below.

\subsection{Stratified Composition Models}

\begin{figure}[bp]
\centering
\includegraphics[width=0.8\linewidth]{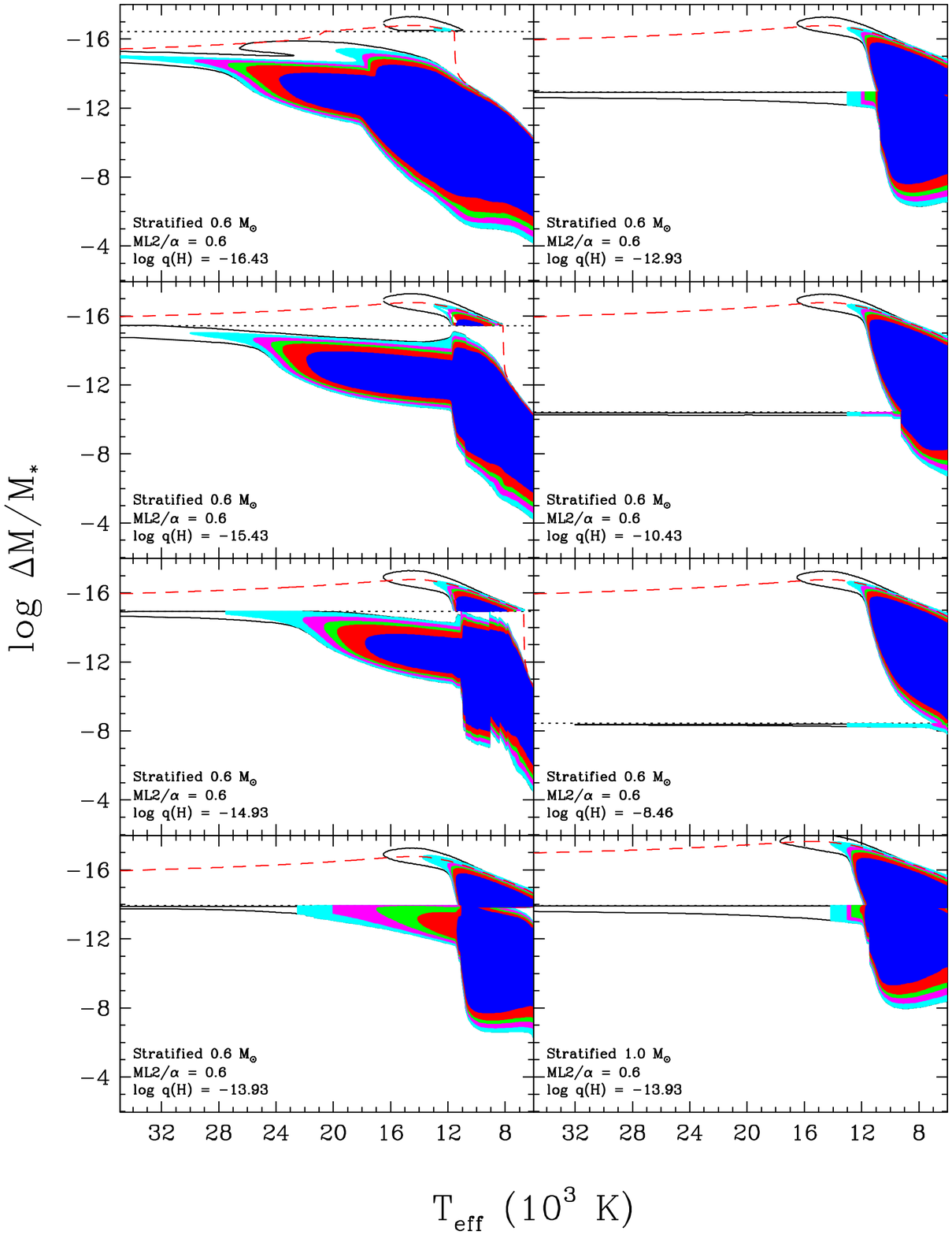}
\caption{Same as Figure \ref{envelopes_hom_ML2} but for stratified
  models. The models illustrated here are for 0.6 \msun\ with
  increasing thickness of the hydrogen layer (from upper left to
  bottom right) expressed as $\logqh=\log\mh/M_\star$, with the
  exception of the bottom right panel that shows the results at 1.0
  \msun\ and $\logqh=-13.93$ --- i.e., the same value of $\logqh$ as
  the panel to the left. The value of $\logqh$ is indicated by a
  black dotted line in each panel. The results shown here assume a
  ML2/$\alpha=0.6$ parameterization of the convective efficiency.
\label{envelopes_str_ML2}}
\end{figure}

\begin{figure}[bp]
\centering
\includegraphics[width=0.8\linewidth]{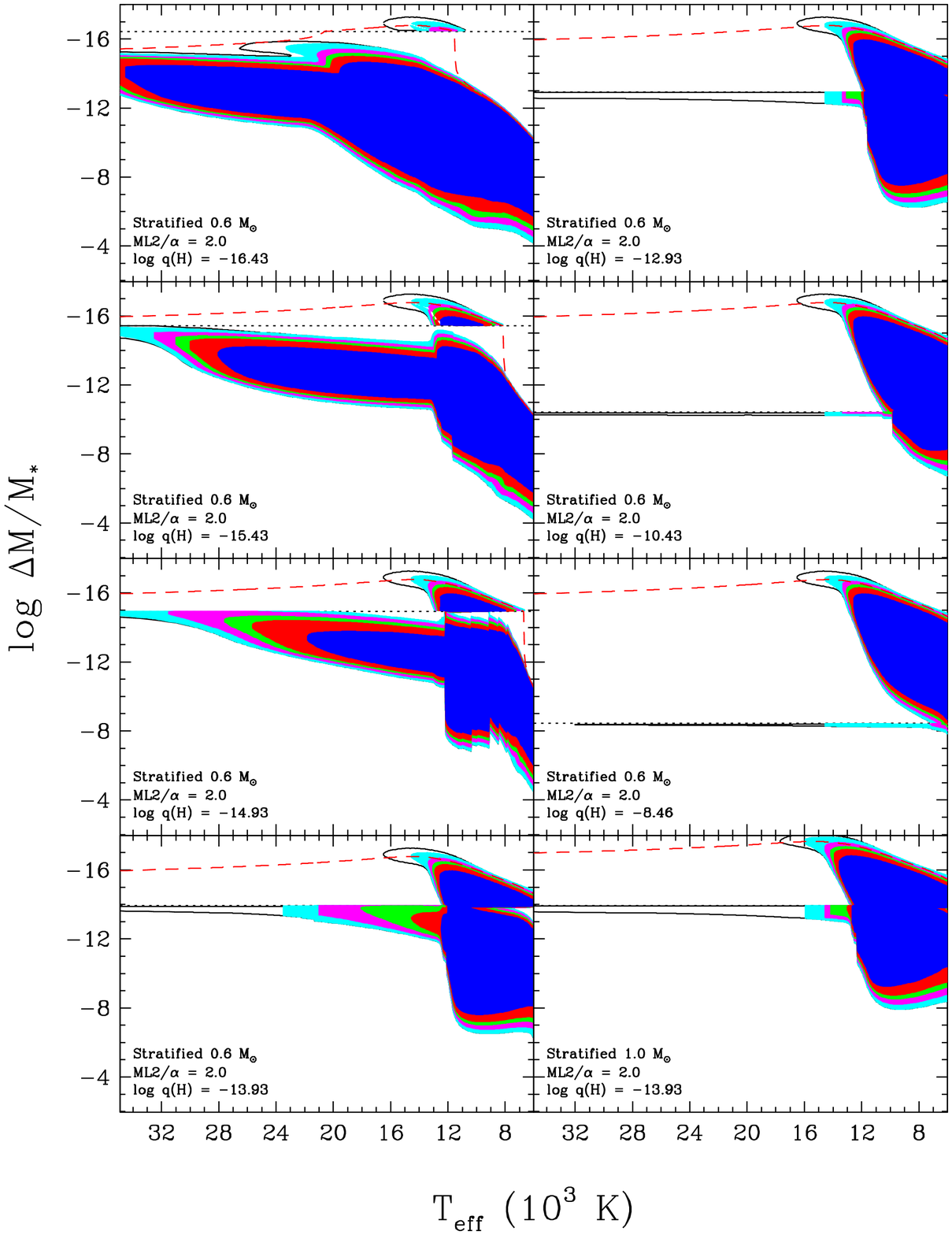}
\caption{Same as Figure \ref{envelopes_str_ML2} but for models
  assuming a ML2/$\alpha=2$ parameterization of the convective
  efficiency.\label{envelopes_str_ML3}}
\end{figure}

Our static, stratified composition models are characterized again by
thick stellar envelopes of $q({\rm env})=10^{-2}$, but composed this
time of a pure hydrogen envelope in diffusive equilibrium on top of a
deeper helium mantle. This stratified model grid covers the same range
of effective temperature as before, and the hydrogen layer mass varies
between $\logqh=\log\mh/M_\star=-17.4$ and $-4.0$ by steps of 0.5
dex. Examples of these stratified models are displayed in Figures
\ref{envelopes_str_ML2} and \ref{envelopes_str_ML3} for various values
of $\qh$, stellar masses, and convective efficiencies.  Note that in
these models, the hydrogen layer is forced to remain in diffusive
equilibrium on top of the helium layer, and is thus never allowed to
mix with the underlying helium envelope, which of course may not be
physically realistic in some cases.

In the thinnest hydrogen layer sequence displayed here, $\logqh
=-16.43$, the structure of the helium convection zone above
$\Te\sim20,000$~K is nearly identical to the almost pure helium
sequences shown in Figures \ref{envelopes_hom_ML2} and
\ref{envelopes_hom_ML3}. In cooler models, however, even the presence
of a small amount of hydrogen at the surface of the star affects the
location of the bottom of the helium convection zone, although such
thin hydrogen layers would have certainly been convectively diluted by
the helium convection zone at higher effective temperatures. As the
thickness of the hydrogen layer is increased, the extent of the helium
convection zone is significantly reduced at high temperatures
($\Te\gtrsim15,000$~K), to the point that it is conceivable that the
hydrogen layer in these models always remains in diffusive equilibrium
on top of the helium envelope --- when $\logqh =-13.93$ for instance
--- at least until the star cools down to $\Te\sim12,000$~K or so,
where {\it convective mixing} might occur.

Indeed, below $\Te\sim16,000$~K, a small hydrogen convection zone
starts to develop at the surface of these models. At lower effective
temperatures, the bottom of the hydrogen convection zone becomes
deeper, eventually reaching the underlying helium convection zone. At
this point, it is believed that the hydrogen layer will be thoroughly
mixed with the deeper and more massive helium convection zone,
resulting in homogeneous H/He convective envelope structures, such as
those illustrated in Figures \ref{envelopes_hom_ML2} and
\ref{envelopes_hom_ML3}. Furthermore, the temperature at which this
mixing process occurs, and the resulting hydrogen-to-helium abundance
ratio at the photosphere, will be a strongly dependent function of the
thickness of the hydrogen layer of the DA progenitor --- the thicker
the hydrogen envelope, the lower the mixing temperature. Note that
according to the models shown here, a DA star is never expected to mix
if $\logqh \gtrsim-6$, which corresponds to the maximum depth of
the hydrogen convective zone.

As for the homogeneous models, both the hydrogen and the helium
convection zones extend much deeper at a given effective temperature
in the models assuming the ML2/$\alpha=2$ convective efficiency, shown
in Figure \ref{envelopes_str_ML3}. The most important consequence in
the present context is that the convective mixing process will occur
at higher effective temperatures in more efficient models. Finally,
the effect of mass is illustrated in the bottom right panel of Figures
\ref{envelopes_str_ML2} and \ref{envelopes_str_ML3} where we show the
results at 1.0 \msun\ and $\logqh=-13.93$, which can be directly
compared with the left panel at 0.6 \msun\ with the same value of
$\logqh$. For a fixed value of $\logqh$, all convection zones are
shifted upward in the more massive models; remember that
$\logqh\equiv\log\mh/M_\star$ is scaled with respect to the mass of
the star, so a given value of $\logqh$ implies a more massive hydrogen
layer in a more massive star. Consequently, the extent of the helium
convection zone at high temperatures is much smaller, but more
importantly in the present context, the effective temperature at which
the hydrogen convection zone connects with the helium convective
envelope is about $\sim$500~K higher.

\subsection{Total Hydrogen Mass}\label{sec:toth}

We now attempt to estimate the total mass of hydrogen present in a
given helium-atmosphere white dwarf, after the superficial hydrogen layer
has been thoroughly diluted --- or convectively mixed --- with the
underlying helium envelope. Our working assumption is that $all$
available hydrogen is found in a region covering the superficial
convection zone extended below by a diffusion tail that must be created
as the result of ordinary diffusion. As discussed in Section \ref{sec:homo},
our homogeneously mixed models are perfectly suitable to estimate the
mass of hydrogen contained in the convection zone. We recall in this
context that the presence of hydrogen below the mixed H/He convection
zone --- and in particular the way it is distributed in these regions
--- does not affect in any way the location of the bottom of this
convection zone in our structures.

Next, the ratio $R$ of the mass of hydrogen contained in the diffusive tail
over that contained in the convection zone can be estimated analytically
following the approach of \citet{ven88}. Under the assumptions
that 1) diffusive equilibrium has been reached, and 2) hydrogen is a
trace species (H $\ll$ He) in the convection zone, one can show that,

\begin{equation}
R=\frac{A_1}{A_1Z_2-A_2(Z_1+1)},
\end{equation}

\noindent where $A_1$ ($Z_1$) is the atomic weight (average charge) of the
dominant element (helium here), and $A_2$ ($Z_2$) is the atomic weight (average
charge) of hydrogen at the bottom of the convection zone. Under
conditions of most interest here, hydrogen is completely ionized at
the bottom of the convection zone ($Z_2=1$) while helium is nearly so 
($Z_1 \simeq 2$). Taking $A_1 = 4$ and $A_2 = 1$, one finds $R \simeq
4$, i.e., there is four times more hydrogen ``hidden'' in the diffusion
tail than present in the convection zone. 

This value of $R$ is necessarily an upper limit because of the following
circumstances. First, if helium is not completely ionized, the diffusion
tail is steeper and contains less hydrogen. For instance, assuming that
$Z_1 = 1$ (He II), one finds $R = 2$. Second, the neglect of thermal
diffusion in our derivation leads also to an overestimate of $R$ as the
diffusion tail would again be steeper otherwise. Third, the assumption
of complete diffusive equilibrium may not be fully justified in the
deeper regions of the diffusion tail as the diffusion timescale there 
may not be negligible anymore in front of the cooling timescale. And
fourth, residual nuclear burning around log $q \sim -4$ also limits the
extension of the diffusion tail and its hydrogen content. In practice,
we assume somewhat arbitrarily in the remainder of our analysis that
$R=2$. Hence, the total amount of hydrogen, $\mh$, is equal to that
measured in the convection zone with our uniformly mixed models
multiplied by ($1 + R$).

\begin{figure}[bp]
\centering
\includegraphics[width=0.8\linewidth]{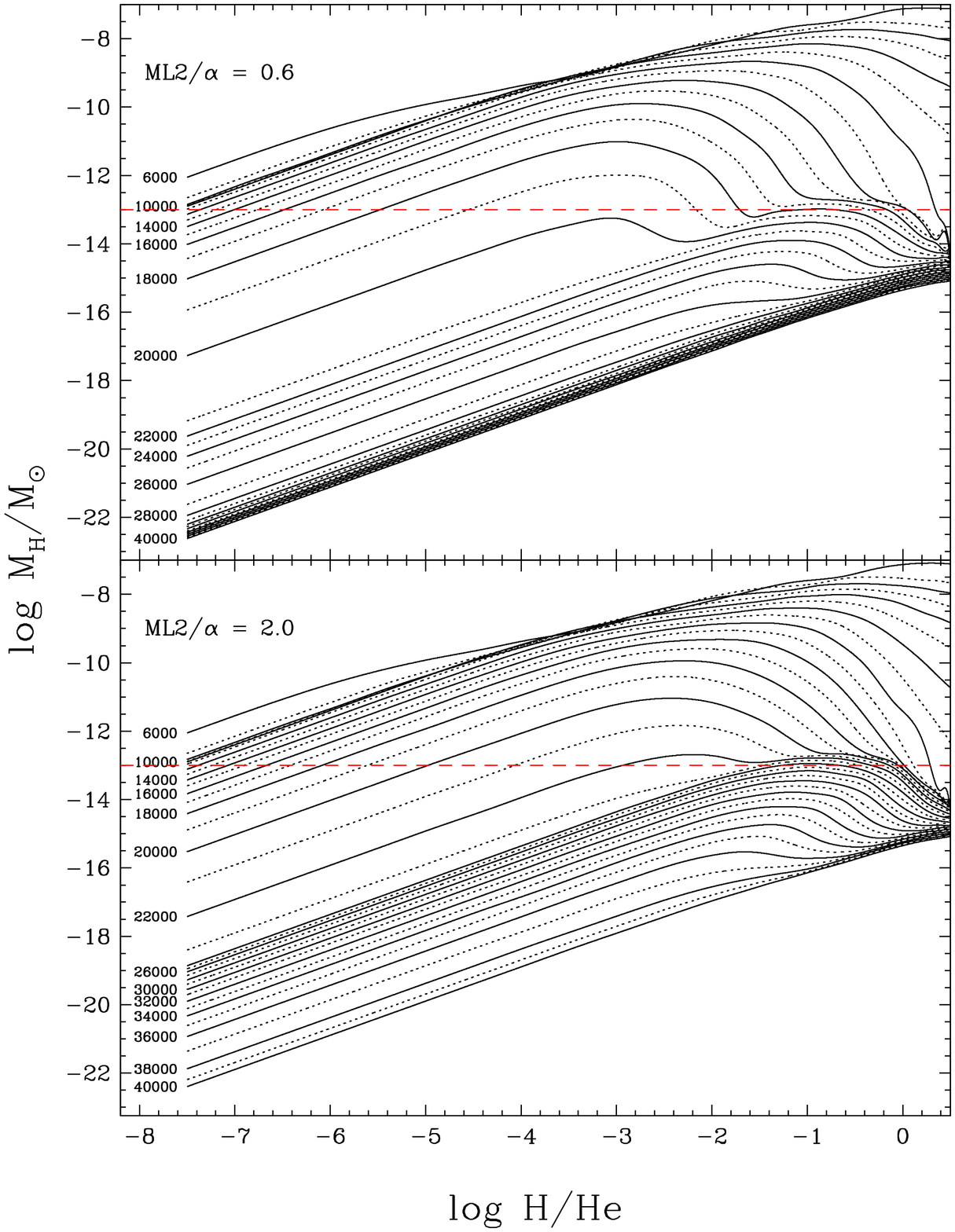}
\caption{Total hydrogen mass contained in homogeneously mixed H/He
  envelope models at 0.6 \msun\ for various effective temperatures
  (labeled on each curve) as a function of the observed photospheric
  hydrogen abundance (H/He). The results are shown for two different
  convective efficiencies. An illustrative example for a total
  hydrogen mass of $\log\mh/M_\odot=-13$ is indicated by a red dashed
  line in each panel.\label{phase_ym}}
\end{figure}

Under these assumptions, it is now possible to calculate the total
mass of hydrogen contained in our grid of homogeneous models at
various effective temperatures as a function of the observed
photospheric hydrogen abundance H/He. The results are summarized in
Figure \ref{phase_ym} where we show, for the two convective
efficiencies explored in our analysis, the total hydrogen mass
contained in the model as a function of H/He at various effective
temperatures ranging from $\Te=6000$~K to 40,000~K. An illustrative
example for a total hydrogen mass of $\mh=10^{-13}$ \msun\ is also
indicated by a red dashed line. For this particular mass value (but
other values as well), we can see that at certain effective
temperatures ($\Te=18,000$~K for instance), there are multiple values
of H/He possible\footnote{Note that \citet{MV91} find more solutions
  than we find here for some $\Te$ values because their grid includes
  models where helium is considered a trace element in diffusive
  equilibrium within the superficial hydrogen-rich layer (see their
  Figure 6).}, generally separated by orders of magnitude, for the
same total hydrogen mass.  This degeneracy reflects the possibility of
mixing the same total amount of hydrogen in a deep, or in a shallow,
helium convection zone (see also \citealt{MV91}).

A careful analysis of the results shown in Figure \ref{phase_ym} also
provides valuable information on the evolution of white dwarfs with
homogeneously mixed H/He compositions. For instance, there is no
homogeneously mixed envelope structure above $\Te\sim 20,000$~K
with a hydrogen mass of $\log\mh/M_\odot=-13$ for 
ML2/$\alpha=0.6$ models (or above $\Te\sim 25,000$~K with $\alpha=2$).
Envelope structures with such large hydrogen masses
can only be stratified (see Figure \ref{envelopes_str_ML2}), corresponding to
DA star configurations. These considerations will thus define an area
in the $\Te$ -- H/He parameter space inaccessible via normal white
dwarf evolution with a constant hydrogen mass, as discussed further
below.

\section{EVOLUTIONARY SCENARIOS}\label{sec:evol}

\subsection{Results from MacDonald \& Vennes (1991)}\label{sec:mv91}

Before discussing our own results, it is worth here summarizing some
of the calculations from \citet{MV91} most relevant to our
study. MacDonald \& Vennes investigated stratified hydrogen/helium
envelope models that are in full diffusive equilibrium for effective
temperatures between 10,000 and 80,000~K, and including convective
mixing using both Schwarschild and Ledoux criteria with different
convective efficiencies. Their results can be best summarized by
examining their Figure 1 (Schwarzschild convection with $\alpha=1$)
where contours of constant total hydrogen mass are shown for
$\log\mh/M_\odot=-16$ to $-10$ (by steps of 1.0 dex) in a diagram of
He/H ratios (measured at a Rosseland optical depth of 2/3) as a
function of effective temperature. To avoid further confusion, it is
important to mention that MacDonald \& Vennes refer to
helium-to-hydrogen ratios (He/H) measured in {\it mass}, while we use
throughout our analysis the inverse ratio (H/He) measured in {\it
  number}. Note also that since these are models in full equilibrium,
all the hydrogen present in the envelope has already reached the
surface in the hottest models ($\Te=80,000$~K). By examining this
figure, one can see that at a given effective temperature, there can
be multiple envelope structures (up to five in some cases) that have
the same total mass of hydrogen. This is further illustrated in their
Figure 6 where the He/H ratio is shown as a function of mass depth for
five envelope models with $\log\mh/M_\odot=-13$ at $\Te=15,000$~K. As
discussed above, such multiple solutions reflect the possibility of
mixing hydrogen and helium in convection zones of various thicknesses
and depths (shown by the flat He/H profiles in their Figure 6). In the
same figure, model A (a typical DA structure with almost all the
hydrogen floating at the surface) and model E (a typical DB structure
with hydrogen being diluted in the deep helium convective envelope)
correspond approximately to our own stratified and homogeneous
envelope models, respectively. But there are also intermediate
solutions (see also our Figure \ref{phase_ym}), which as discussed by
MacDonald \& Vennes, are either unstable (dotted lines in their Figure
1) or unlikely to occur in nature.

As discussed by the authors, the contours in their Figure 1 are not to
be interpreted as evolutionary tracks, but the evolution of white
dwarfs with fixed hydrogen mass can still be determined by studying
the appropriate contour in the following way. For instance, MacDonald
\& Vennes (see their Section 3) discuss a particular example at
$\log\mh/M_\odot=-13$. As the star cools from $\Te=80,000$~K down to
$\sim$35,000~K, the photospheric helium abundance --- always a trace
element --- remains nearly constant, and then starts to decrease
(following the lower branch in their Figure 1) since radiative
acceleration no longer supports helium below 35,000~K.  The He/H
ratio reaches a minimum value near $\Te=14,000$~K, and then starts to
increase steadily again due to the onset of the hydrogen convection
zone, until a minimum in effective temperature is reached at
$\Te=11,700$~K, which corresponds to the point where both helium and
hydrogen convection zones connect. At this point, the He/H ratio
discontinuously jumps to the upper branch of the contour. In our own
terminology, this corresponds to the {\it convective mixing
  scenario} (see Figures \ref{envelopes_str_ML2} and \ref{envelopes_str_ML3}).

Another example worth considering is the case with
$\log\mh/M_\odot=-14$. Again, as the star cools, the photospheric
helium abundance decreases steadily, eventually reaches a minimum
value, and starts to rise slowly. However, for this particular total
hydrogen mass value, the coolest model on the lower branch is at
$\Te=17,900$~K, that is, cooler stratified hydrogen/helium envelope
models in diffusive equilibrium where almost all the hydrogen floats
on top of the star do not exist within their theoretical framework. So
again here, the He/H ratio discontinuously jumps to the upper branch
of the contour, reaching the DB star configurations. This corresponds
to what we referred to as the {\it convective dilution scenario}. Note
that MacDonald \& Vennes refer to both convective mixing and
convection dilution scenarios as {\it convective dredge-up}. In
particular, their Table 1 provides effective temperatures at which the
so-called convective dredge-up occurs ($T_{\rm ed}$) for different
total hydrogen masses and assumed convection models, but it is
important to realize that these $T_{\rm ed}$ values include both
convective dilution (above $\Te\sim13,000$~K) and convective mixing
(below $\Te\sim13,000$~K) processes.

More importantly, we want to emphasize here that the effective
temperatures at which the convection dilution process occurs in the
study of MacDonald \& Vennes are based on static equilibrium
models. In other words, the convective dilution process itself is not
modeled in any way. Following the example above with
$\log\mh/M_\odot=-14$, no particular event occurs below
$\Te\sim17,900$~K. This is just the temperature below which no
envelope models can be found within the framework assumed by the
authors. For instance, for the same value of $\mh$, our stratified
envelope models extend to much lower temperatures (see bottom left
panel of Figure \ref{envelopes_str_ML2}) because we simply forced hydrogen to
remain in equilibrium on top of the helium convection zone. Hence the
real question is whether the underlying helium convection zone becomes
efficient enough to dilute the superficial hydrogen layer, and if so,
at which temperature. This is a {\it dynamical process}, which, to our
knowledge, has never been modeled properly. With these considerations
in mind, we now present the results of our own simulations.

\subsection{Convective Dilution Scenario}\label{sec:dilution}

\begin{figure}[bp]
\centering
\includegraphics[width=0.8\linewidth]{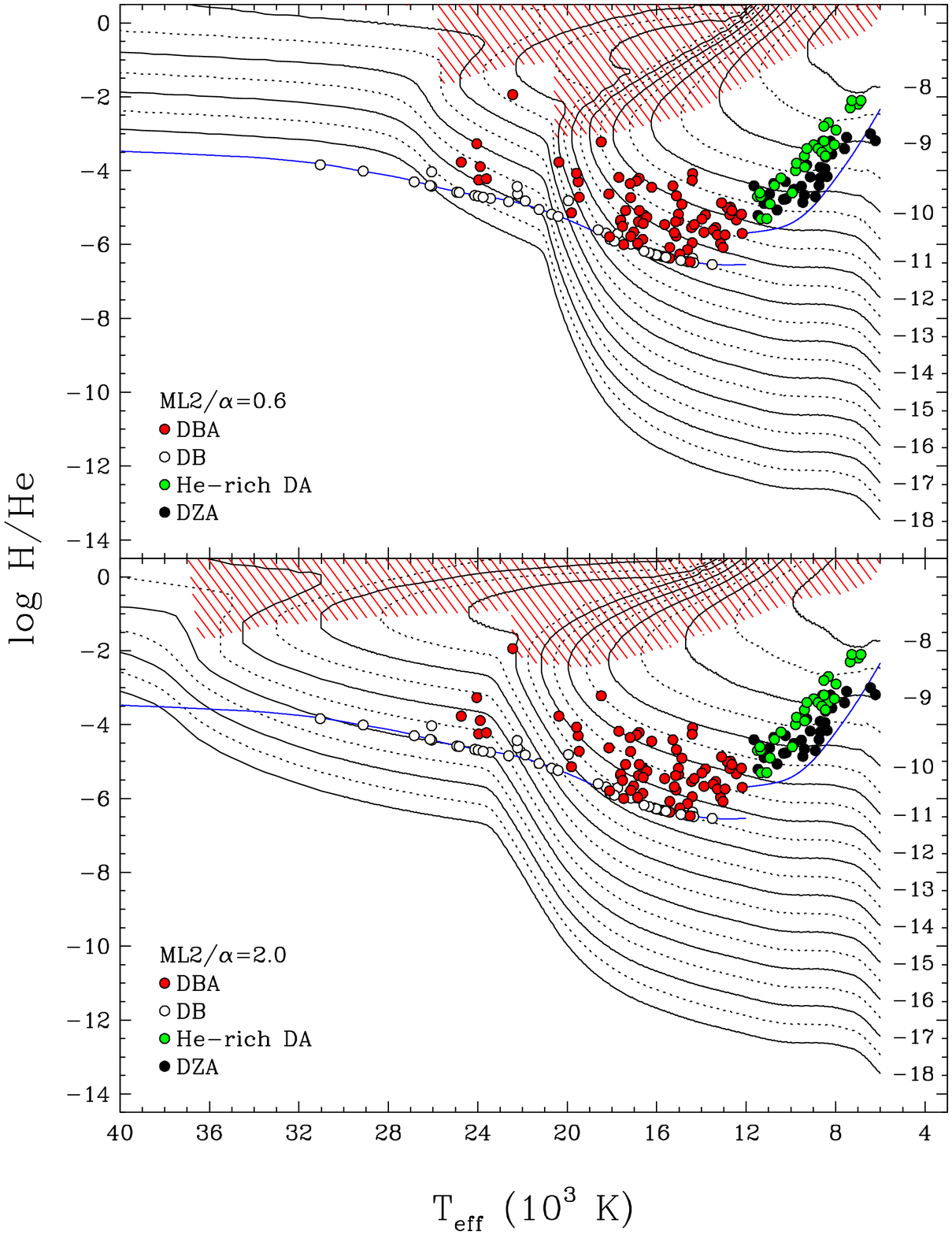}
\caption{Results of our simulations for homogeneously mixed models at
  0.6 \msun\ for both the ML2/$\alpha=0.6$ (upper panel) and
  $\alpha=2$ (lower panel) versions of the mixing-length theory. Each
  curve is labeled with the corresponding value of $\log
  \mh/M_\odot$. Results from Figure \ref{correlty_all} are also
  reproduced; limits on the hydrogen abundance set by our
  spectroscopic observations are shown by the solid blue lines. The
  red hatched regions represent the forbidden region through which
  white dwarfs cannot evolve continuously with a constant total
  hydrogen mass (see text).\label{seq_dilution}}
\end{figure}

We attempt in this section to interpret the hydrogen abundance pattern
observed in DB/DBA white dwarfs and cool, He-rich DA/DZA stars, as
depicted in Figure \ref{correlty_all}.  We first begin by exploring
the scenario where a thin, superficial hydrogen layer of a given mass
has been convectively diluted within the helium envelope, resulting in
a homogeneously mixed H/He convection zone, with some of this hydrogen
lying below the convection zone, as discussed in Section
\ref{sec:toth}. More specifically, we {\it assume} the hydrogen layer
has been convectively diluted, and we do not pay attention to the
dilution process, for the moment.  The results of our simulations for
homogeneously mixed models at 0.6 \msun\ are presented in Figure
\ref{seq_dilution} for both the ML2/$\alpha=0.6$ and $\alpha=2$
versions of the mixing-length theory, together with the observed
hydrogen abundance pattern reproduced from Figure
\ref{correlty_all}. Each curve in this plot represents the location of
white dwarf stars with a constant value of $\log\mh/M_\odot$, labeled
in the figure. These results are similar to those presented in Figures
1 and 2 of \citet{MV91}, although our calculations are restricted to
${\rm H/He} < 1$ (i.e.~the upper portions of their figures). For the
models with $\log\mh/M_\odot\lesssim-14$, the sudden change of slope
near 20,000~K for the ML2/$\alpha=0.6$ models ($\sim$23,000~K for the
$\alpha=2$ models) corresponds to the temperature where the bottom of
the helium convection zone sinks deep into the star (see the top left
panel of Figures \ref{envelopes_hom_ML2} and \ref{envelopes_hom_ML3},
where hydrogen is considered a trace element).

The hottest DBA stars in our sample near $\Te\sim24,000$~K have
inferred total hydrogen masses around $\log\mh/M_\odot\sim-16.5$ with
ML2/$\alpha=0.6$ models, and around $-15$ with $\alpha=2$
models. Larger hydrogen masses are required with $\alpha=2$ models to
produce the same photospheric hydrogen abundance since the convection
zone is much deeper in these models (see Figures
\ref{envelopes_hom_ML2} and \ref{envelopes_hom_ML3}).  Notice that
these hot DBA white dwarfs will rapidly evolve as DB stars at lower
effective temperatures --- below our \ha\ detection threshold --- when
the hydrogen content becomes increasingly more diluted within the
growing helium convection zone, a conclusion also reached by
\citet[][see their Section 5.1]{KK15}. The bulk of the DBA stars in
our sample, however, is found at lower temperatures
($\Te\lesssim20,000$~K). \citet{bergeron11} showed that this
corresponds to the temperature range where the DB/DA ratio reaches a
value of 25\%, based on the luminosity function obtained from the
subset of white dwarfs identified in the Palomar-Green survey, while
this fraction drops to only half this value above
$\Te\sim20,000$~K. As mentioned above, this corresponds also to the
temperature at which the bottom of the helium convection zone sinks
rapidly into the stellar envelope, strongly suggesting that the
convective dilution model is the most likely scenario responsible for
the transformation of some DA white dwarfs into DBA stars.

The results of Figure \ref{seq_dilution} indicate that the bulk of the
DBA stars in our sample can be explained in terms of homogeneous
models with total hydrogen masses between $\log\mh/M_\odot=-13$ and
$-10$ according to our ML2/$\alpha=2$ models. Note how the DBA
abundance determinations are well contained within these two
boundaries. In particular, a DBA star in this temperature range
($\Te\lesssim20,000$~K) is expected to show H$\alpha$ almost all the
way down to $\sim$12,000~K when the helium lines vanish. This
is not necessarily the case with the ML2/$\alpha=0.6$ models, which
require thinner hydrogen layers of the order of
$\log\mh/M_\odot\sim-15$ to account for the DBA stars around 20,000~K;
hydrogen features in these stars would rapidly become undetectable as
they cool off by only $\sim$2000~K or so.

The cool, He-rich DA/DZA white dwarfs in Figure \ref{seq_dilution}
require much larger hydrogen masses, ranging from
$\log\mh/M_\odot=-11$ to $-8$, regardless of the assumed convective
efficiency. Clearly, the progenitors of these objects are not DBA
white dwarfs, which have much lower hydrogen content; DBA stars are
likely to evolve instead into DC stars below $\Te\sim12,000$~K since
their expected hydrogen abundances will be below our H$\alpha$
detection threshold in this temperature range. Note, however, that the
distinction between these two populations cannot be easily
made at the boundary near 12,000~K, and some of the He-rich
DA/DZA white dwarfs with the lowest hydrogen abundances can probably
be interpreted as cooled off DBA stars. We can also see that under the
assumption of a constant total hydrogen mass, a given He-rich DA/DZA
star will evolve at an almost constant photospheric hydrogen abundance,
and will eventually (and rather quickly) turn into a DC star, that is,
below our H$\alpha$ detection threshold. Because of the large differences in
total hydrogen mass between the DBA white dwarfs and the cool, He-rich
DA/DZA stars, we must conclude that the latter have a different
origin, most likely resulting from the mixing of the convective
hydrogen layer with the deeper helium convection zone, a scenario we
explore further in Section \ref{sec:mixing}.

We now turn our attention to the convective dilution process more
specifically. In order for a DBA star to cool off with a constant
total mass of hydrogen already homogeneously mixed within the
convective layer, it must be able to evolve {\it continuously} from
the left to the right in Figure \ref{seq_dilution} along a single
sequence with a given value of $\log \mh/M_\odot$. In other words, the
sequence cannot turn back towards higher temperatures at any point
(see also \citealt{MV91}). These considerations thus allow us to
define a region in the $\Te$ -- H/He parameter space --- represented
by the red hatched regions in Figure \ref{seq_dilution} --- through
which white dwarfs cannot evolve continuously with a constant hydrogen
mass.  For instance, we already presented an example in Figure
\ref{phase_ym} where we showed that homogeneously mixed stellar
envelopes with a total hydrogen mass of $\log\mh/M_\odot=-13$ could
not exist above $\Te\sim 20,000$~K with ML2/$\alpha=0.6$ (or above
$\Te\sim 25,000$~K with $\alpha=2$), in agreement with the results of
Figure \ref{seq_dilution}. White dwarfs containing such large amounts
of hydrogen can only exist as chemically stratified white dwarfs above
these temperatures, with hydrogen floating in diffusive equilibrium on
top of the helium envelope, corresponding to a DA star
configuration. Similar conclusions can be reached from an examination
of the results displayed in Figures 1 and 2 of \citet{MV91}.

The results presented in Figure \ref{seq_dilution} indicate that some
white dwarfs can indeed evolve with a constant hydrogen mass diluted within
the convection zone, but only if the total hydrogen mass is very small,
i.e.~$\log \mh/M_\odot\lesssim-15.5$ for ML2/$\alpha=0.6$ models,
and $\log \mh/M_\odot\lesssim-16$ for $\alpha=2$
models. Incidentally, the hottest DBA stars in our sample near
$\Te\sim24,000$~K can be explained by this scenario, but only if the
convective efficiency is low. All the cooler DBA stars in our sample can
only be explained by some kind of {\it dynamical transformation}, such
as the convective dilution scenario, where the superficial hydrogen
layer of a chemically stratified DA white dwarf is convectively
diluted by the underlying helium convective envelope. In other words,
this convective dilution process will allow a given DA star to cross
the red-hatched region in Figure \ref{seq_dilution}, directly into the
region where DBA stars are found. The question is, under which
physical circumstances?

As discussed in Section \ref{sec:mv91}, \citet{MV91} concluded that DA
stars could be transformed into DB white dwarfs near $\Te\sim18,000$~K
if the hydrogen layer mass was of the order of
$\log\mh/M_\odot\sim-14$. We emphasize that this so-called convective
dredge-up temperature ($T_{\rm ed}$) given in their Table 1
corresponds simply to the coolest stratified DA model in their grid
for this particular hydrogen layer mass (i.e., the coolest point on
the lower branch in their Figure 1).  If we now take these results at
face value, this implies --- according to our results displayed in
Figures \ref{envelopes_str_ML2} and \ref{envelopes_str_ML3} --- that
the convective dilution of a hydrogen layer with
$\log\mh/M_\odot\sim-14$ will occur near $\Te\sim20,000$~K if at least
$\sim$50\% of the total energy flux (the magenta contours) is
transported by convection\footnote{Note that the hydrogen layer masses
  in Figures \ref{envelopes_str_ML2} and \ref{envelopes_str_ML3} are
  given in terms of $\log\qh\equiv
  \log\mh/M_{\star}=\log\mh/M_\odot-0.22$ for a 0.6 \msun\ white
  dwarf.}. If we adopt arbitrarily this fraction of the total flux for
the convective dilution process to occur, we find that the DA-to-DB
transition will take place at $\Te\sim 32,000$~K for $\alpha=2$ models
with $\log \mh/M_\odot\sim-15$, a temperature that is entirely
consistent with the results of MacDonald \& Vennes (see the S2 results
in their Table 1). More importantly, however, for hydrogen layers
between $\log\mh/M_\odot=-13$ and $-10$, where the bulk of the DBA
stars in our sample are found, the underlying helium convection zone
is almost completely inhibited, and the thin convection zone still
present in our models is most certainly too inefficient (less than 1\%
of the total flux) to dilute the superficial hydrogen layer.

We must therefore conclude that most --- but not all ---
helium-atmosphere white dwarfs below $\Te\sim30,000$~K that contain
traces of hydrogen cannot be explained in terms of a convective
dilution scenario. The total amount of hydrogen present in these white
dwarfs implies that their DA progenitors had hydrogen layers that were
far too thick to allow the convective dilution process to occur. A
similar conclusion has also been reached by \citet{MV91}.  The most
common solution proposed to solve this problem is to assume that a
significant fraction of DB stars are indeed the result of a convective
dilution scenario, with DA progenitors having very thin hydrogen
layers ($\log\mh/M_\odot\lesssim-15$).  After the DA-to-DB transition,
accretion of hydrogen from the interstellar medium or other external
bodies (comets, disrupted asteroids, etc.) increases the hydrogen
content in the stellar envelope, up to the level observed in DBA stars
(see, e.g., \citealt{MV91}). We explore this scenario more
quantitatively in the next section.

\subsection{Accretion of Hydrogen from External Sources}

Our results from the previous section strongly suggest that a simple
convective dilution model with a constant hydrogen mass is an unlikely
evolutionary scenario for the origin of DBA and cool, He-rich DA/DZA
white dwarfs. We explore here the possibility of accretion from
external sources, either from the interstellar medium, or from other
bodies such as comets, disrupted asteroids, or small planets. To
model this process, the various episodes of accretion occurring during
the white dwarf evolution are averaged with a constant accretion
rate. We first compute the total accreted mass of hydrogen for various
rates ranging from $\log\mh/M_\odot=-27.5$ to $-17.0$ per year by
steps of 0.5 dex using the cooling times of a typical DB star at 0.6
\msun\ (see Section \ref{sec:res}). For a given effective temperature,
these hydrogen masses are then converted into hydrogen-to-helium
abundance ratios using our $\Te$ -- H/He parameter space map (Figure
\ref{phase_ym}). We assume here for simplicity that the material has
been accreted on top of a pure helium atmosphere; results obtained
with a small initial hydrogen mass of $\log\mh/M_\odot\sim-15$ are
almost identical to those presented here since such a small amount of
hydrogen yields photospheric abundances of only ${\rm H/He}\sim
10^{-10}-10^{-8}$ in the temperature range where most of the DBA stars
are found (see Figure \ref{seq_dilution}).

\begin{figure}[bp]
\centering
\includegraphics[width=0.8\linewidth]{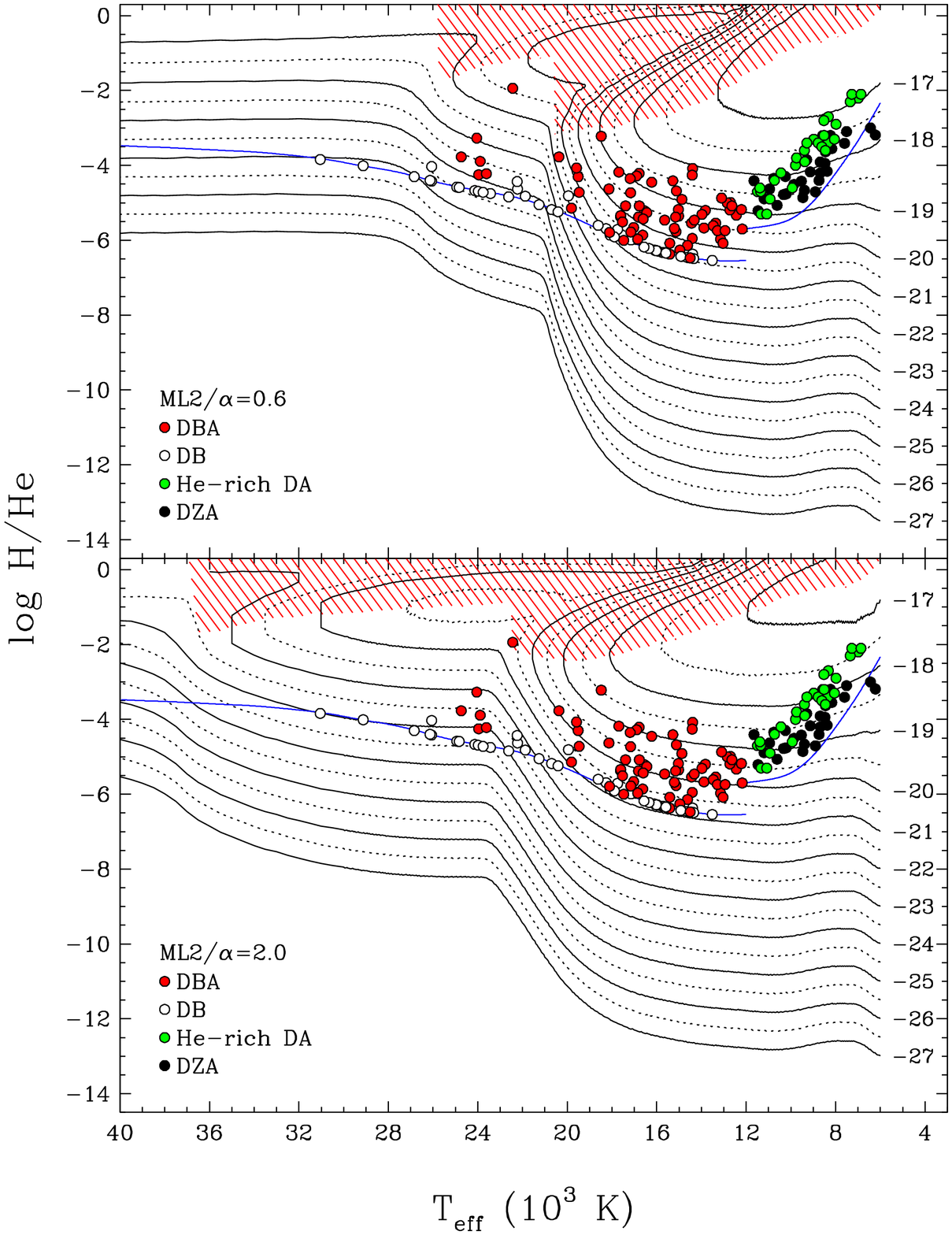}
\caption{Results of our accretion simulations for models at 0.6
  \msun\ subject to constant accretion rates of hydrogen.  Each curve
  is labeled with the corresponding total mass of hydrogen accreted in
  $M_\odot$ per year on a logarithmic scale. Calculations are shown
  for both the ML2/$\alpha=0.6$ (upper panel) and $\alpha=2$ (lower
  panel) versions of the mixing-length theory. Results from Figure
  \ref{correlty_all} are also reproduced.  The red hatched regions
  represent the forbidden region through which white dwarfs cannot
  evolve continuously (see text).\label{seq_accretion}}
\end{figure}

Results of these simulations are displayed in Figure
\ref{seq_accretion} for both prescriptions of the mixing-length theory
considered in this study. Our results indicate that the amount of
hydrogen observed in the bulk of DBA white dwarfs in our sample can be
accounted for with average accretion rates ranging from $10^{-22}$ to
$10^{-19}$ \msun\ yr$^{-1}$, and from $10^{-20}$ to $10^{-17}$
\msun\ yr$^{-1}$ for the cool, He-rich DA/DZA stars. These rates are totally
compatible with those estimated in previous studies
\citep{MV91,dufour07,Voss07,bergeron11,KK15}. The fundamental problem
with this accretion scenario, however, is that for such a range of
accretion rates, our simulations at higher effective temperatures predict
hydrogen abundances in the $\Te$ -- He/H plane where homogeneously
mixed models cannot evolve in a continuous fashion. In other words,
for the accretion model to be valid, the evolutionary tracks would
have to cross the ``forbidden'' red hatched
regions in Figure \ref{seq_accretion}, as was the case for the
convective dilution scenario. 

As a simple example, a pure DB star will take roughly $\sim$$10^7$
years to cool down to $\Te=30,000$~K (see Figure \ref{correltm}), and
even for an accretion rate as low as $10^{-20}$ \msun\ yr$^{-1}$, will
have accumulated $\sim$$10^{-13}$ \msun\ of hydrogen during this
period. By referring to the results shown in Figure \ref{phase_ym},
one can see that this configuration is impossible as a homogeneously
mixed white dwarf, with either version of the mixing-length
theory. Such an object can only exist as a DA star, with all the
hydrogen floating in diffusive equilibrium on top of the helium
envelope, which according to Figures \ref{envelopes_hom_ML2} and
\ref{envelopes_hom_ML3} (right panels), will not mix until it reaches
$\Te\sim 11,000$~K. We are thus forced to conclude that the hydrogen
abundances measured in DBA white dwarfs, and cool He-rich DA/DZA stars
as well, cannot be accounted for by any kind of accretion mechanism
onto a pure helium DB star. Our conclusion remains the same even if we
allow for an initial hydrogen mass of $\log\mh/M_\odot\sim-15$ instead
of a DB white dwarf with a pure helium atmosphere.

Note that it is always possible to invoke the accretion of large
bodies such as comets, disrupted asteroids, or small planets as the
source of hydrogen in DBA stars if the accretion process begins only
{\it after} the white dwarf has evolved through the forbidden red
hatched region in Figure \ref{seq_accretion} --- either as a pure
helium-atmosphere DB star or as a DA star with a very thin hydrogen
layer --- but this would require extraordinary circumstances for such
a process to occur precisely below 20,000~K for a significant fraction
of DB stars.

\subsection{Convective Mixing Scenario}\label{sec:mixing}

At lower effective temperatures ($\Te\lesssim13,000$~K), DA white
dwarfs with thin enough hydrogen layers may get a second opportunity
to turn into helium-dominated atmospheres as a result of {\it
  convective mixing}, which occurs when the bottom of the superficial
hydrogen convective envelope sinks into the star, and eventually
connects with the underlying and more massive helium convection zone
(see Figures \ref{envelopes_str_ML2} and \ref{envelopes_str_ML3}).  At
this point, it is generally assumed that both hydrogen and helium
convection zones merge, with the total hydrogen content homogeneously
mixed within this H/He convective layer. As discussed in the
Introduction, convective mixing is the most likely explanation to
account for the significant increase in the ratio of non-DA to DA
stars below $\Te\sim 10,000$~K. After convective mixing occurs, the
star will continue its evolution with a homogeneously mixed envelope
with constant total hydrogen mass, a scenario already described in
Section \ref{sec:dilution}. Our stratified model structures, displayed
in Figures \ref{envelopes_str_ML2} and \ref{envelopes_str_ML3},
indicate that this mixing process can occur if the mass of the
hydrogen layer is in the range $\log\mh\sim 10^{-15}$ to
$10^{-6}\ M_\star$, where the upper limit is set by the maximum depth
reached by the bottom of the hydrogen convection zone near
$\Te\sim5000$~K (see also Figure 40 of \citealt{BRL97}). However, for
hydrogen layers thinner than $\mh\sim 10^{-14}\ M_\star$, the
convective dilution process discussed in Section \ref{sec:dilution} is
most likely to occur at much higher temperatures
($\Te\gtrsim20,000$~K), hence a more realistic lower limit for the
occurrence of convective mixing is set here at
$\mh=10^{-14}\ M_\star$. To model the convective mixing scenario, we
thus calculated the effective temperature at which the hydrogen and
helium convection zones connect in a 0.6 \msun\ stratified envelope
model, for a given value of the total hydrogen mass $\mh$. From that
point on, we assume complete mixing, and follow the evolution at lower
effective temperatures using the homogeneous sequences with the
corresponding value of $\mh$, as described in Section
\ref{sec:dilution}.

\begin{figure}[bp]
\centering
\includegraphics[width=0.8\linewidth]{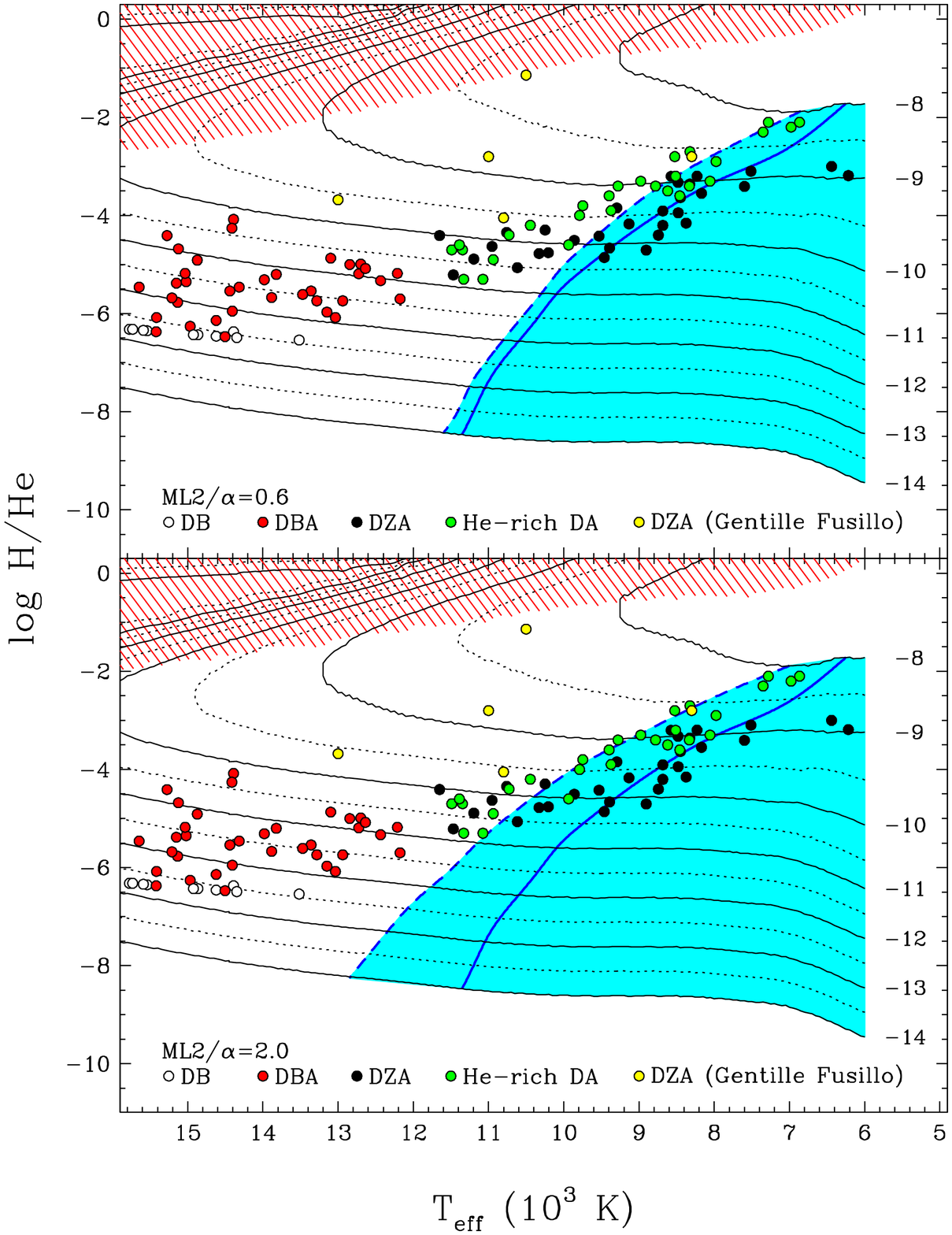}
\caption{Results of our convective mixing simulations for models at
  0.6 \msun, assuming that after mixing occurs, the total hydrogen
  mass is distributed in the way described in Section
  \ref{sec:toth}. Each curve is labeled with the corresponding value
  of $\log \mh/M_\odot$. The solid blue line shows the predicted
  hydrogen-to-helium abundance ratio as a function of the temperature
  at which mixing occurs, while the dashed blue line allows for
  convective overshooting over one pressure scale height; the filled
  cyan area represents the region where white dwarfs will evolve after
  mixing has occurred. Calculations are shown for both the
  ML2/$\alpha=0.6$ (upper panel) and $\alpha=2$ (lower panel) versions
  of the mixing-length theory.  Results from Figure \ref{correlty_all}
  are also reproduced, together with the objects discussed in
  \citet[][see text]{fusillo17}.
\label{seq_mixing}}
\end{figure}

\begin{figure}[bp]
\centering
\includegraphics[width=0.8\linewidth]{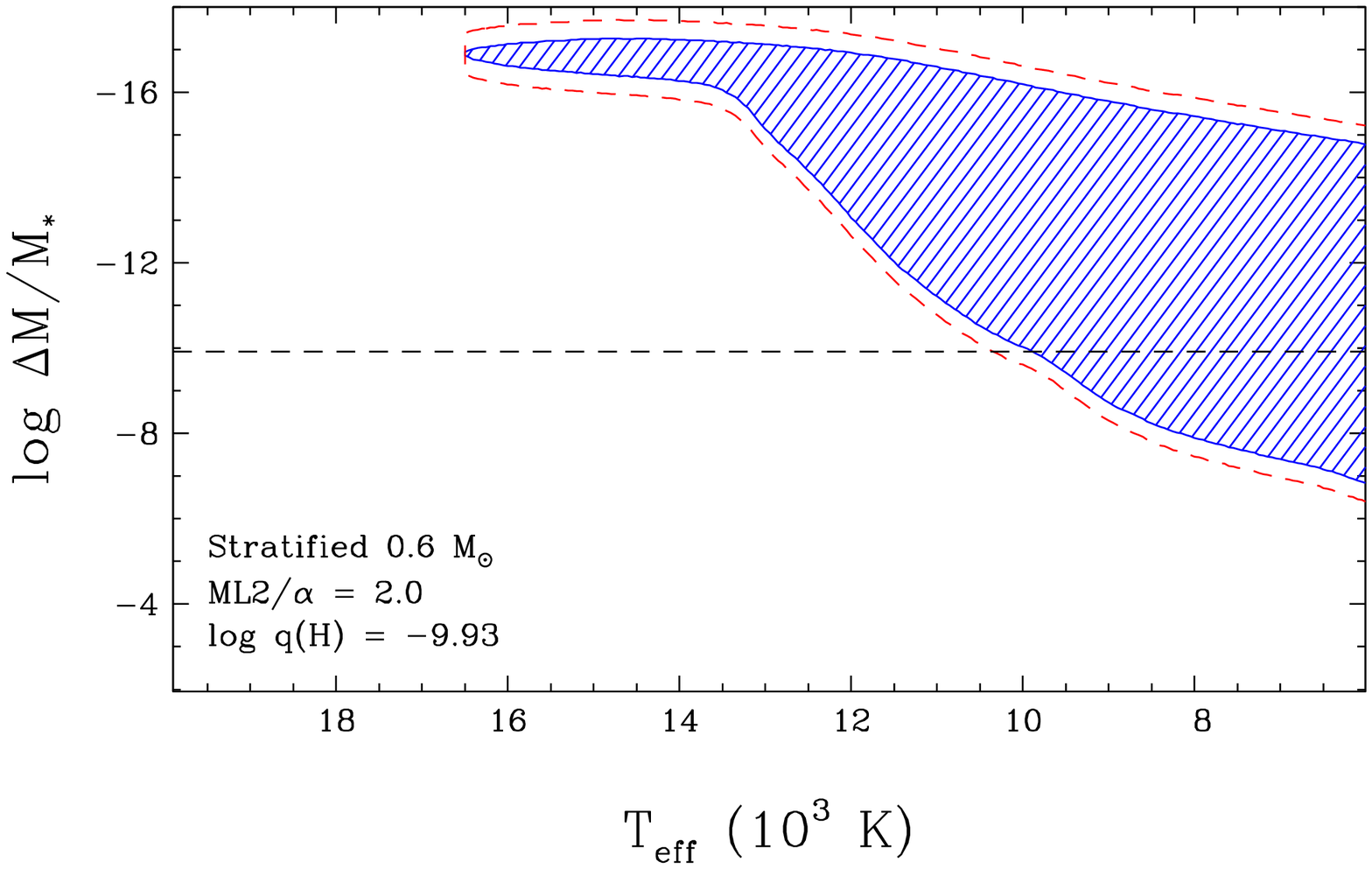}
\caption{Example of envelope structures, as a function of
  effective temperature, for chemically stratified white dwarf models,
  with parameters given in the figure.  The convection zones are shown
  by the hatched region, while the red dashed line shows the extent of
  the hydrogen and helium convection zones allowing for convective
  overshooting over a distance of one pressure scale
  height.\label{overshoot}}
\end{figure}

Results of our convective mixing simulations are displayed in Figure
\ref{seq_mixing} for both prescriptions of the mixing-length theory
considered is this study. The blue solid line in this figure indicates
the effective temperature at which mixing occurs, and the predicted
H/He abundance ratio upon mixing. After mixing, the white dwarf
evolves at a constant value of $\mh$ in the region represented by the
cyan area in Figure \ref{seq_mixing}.  The particular behavior of the
mixing temperature as a function of H/He can be explained
qualitatively in the following way. Since the bottom of the hydrogen
convection zone gets deeper as the white dwarf cools off (see right
panels of Figures \ref{envelopes_str_ML2} and
\ref{envelopes_str_ML3}), the effective temperature at which
convective mixing occurs will depend strongly on the thickness of the
hydrogen layer --- the thicker the hydrogen envelope, the lower the
mixing temperature. Furthermore, since the depth of the mixed H/He
convection zone remains almost constant in this temperature range (see
Figures \ref{envelopes_hom_ML2} and \ref{envelopes_hom_ML3}), the
predicted H/He abundance ratio upon mixing increases with decreasing
mixing temperature, as shown by our simulations in Figure
\ref{seq_mixing}. This mixing temperature can also be made
significantly hotter in our calculations if we allow for even a modest
convective overshooting. For instance, we show in Figure
\ref{overshoot} the extent of the convection zones in a typical
sequence of stratified models from our grid, by allowing both hydrogen
and helium convection zones to overshoot over a distance of one
pressure scale height, a completely reasonable assumption \citep[see,
  for instance,][]{tremblay15}. In the example displayed in Figure
\ref{overshoot}, this simple prescription increases the mixing
temperature by $\sim 500$~K. The results of convective overshooting
applied to all our models are indicated by the dashed blue line in
Figure \ref{seq_mixing}.

Also reproduced in Figure \ref{seq_mixing} are the hydrogen abundances
measured in cool, He-rich DA/DZA stars, which are the objects of
interest in the present context.  As discussed in Section
\ref{sec:dilution}, under the assumption of a constant total hydrogen
mass, He-rich DA/DZA stars will evolve at an almost constant
photospheric hydrogen abundance, and will eventually turn into DC (or
DZ) stars, that is, below our H$\alpha$ detection threshold (see
Figure \ref{correlty_all}). Due to these observational limitations,
the observed sequence in Figure \ref{seq_mixing} defines in fact only
the blue edge of a region in $\Te$ and H/He where these objects could
be found; higher signal-to-noise observations should allow the
detection of hydrogen in helium-rich atmospheres at even lower
temperatures.

Interestingly enough, the overall trend of the H/He abundance ratio
predicted by the convective mixing scenario, as a function of the
mixing temperature, represents an excellent match to the blue edge
defined by the measured hydrogen abundances in cool, He-rich DA and
DZA stars, particularly if we allow for a more convective efficiency
(ML2/$\alpha=2$) and a modest convective overshooting (cyan region
defined by the dashed blue line in the bottom panel of Figure
\ref{seq_mixing}). More specifically, the convective mixing scenario
predicts higher hydrogen abundances in cooler white dwarfs, as
observed here. Also, note the absence of white dwarfs at
$\Te\sim10,000$~K in our sample with large hydrogen abundances, as
predicted by the models. The overall agreement between the location of
the cool, He-rich DA/DZA stars and the region predicted by our
simulations (shown in cyan) clearly demonstrates that the convective
mixing scenario is the most plausible interpretation for the presence
of traces of hydrogen in cool, helium-atmosphere white dwarfs.

In the same context, we also show in Figure \ref{seq_mixing} the five
helium-atmosphere white dwarfs with exceptionally high hydrogen
abundances discussed in \citet{fusillo17} and references therein;
these are, from hottest to coolest, SDSS J124231.07+522626.6, GD 16,
PG 1225$-$079, GD 362, and GD 17. While the amount of hydrogen in
these objects has been considered exceptional, thus requiring large
external sources of hydrogen --- from the accretion of water-bearing
planetesimals for instance --- our results indicate that even though
this is probably true for SDSS J124231.07+522626.6, GD 16, and GD 362,
{\it convective mixing alone} can provide sufficient amounts of
hydrogen without invoking additional sources in the case of PG
1225$-$079 and GD 17.

We end this section by summarizing in Table \ref{table_mixingT} the
effective temperatures at which the transition from a
hydrogen-atmosphere to a helium-atmosphere white dwarf occurs as a
function of the thickness of the hydrogen layer ($\log \mh$/\msun),
for both the convective dilution and convective mixing scenarios, and
for both prescriptions of the mixing-length theories assumed in our
study. We also provide the predicted hydrogen-to-helium abundance
ratios when the transition occurs. As discussed in Section
\ref{sec:dilution}, we simply assume here that the convection dilution
process occurs when $\sim$50\% of the total energy flux is transported
by convection, and for the convective mixing process, we allow both
hydrogen and helium convection zones to overshoot over a distance of
one pressure scale height, as discussed above. When comparing our
results with those presented in Table 1 of \citet{MV91}, we find that
our transition temperatures compare remarkably well with their
convective dredge-up temperatures, both qualitatively and
quantitatively, especially given the many different assumptions made
in both studies. And once again here, we want to emphasize that
neither of these two analyses have properly modeled the convective
dilution process, which is a time-dependent {\it dynamical process}.

\section{DISCUSSION}\label{sec:concl}

\subsection{The origin of hydrogen in DBA stars}

The existence of a DB-gap --- or DB-deficiency --- in the 30,000 to
45,000~K temperature range clearly demonstrates that the DA-to-DB
transition must necessarily occur in nature below 30,000 K or so. The
only viable physical mechanism for this transition to occur in this
temperature regime is the convective dilution of a thin radiative
hydrogen layer at the surface of a DA star with the deeper and more
massive convective helium zone. \citet{bergeron11} showed, however,
that the significant increase in the number of DB stars in this
temperature range occurs only below $\Te\sim20,000$~K, rather than the
canonical 30,000~K, where the hottest DB white dwarfs have been
identified in the PG survey. This makes perfect sense when looking at
our Figures \ref{envelopes_str_ML2} and \ref{envelopes_str_ML3}, since
the helium convection zone becomes efficient only at much lower
temperatures in stratified H/He atmospheres. Since most, but not all,
DB white dwarfs below $\sim$20,000~K in our sample now show traces of
hydrogen, with respect to our previous analysis \citep{bergeron11}, it
is reasonable to conclude that hydrogen must have a primordial origin.

Actually, 63\% of all DB white dwarfs in our enlarged sample are DBA
stars, but this fraction increases to 75\% if we consider only the
objects below 20,000~K. We note that pure DB stars --- at least in the
optical --- still exist in this temperature range, with limits as low
as ${\rm H/He} \lesssim10^{-6}$. Because of the higher S/N of our
sample, these limits are much more stringent than those reported by
\citet{KK15} based on SDSS spectra. We can identify at least two, or
perhaps three, channels that could account for the existence of these
hydrogen-deficient DB stars. The first one corresponds to the hot DB
white dwarfs in the gap. These, apparently, did not have enough
hydrogen left in their envelope to build a hydrogen atmosphere and to
become DA stars over time, and they will most likely evolve as
helium-rich atmospheres throughout their lifetime.  The hottest DB
stars in our sample above $\Te\sim25,000$~K probably belong to this
category. A second channel corresponds to DA stars with very thin
hydrogen layers --- of the order of $\mh\sim10^{-15}$ \msun\ --- that
would turn into DBA stars at $\Te\sim24,000$~K according to our
results displayed in Figure \ref{seq_dilution}. But as shown in the
same figure, such objects would rapidly turn into pure DB stars as the
small amount of hydrogen present in these stars is further diluted in
the growing helium convection zone at lower effective
temperatures. Finally, there is probably a third, but numerically less
important channel producing hydrogen-deficient DB stars at low
temperatures, the so-called Hot DQ stars \citep{dufourNature07}. As
discussed in \citet[][see also \citealt{dufour08}]{bergeron11},
because the coolest Hot DQ star currently known has a temperature near
18,000~K, these must somehow turn into DB white dwarfs at lower
effective temperatures, through a process currently unknown. But the
number of known Hot DQ stars is so small, that this particular channel
is certainly negligible from an ensemble point of view.

The convective dilution scenario alone can probably account for the
amount of hydrogen in the few hottest DBA stars in our sample near
$\Te\sim24,000$~K. It is worth mentioning in this context that the hot
($\Te\sim30,000$~K) DBA star SDSS 1509$-$0108, which \citet[][see
  their Figure 14]{manseau16} interpreted as a chemically stratified
white dwarf with $\log\mh/M_\odot\sim-16.7$, represents an obvious
progenitor of these hottest DBA stars in our sample. For the bulk of
the DBA white dwarfs, however, the total amount of hydrogen inferred
from the photospheric hydrogen abundance is simply too large. DA
progenitors with such large hydrogen masses --- of the order of
$\mh\sim10^{-13}$ \msun --- would have stratified atmospheres with
hydrogen layers so thick that they would not stand a chance to turn
into helium-atmosphere DB stars in the appropriate temperature range.
We thus conclude that the total mass of hydrogen estimated in DBA
stars, assuming complete mixing within the stellar envelope, is
too large, and incompatible with a scenario involving the
transformation of a DA star progenitor into a DB white
dwarf through the convective dilution of a thin hydrogen atmosphere
with the deeper and more massive helium convection zone. DA stars with
such massive hydrogen layers would not mix until they reach
temperatures that are significantly cooler than the entire DBA
population.

Accretion of hydrogen from external sources has thus been invoked
repeatedly to explain the total mass of hydrogen determined in DBA
stars. One of the most recent studies by \citet{veras17} suggested,
for instance, the gradual accretion of hydrogen from exo-Oort cloud
comets. However, we found in our study that the required amount of
accreted material, with even a moderate accretion rate, would build a
superficial hydrogen layer thick enough by the time the white dwarf
reaches a temperature of $\Te\sim30,000$~K, that this object ---
presumably a DA star --- would never turn into a helium-atmosphere DB
star. Another obvious problem with the accretion scenario is that even
modest amounts of hydrogen accreted at the surface of hot
helium-atmosphere white dwarfs in the $\sim$30,000~K temperature range
would easily show up spectroscopically since the extent of the helium
convection zone at these temperatures is too small to allow any
significant dilution of the accreted material into the deeper envelope
(see Figures \ref{envelopes_hom_ML2} and \ref{envelopes_hom_ML3}). We
thus conclude that the hydrogen abundances measured in DBA stars
cannot be accounted for by any kind of accretion mechanism onto a pure
helium DB star progenitor.

Hence we are left with no satisfactory explanation for the presence of
hydrogen in the bulk of DBA white dwarfs at the observed abundance
level. One explanation proposed by \cite{G-BB16} is that perhaps
hydrogen tends to float in the radiative zone on top of the
photosphere rather than being completely mixed within the helium
convection zone. They explored this possibility by calculating spectra
with an abundance profile given by the diffusive equilibrium
approximation in the radiative layers above the convection zone, as
illustrated, for instance, in Figure 6 of \citet{MV91}. Based on their
preliminary calculations, Genest-Beaulieu \& Bergeron suggested that
the hydrogen-to-helium abundance ratios measured in DBA white dwarfs
could be overestimated by perhaps 2 orders of magnitude due to the
inhomogeneous H and He abundance profiles in the upper radiative
atmosphere. The inhomogeneous profile was only used in their
calculation of the synthetic spectrum, however, and the next step,
currently underway, is to implement this inhomogeneous hydrogen
abundance profile in the calculation of the atmospheric structure
itself to get a self-consistent solution.

Yet another possibility, not envisaged explicitly in our analysis, is
that primordial hydrogen, highly diluted in a post-born-again PG1159
progenitor, may not have had the time to diffuse upward completely
during the cooling process, contrary to what is generally assumed. In
that case, not all of the hydrogen would find itself distributed in
the convection zone of a DBA star (including a diffusion tail), so
that the $observable$ hydrogen content could be much less than the
actual amount of that element in a given star. With passing time, more
hydrogen would enrich the convection zone of a DBA star from below (as
opposed to from above as in the case of accretion). Such a process
could potentially explain why a DBA star at $\Te\sim18,000$~K, say,
with a relatively large quantity of hydrogen in its outer convection
zone, could still be the descendant of a $\Te\sim25,000$~K star rather
characterized by a much smaller amount of hydrogen pollution in its
convective atmosphere-envelope.  To go further in that direction,
however, requires demanding and detailed time-dependent calculations
combining evolution, convective mixing, and diffusion. This is beyond
the scope of the present paper but deserves consideration for the
future.

\subsection{The origin of hydrogen in cool, He-rich DA/DZA stars}

We have shown that the presence of hydrogen in cool
($\Te\lesssim12,000$~K), helium-rich atmosphere white dwarfs
discovered in the SDSS is a common phenomenon, although the exact
fraction showing H$\alpha$ remains undetermined. The hydrogen
abundances determined in these objects define a sequence in the
H/He -- $\Te$ diagram --- with cooler objects showing larger hydrogen
abundances --- which overlaps with the abundances measured
in DZA white dwarfs, suggesting that the only difference between these
two populations is the presence or not of a source of accreted
material such as comets, disrupted asteroids, small planets, etc. The
cool edge of the sequence is probably just a selection effect due to
the increasing difficulty of detecting H$\alpha$ at low temperatures
(see the detection threshold in Figure \ref{correlty_all}). The blue
edge of the sequence is well defined, however, and objects to the left
of this sequence could easily be detected; they are thus obviously
rare, with a few exceptions discussed in \citet{fusillo17} and in
Section \ref{sec:mixing} (see Figure \ref{seq_mixing}). The blue edge
of the sequence must therefore have an astrophysical origin.

Our envelope models with constant total hydrogen mass (see Figure
\ref{seq_dilution}) clearly show that DBA white dwarfs will not turn
into cool, He-rich DA stars, but will instead become DC stars when the
hydrogen abundances fall below the detection threshold at low
temperatures ($\Te\lesssim12,000$~K). Instead our simulations indicate
that convective mixing of the thin hydrogen layer with the deeper
helium convection zone is the most likely explanation for the presence
of hydrogen in cool, He-rich DA/DZA white dwarfs.  After mixing, these
stars will evolve at an almost constant hydrogen abundance (see Figure
\ref{seq_mixing}), eventually turning into DC or DZ stars when
H$\alpha$ falls below the detection threshold.  With these
considerations in mind, if we assume that the non-DA to DA ratio below
$\sim$10,000~K is near unity (see, e.g., \citealt{FW87}) and that
$\sim$20\% of white dwarfs are DB/DBA stars in the appropriate
temperature range \citep{bergeron11}, we can estimate that convective
mixing eventually occurs for 40\% of cool DA stars.

It is also interesting in this context to speculate about the origin
of cool DQ stars. Since practically none of them show traces of
hydrogen (with a few notable exceptions where a CH feature is
present), perhaps the progenitors of these stars are the pure DB stars
that show no traces of hydrogen either. Then the reason why some cool
non-DA white dwarfs show carbon features while others do not could be
related to the same reason why some DB stars appear to have very
little, or no hydrogen at all. In the latter case, hydrogen has
probably been completely depleted during the earlier born-again
post-AGB evolutionary phases. Perhaps then, the overall stellar
structure of the progenitor has been affected in such a way to
facilitate the carbon dredge-up from the core at low effective
temperatures. This could even explain the absence of DQZ stars if
these earlier post-AGB phases have somehow managed to wipe out any
material surrounding the white dwarf progenitor, preventing any
further accretion of heavy elements during the course of its
evolution.

We also note that the highest mixing temperature in Figure
\ref{seq_mixing} is $\Te\sim13,000$~K, which occurs for $\mh=10^{-14}$
\msun\ in ML2/$\alpha=2$ models, that is {\it above} the blue edge of
the ZZ Ceti instability strip (see for instance Figure 33 of
\citealt{gianninas11}), while the red edge of the strip at
$\Te\sim11,000$~K corresponds to a mixing temperature of models with
$\mh\sim10^{-11}$. Hence it is possible that DA stars mix above, or
even within, the ZZ Ceti instability strip, implying that
asteroseismological analyses of ZZ Ceti stars may not be sampling the
entire range of hydrogen layer masses in DA stars.

Many of the ideas and speculations presented in this paper can be
studied further by performing statistical analyses of large white
dwarf samples such as the SDSS sample. Unfortunately, there are many
selection effects in the SDSS, and it is therefore difficult to define
a statistically meaningful sample to study the spectral evolution of
white dwarf stars. To do things properly, one would require accurate
distances for these white dwarfs, a situation that will greatly be
improved with the trigonometric parallax measurements from the {\it
  Gaia} mission that will be released in the very near future.

\acknowledgements We would like to thank the director and staff of
Steward Observatory and Kitt Peak National Observatory for providing
observing time for this project. We are also grateful ro
M.-M.~Limoges, N.~Giammichele, L.~S\'eguin-Charbonneau, and
E.~M. Green for acquiring some of the spectra used in our
analysis. This work was supported in part by the NSERC Canada and by
the Fund FRQ-NT (Qu\'ebec).

\clearpage
\bibliography{ms}{}
\bibliographystyle{apj}

\clearpage
\begin{deluxetable}{lllrrlcccccc}
\tabletypesize{\scriptsize}
\tablecolumns{12}
\tablewidth{0pt}
\rotate
\tablecaption{Atmospheric Parameters of DB and DBA White Dwarfs\label{table_fits_DBA}}
\tablehead{
\colhead{WD} &
\colhead{Name} &
\colhead{$\Te$ (K)} &
\colhead{log $g$} &
\colhead{log H/He} &
\colhead{$M/$\msun} &
\colhead{$M_V$} &
\colhead{log $L/$\lsun} &
\colhead{$V$} &
\colhead{$D ({\rm pc})$} &
\colhead{log $\tau$} &
\colhead{Notes}}
\startdata
0000$-$170  & G266-32           & 13,880 ( 361) & 8.63 (0.12) & $ -$5.67 (0.52) & 0.98 (0.07) & 12.51 & $-$2.67 & 14.69 &  27 & 8.86 &     \\
0002$+$729  & GD 408            & 14,410 ( 351) & 8.27 (0.09) & $ -$5.95 (0.79) & 0.76 (0.06) & 11.79 & $-$2.36 & 14.33 &  32 & 8.52 &     \\
0017$+$136  & Feige 4           & 18,130 ( 438) & 8.08 (0.05) & $ -$4.63 (0.21) & 0.65 (0.03) & 10.98 & $-$1.85 & 15.37 &  75 & 8.07 &     \\
0025$-$032  & PB 8252           & 18,480 ( 437) & 8.20 (0.04) & $ -$3.22 (0.04) & 0.72 (0.02) & 11.12 & $-$1.89 & 15.69 &  82 & 8.14 &     \\
0031$-$186  & KUV 00312$-$1837  & 15,020 ( 396) & 8.43 (0.11) & $ -$5.35 (0.34) & 0.86 (0.07) & 11.96 & $-$2.39 & 16.66 &  87 & 8.60 &     \\
0100$-$068  & G270-124          & 19,820 ( 531) & 8.06 (0.04) & $ -$5.14 (1.06) & 0.64 (0.03) & 10.78 & $-$1.68 & 13.95 &  43 & 7.91 &     \\
0112$+$104  & EGGR 409          & 31,040 (1056) & 7.83 (0.03) & $<-$3.84 (0.84) & 0.53 (0.02) &  9.89 & $-$0.75 & 15.36 & 123 & 7.20 & 1   \\
0119$-$004  & G271-47A          & 16,060 ( 404) & 8.07 (0.06) & $ -$6.27 (1.70) & 0.63 (0.04) & 11.23 & $-$2.05 & 16.00 &  89 & 8.24 &     \\
0125$-$236  & G274-39           & 16,550 ( 436) & 8.24 (0.07) & $ -$5.21 (0.32) & 0.74 (0.05) & 11.43 & $-$2.11 & 15.38 &  61 & 8.33 &     \\
0129$+$246  & PG 0129$+$247     & 16,450 ( 461) & 8.27 (0.09) & $ -$5.26 (0.40) & 0.76 (0.06) & 11.49 & $-$2.13 & 16.09 &  83 & 8.35 &     \\
0158$-$160  & G272-B2A          & 24,130 (1369) & 7.94 (0.03) & $<-$4.67 (1.39) & 0.57 (0.02) & 10.40 & $-$1.26 & 14.38 &  62 & 7.38 &     \\
0203$-$181  & HE 0203$-$180     & 12,180 ( 652) & 8.90 (0.42) & $ -$5.70 (2.43) & 1.14 (0.24) & 13.35 & $-$3.12 & 16.00 &  33 & 9.16 &     \\
0211$+$646  & Lan 150           & 20,700 ( 719) & 8.00 (0.04) & $<-$5.18 (2.18) & 0.60 (0.03) & 10.62 & $-$1.57 & 17.43 & 230 & 7.77 &     \\
0214$+$699  & Lan 158           & 29,130 (1329) & 7.88 (0.05) & $<-$4.01 (0.97) & 0.55 (0.02) & 10.07 & $-$0.89 & 16.60 & 202 & 6.96 &     \\
0215$-$024  & PB 6822           & 16,870 ( 402) & 8.12 (0.09) & $ -$5.96 (1.35) & 0.66 (0.06) & 11.19 & $-$2.00 & 16.13 &  97 & 8.20 &     \\
0220$+$480  & GD 27             & 16,570 ( 403) & 8.33 (0.05) & $<-$5.12 (0.21) & 0.80 (0.03) & 11.57 & $-$2.16 & 15.11 &  51 & 8.39 & 2   \\
0224$+$683  & Lan 142           & 18,270 ( 471) & 8.23 (0.10) & $<-$5.69 (2.42) & 0.74 (0.07) & 11.18 & $-$1.92 & 17.78 & 208 & 8.18 &     \\
0244$+$414  & PM J02478$+$4138  & 17,170 ( 439) & 8.28 (0.10) & $ -$4.73 (0.27) & 0.77 (0.06) & 11.41 & $-$2.07 & 17.40 & 157 & 8.31 &     \\
0249$+$346  & KUV 02499$+$3442  & 13,360 ( 436) & 9.02 (0.21) & $ -$5.54 (1.16) & 1.20 (0.11) & 13.36 & $-$3.05 & 16.40 &  40 & 9.15 &     \\
0249$-$052  & KUV 02498$-$0515  & 17,630 ( 549) & 8.15 (0.08) & $ -$5.37 (0.43) & 0.68 (0.05) & 11.13 & $-$1.94 & 16.60 & 123 & 8.16 &     \\
0258$+$683  & Lan 143           & 14,390 ( 364) & 8.14 (0.10) & $ -$4.08 (0.05) & 0.68 (0.07) & 11.60 & $-$2.29 & 16.80 & 109 & 8.44 &     \\
0300$-$013  & GD 40             & 14,620 ( 399) & 7.99 (0.12) & $ -$6.14 (1.58) & 0.58 (0.07) & 11.34 & $-$2.17 & 15.56 &  69 & 8.32 &     \\
0308$-$565  & L175-34           & 22,840 (2016) & 8.07 (0.05) & $<-$4.82 (3.04) & 0.64 (0.03) & 10.60 & $-$1.43 & 14.07 &  49 & 7.63 & 2   \\
0336$+$625  & Lan 174           & 23,960 (2532) & 8.09 (0.05) & $ -$4.25 (1.40) & 0.66 (0.03) & 10.60 & $-$1.36 & 17.15 & 203 & 7.57 &     \\
0349$+$015  & KUV 03493$+$0131  & 24,860 (1936) & 7.95 (0.05) & $<-$4.59 (1.79) & 0.58 (0.03) & 10.39 & $-$1.22 & 17.20 & 230 & 7.32 &     \\
0414$-$045  & HE 0414$-$043     & 13,470 ( 334) & 8.14 (0.10) & $ -$5.61 (0.32) & 0.68 (0.07) & 11.76 & $-$2.40 & 15.70 &  61 & 8.53 &     \\
0418$-$539  & BPM 17731$ $      & 19,090 ( 464) & 8.10 (0.03) & $<-$4.57 (0.20) & 0.66 (0.02) & 10.90 & $-$1.77 & 15.32 &  76 & 8.00 & 2   \\
0423$-$145  & HE 0423$-$143     & 16,900 ( 401) & 8.08 (0.07) & $<-$5.98 (1.23) & 0.64 (0.04) & 11.12 & $-$1.97 & 16.21 & 104 & 8.17 &     \\
0429$-$168  & HE 0429$-$165     & 15,540 ( 415) & 7.99 (0.15) & $<-$6.35 (3.11) & 0.59 (0.09) & 11.20 & $-$2.07 & 15.82 &  83 & 8.24 &     \\
0435$+$410  & GD 61             & 16,790 ( 408) & 8.18 (0.08) & $ -$4.21 (0.07) & 0.70 (0.05) & 11.30 & $-$2.04 & 14.86 &  51 & 8.26 &     \\
0437$+$138  & LP 475-242        & 15,120 ( 361) & 8.25 (0.07) & $ -$4.68 (0.06) & 0.75 (0.04) & 11.65 & $-$2.27 & 14.92 &  45 & 8.45 &     \\
0503$+$147  & KUV 05034$+$1445  & 15,640 ( 382) & 8.09 (0.06) & $ -$5.46 (0.28) & 0.65 (0.04) & 11.33 & $-$2.11 & 13.80 &  31 & 8.29 &     \\
0513$+$260  & KUV 05134$+$2605  & 24,740 (1334) & 8.21 (0.03) & $ -$3.77 (0.34) & 0.74 (0.02) & 10.76 & $-$1.38 & 16.70 & 154 & 7.67 & 1   \\
0517$+$771  & GD 435            & 13,150 ( 337) & 8.13 (0.12) & $ -$5.97 (0.76) & 0.67 (0.08) & 11.80 & $-$2.44 & 16.01 &  69 & 8.55 &     \\
0615$-$591  & L182-61           & 15,770 ( 373) & 8.04 (0.04) & $<-$6.32 (1.08) & 0.61 (0.03) & 11.23 & $-$2.07 & 13.92 &  34 & 8.25 &     \\
0716$+$404  & GD 85             & 17,150 ( 408) & 8.08 (0.06) & $<-$5.99 (1.28) & 0.64 (0.04) & 11.09 & $-$1.95 & 14.94 &  58 & 8.16 &     \\
0825$+$367  & CBS 73            & 16,100 ( 443) & 8.10 (0.09) & $<-$6.26 (2.47) & 0.65 (0.06) & 11.27 & $-$2.07 & 17.00 & 139 & 8.26 &     \\
0835$+$340  & CSO 197           & 22,230 (1348) & 8.25 (0.05) & $<-$4.63 (1.83) & 0.76 (0.03) & 10.89 & $-$1.59 & 16.00 & 105 & 7.90 &     \\
0838$+$375  & CBS 78            & 13,520 ( 553) & 8.20 (0.49) & $<-$6.54 (5.89) & 0.71 (0.31) & 11.83 & $-$2.43 & 17.71 & 149 & 8.56 &     \\
0840$+$262  & TON 10            & 17,700 ( 420) & 8.28 (0.04) & $ -$4.18 (0.06) & 0.77 (0.03) & 11.33 & $-$2.01 & 14.78 &  49 & 8.26 &     \\
0840$+$364  & CBS 82            & 21,260 ( 863) & 8.15 (0.05) & $<-$5.05 (2.60) & 0.69 (0.03) & 10.80 & $-$1.61 & 17.03 & 176 & 7.86 &     \\
0845$-$188  & L748-70           & 17,470 ( 418) & 8.15 (0.06) & $ -$6.00 (1.55) & 0.69 (0.04) & 11.16 & $-$1.95 & 15.55 &  75 & 8.18 &     \\
0900$+$142  & PG 0900$+$142     & 14,860 ( 351) & 8.07 (0.09) & $<-$6.43 (1.30) & 0.63 (0.06) & 11.43 & $-$2.19 & 16.48 & 102 & 8.35 &     \\
0902$+$293  & CBS 3             & 18,610 ( 502) & 8.02 (0.07) & $<-$5.60 (2.22) & 0.60 (0.04) & 10.82 & $-$1.76 & 17.00 & 171 & 7.98 &     \\
0906$+$341  & CBS 94            & 17,750 ( 480) & 8.12 (0.13) & $<-$5.71 (2.35) & 0.67 (0.09) & 11.08 & $-$1.91 & 17.00 & 152 & 8.14 &     \\
0921$+$091  & PG 0921$+$092     & 19,470 ( 522) & 8.01 (0.04) & $ -$4.72 (0.43) & 0.60 (0.03) & 10.73 & $-$1.68 & 16.19 & 123 & 7.90 &     \\
0948$+$013  & PG 0948$+$013     & 16,810 ( 430) & 8.09 (0.05) & $ -$5.38 (0.29) & 0.65 (0.03) & 11.16 & $-$1.99 & 15.59 &  76 & 8.19 &     \\
0954$+$342  & CBS 114           & 26,060 (1797) & 7.98 (0.06) & $<-$4.03 (0.49) & 0.60 (0.03) & 10.38 & $-$1.15 & 17.20 & 231 & 7.25 & 1   \\
1006$+$413  & KUV 10064$+$4120  & 15,030 ( 465) & 8.80 (0.19) & $ -$5.18 (0.49) & 1.08 (0.11) & 12.66 & $-$2.67 & 17.83 & 108 & 8.91 &     \\
1009$+$416  & KUV 10098$+$4138  & 16,600 ( 456) & 8.67 (0.07) & $ -$5.43 (0.47) & 1.01 (0.04) & 12.17 & $-$2.40 & 16.33 &  67 & 8.70 &     \\
1011$+$570  & GD 303            & 17,610 ( 475) & 8.16 (0.05) & $ -$5.34 (0.28) & 0.69 (0.03) & 11.16 & $-$1.95 & 14.57 &  48 & 8.18 &     \\
1026$-$056  & PG 1026$-$057     & 18,080 ( 425) & 8.11 (0.07) & $<-$5.86 (1.32) & 0.66 (0.04) & 11.03 & $-$1.87 & 16.94 & 152 & 8.10 &     \\
1038$+$290  & Ton 40            & 16,630 ( 390) & 8.10 (0.07) & $ -$5.86 (0.82) & 0.66 (0.05) & 11.20 & $-$2.01 & 16.94 & 140 & 8.22 &     \\
1046$-$017  & GD 124            & 14,620 ( 352) & 8.15 (0.12) & $<-$6.46 (1.62) & 0.68 (0.08) & 11.57 & $-$2.26 & 15.81 &  70 & 8.42 &     \\
1056$+$345  & G119-47           & 12,440 ( 336) & 8.23 (0.14) & $ -$5.33 (0.23) & 0.73 (0.10) & 12.09 & $-$2.60 & 15.58 &  49 & 8.69 &     \\
1107$+$265  & GD 128            & 15,130 ( 357) & 8.11 (0.06) & $ -$5.77 (0.46) & 0.65 (0.04) & 11.43 & $-$2.18 & 15.89 &  78 & 8.35 &     \\
1115$+$158  & PG 1115$+$158     & 23,890 (1726) & 7.91 (0.05) & $ -$3.89 (0.46) & 0.56 (0.03) & 10.36 & $-$1.26 & 16.12 & 142 & 7.37 & 1   \\
1129$+$373  & PG 1129$+$373     & 13,040 ( 358) & 8.16 (0.16) & $ -$6.08 (1.25) & 0.69 (0.10) & 11.87 & $-$2.47 & 16.23 &  74 & 8.58 &     \\
1144$-$084  & PG 1144$-$085     & 15,730 ( 377) & 8.06 (0.06) & $<-$6.32 (1.37) & 0.63 (0.04) & 11.28 & $-$2.09 & 15.95 &  86 & 8.27 &     \\
1148$+$408  & KUV 11489$+$4052  & 17,530 ( 615) & 8.34 (0.10) & $ -$5.51 (0.88) & 0.81 (0.06) & 11.45 & $-$2.06 & 17.33 & 150 & 8.33 &     \\
1149$-$133  & PG 1149$-$133     & 20,370 ( 574) & 8.30 (0.03) & $ -$3.77 (0.13) & 0.78 (0.02) & 11.08 & $-$1.78 & 16.29 & 109 & 8.08 &     \\
1200$+$249  & PM J12033$+$2439  & 13,820 ( 363) & 8.22 (0.13) & $ -$5.20 (0.20) & 0.73 (0.09) & 11.82 & $-$2.41 & 18.00 & 171 & 8.55 &     \\
1240$+$212  & PM J12430$+$2057  & 14,390 ( 364) & 8.06 (0.11) & $<-$6.37 (1.68) & 0.62 (0.07) & 11.48 & $-$2.24 & 17.37 & 150 & 8.39 &     \\
1252$-$289  & EC 12522-2855     & 21,880 ( 756) & 8.03 (0.03) & $<-$4.82 (1.17) & 0.62 (0.02) & 10.59 & $-$1.49 & 15.85 & 112 & 7.69 &     \\
1311$+$129  & LP 497-114        & 22,440 ( 584) & 7.90 (0.04) & $ -$1.94 (0.11) & 0.55 (0.02) & 10.39 & $-$1.37 & 16.26 & 149 & 7.51 &     \\
1326$-$037  & PG 1326$-$037     & 19,950 ( 533) & 8.03 (0.04) & $<-$4.81 (0.53) & 0.61 (0.02) & 10.71 & $-$1.65 & 15.60 &  94 & 7.87 &     \\
1332$+$162  & PB 3990           & 16,780 ( 419) & 8.17 (0.06) & $ -$5.08 (0.26) & 0.70 (0.04) & 11.28 & $-$2.04 & 15.98 &  86 & 8.25 &     \\
1333$+$487  & GD 325            & 15,420 ( 370) & 8.01 (0.09) & $ -$6.37 (1.66) & 0.60 (0.05) & 11.24 & $-$2.09 & 14.02 &  35 & 8.27 &     \\
1336$+$123  & LP 498-26         & 15,950 ( 405) & 8.01 (0.07) & $ -$6.29 (1.90) & 0.60 (0.04) & 11.17 & $-$2.03 & 14.72 &  51 & 8.22 &     \\
1351$+$489  & PG 1351$+$489     & 26,070 (1522) & 7.91 (0.04) & $<-$4.42 (0.90) & 0.56 (0.02) & 10.28 & $-$1.11 & 16.38 & 166 & 7.18 & 1   \\
1352$+$004  & PG 1352$+$004     & 13,980 ( 340) & 8.05 (0.09) & $ -$5.31 (0.17) & 0.62 (0.06) & 11.54 & $-$2.29 & 15.72 &  68 & 8.42 &     \\
1403$-$010  & G64-43            & 15,420 ( 372) & 8.10 (0.06) & $ -$6.08 (0.91) & 0.65 (0.04) & 11.37 & $-$2.14 & 15.90 &  80 & 8.32 &     \\
1411$+$218  & PG 1411$+$219     & 14,970 ( 369) & 8.02 (0.07) & $ -$6.26 (1.19) & 0.60 (0.04) & 11.32 & $-$2.15 & 14.30 &  39 & 8.31 &     \\
1415$+$234  & PG 1415$+$234     & 17,390 ( 478) & 8.19 (0.06) & $ -$5.08 (0.35) & 0.71 (0.04) & 11.23 & $-$1.99 & 16.80 & 129 & 8.22 &     \\
1416$+$229  & KUV 14161$+$2255  & 17,890 ( 444) & 8.25 (0.12) & $<-$5.91 (2.21) & 0.75 (0.08) & 11.26 & $-$1.97 & 16.60 & 117 & 8.22 &     \\
1419$+$351  & GD 335            & 12,730 ( 620) & 8.77 (0.40) & $ -$5.19 (0.62) & 1.06 (0.24) & 12.98 & $-$2.94 & 16.89 &  60 & 9.05 &     \\
1421$-$011  & PG 1421$-$011     & 16,900 ( 411) & 8.19 (0.07) & $ -$4.28 (0.07) & 0.71 (0.05) & 11.30 & $-$2.04 & 15.97 &  85 & 8.26 &     \\
1425$+$540  & G200-39           & 14,410 ( 341) & 7.89 (0.06) & $ -$4.26 (0.03) & 0.53 (0.04) & 11.24 & $-$2.14 & 15.04 &  57 & 8.29 &     \\
1444$-$096  & PG 1444$-$096     & 17,030 ( 429) & 8.26 (0.07) & $ -$5.66 (0.91) & 0.76 (0.04) & 11.40 & $-$2.07 & 14.98 &  52 & 8.30 &     \\
1445$+$152  & PG 1445$+$153     & 20,420 ( 780) & 8.05 (0.06) & $<-$5.24 (2.74) & 0.63 (0.03) & 10.71 & $-$1.62 & 15.55 &  92 & 7.84 &     \\
1454$-$630  & L151-81A          & 14,030 ( 334) & 7.95 (0.07) & $ -$4.83 (0.06) & 0.56 (0.04) & 11.39 & $-$2.22 & 16.60 & 110 & 8.36 & 2   \\
1456$+$103  & PG 1456$+$103     & 24,050 (1206) & 7.91 (0.06) & $ -$3.27 (0.14) & 0.55 (0.03) & 10.35 & $-$1.25 & 15.89 & 128 & 7.35 & 1   \\
1459$+$821  & G256-18           & 16,020 ( 397) & 8.08 (0.06) & $<-$6.28 (1.59) & 0.64 (0.04) & 11.25 & $-$2.06 & 14.78 &  50 & 8.25 &     \\
1540$+$680  & PG 1540$+$681     & 22,240 (1304) & 7.96 (0.04) & $<-$4.43 (0.89) & 0.58 (0.02) & 10.47 & $-$1.42 & 16.19 & 139 & 7.58 &     \\
1542$+$182  & GD 190            & 22,620 ( 978) & 8.04 (0.02) & $<-$4.84 (1.41) & 0.63 (0.01) & 10.57 & $-$1.44 & 14.72 &  67 & 7.62 &     \\
1542$-$275  & LP 916-27         & 12,700 ( 384) & 9.13 (0.14) & $ -$4.99 (0.64) & 1.26 (0.08) & 13.70 & $-$3.23 & 15.49 &  22 & 9.25 &     \\
1545$+$244  & Ton 249           & 12,850 ( 331) & 8.19 (0.12) & $ -$5.00 (0.11) & 0.70 (0.08) & 11.94 & $-$2.51 & 15.78 &  58 & 8.61 &     \\
1551$+$175  & KUV 15519$+$1730  & 15,280 ( 380) & 7.80 (0.12) & $ -$4.41 (0.08) & 0.48 (0.06) & 10.97 & $-$1.99 & 17.50 & 202 & 8.15 &     \\
1557$+$192  & KUV 15571$+$1913  & 19,510 ( 546) & 8.15 (0.05) & $ -$4.30 (0.26) & 0.69 (0.03) & 10.93 & $-$1.76 & 15.40 &  78 & 8.01 &     \\
1610$+$239  & PG 1610$+$239     & 13,280 ( 332) & 8.13 (0.11) & $ -$5.74 (0.42) & 0.67 (0.07) & 11.77 & $-$2.42 & 15.34 &  51 & 8.53 &     \\
1612$-$111  & GD 198            & 23,430 (1782) & 7.96 (0.04) & $<-$4.75 (2.10) & 0.58 (0.02) & 10.44 & $-$1.33 & 15.53 & 104 & 7.46 &     \\
1644$+$198  & PG 1644$+$199     & 15,210 ( 360) & 8.14 (0.06) & $ -$5.68 (0.39) & 0.68 (0.04) & 11.47 & $-$2.19 & 15.20 &  55 & 8.37 &     \\
1645$+$325  & GD 358            & 24,940 (1114) & 7.92 (0.03) & $<-$4.58 (0.88) & 0.56 (0.01) & 10.34 & $-$1.19 & 13.65 &  45 & 7.28 & 1   \\
1654$+$160  & PG 1654$+$160     & 26,140 (1211) & 7.91 (0.03) & $<-$4.40 (0.64) & 0.56 (0.02) & 10.27 & $-$1.10 & 16.55 & 180 & 7.17 & 1   \\
1703$+$319  & PG 1703$+$319     & 14,440 ( 360) & 8.46 (0.10) & $ -$5.54 (0.37) & 0.88 (0.06) & 12.10 & $-$2.48 & 16.25 &  67 & 8.67 &     \\
1708$-$871  & L7-44             & 23,980 (1686) & 8.05 (0.03) & $<-$4.69 (1.93) & 0.63 (0.02) & 10.55 & $-$1.34 & 14.38 &  58 & 7.51 &     \\
1709$+$230  & GD 205            & 19,590 ( 504) & 8.08 (0.03) & $ -$4.07 (0.14) & 0.65 (0.02) & 10.83 & $-$1.71 & 14.90 &  65 & 7.95 &     \\
1726$-$578  & L204-118          & 14,320 ( 340) & 8.20 (0.06) & $ -$5.46 (0.19) & 0.71 (0.04) & 11.70 & $-$2.33 & 15.27 &  51 & 8.49 &     \\
1822$+$410  & GD 378            & 16,230 ( 383) & 8.00 (0.06) & $ -$4.45 (0.06) & 0.59 (0.04) & 11.11 & $-$2.00 & 14.39 &  45 & 8.19 &     \\
1919$-$362  &                   & 23,610 ( 988) & 8.10 (0.02) & $ -$4.22 (0.44) & 0.66 (0.01) & 10.62 & $-$1.40 & 13.60 &  39 & 7.61 &     \\
1940$+$374  & L1573-31          & 16,850 ( 406) & 8.07 (0.09) & $ -$5.97 (1.50) & 0.64 (0.06) & 11.13 & $-$1.97 & 14.51 &  47 & 8.18 &     \\
2034$-$532  & L279-25           & 17,160 ( 403) & 8.47 (0.05) & $ -$5.78 (0.59) & 0.89 (0.03) & 11.73 & $-$2.19 & 14.46 &  35 & 8.48 &     \\
2058$+$342  & GD 392A           & 12,210 ( 447) & 9.05 (0.21) & $ -$5.18 (1.09) & 1.22 (0.11) & 13.64 & $-$3.23 & 15.68 &  25 & 9.24 &     \\
2129$+$000  & G26-10            & 14,350 ( 349) & 8.25 (0.12) & $<-$6.49 (1.65) & 0.74 (0.08) & 11.77 & $-$2.36 & 15.27 &  50 & 8.52 &     \\
2130$-$047  & GD 233            & 18,110 ( 426) & 8.11 (0.07) & $ -$5.79 (1.36) & 0.66 (0.04) & 11.02 & $-$1.87 & 14.52 &  50 & 8.09 &     \\
2144$-$079  & G26-31            & 16,340 ( 408) & 8.18 (0.05) & $<-$6.22 (1.45) & 0.70 (0.03) & 11.36 & $-$2.09 & 14.82 &  49 & 8.30 &     \\
2147$+$280  & G188-27           & 12,940 ( 399) & 8.86 (0.19) & $ -$5.74 (0.96) & 1.11 (0.10) & 13.10 & $-$2.97 & 14.68 &  20 & 9.08 &     \\
2222$+$683  & G241-6            & 14,920 ( 383) & 8.00 (0.19) & $<-$6.43 (2.84) & 0.59 (0.12) & 11.31 & $-$2.15 & 15.65 &  73 & 8.31 &     \\
2229$+$139  & PG 2229$+$139     & 14,870 ( 352) & 8.15 (0.06) & $ -$4.91 (0.08) & 0.69 (0.04) & 11.55 & $-$2.24 & 15.99 &  77 & 8.41 &     \\
2234$+$064  & PG 2234$+$064     & 23,770 (1770) & 8.07 (0.03) & $<-$4.72 (2.10) & 0.65 (0.02) & 10.58 & $-$1.37 & 16.03 & 122 & 7.56 &     \\
2236$+$541  & KPD 2236+5410     & 15,590 ( 379) & 8.28 (0.07) & $<-$6.34 (1.67) & 0.77 (0.05) & 11.63 & $-$2.23 & 16.19 &  81 & 8.43 &     \\
2237$-$051  & PHL 363           & 13,100 ( 460) & 8.73 (0.24) & $ -$4.87 (0.21) & 1.04 (0.14) & 12.83 & $-$2.85 & 14.00 &  17 & 8.99 &     \\
2246$+$120  & PG 2246$+$121     & 26,840 (1433) & 7.92 (0.04) & $<-$4.30 (0.65) & 0.56 (0.02) & 10.25 & $-$1.06 & 16.73 & 197 & 7.13 & 1   \\
2250$+$746  & GD 554            & 16,560 ( 390) & 8.15 (0.03) & $<-$6.18 (0.68) & 0.69 (0.02) & 11.28 & $-$2.05 & 16.69 & 120 & 8.26 &     \\
2253$-$062  & GD 243            & 17,190 ( 436) & 8.07 (0.09) & $ -$4.35 (0.13) & 0.64 (0.06) & 11.07 & $-$1.93 & 15.06 &  62 & 8.14 &     \\
2310$+$175  & KUV 23103$+$1736  & 15,150 ( 370) & 8.37 (0.07) & $ -$5.38 (0.23) & 0.82 (0.04) & 11.84 & $-$2.34 & 15.88 &  64 & 8.54 &     \\
2316$-$173  & G273-13           & 12,640 ( 421) & 9.11 (0.18) & $ -$5.08 (0.87) & 1.25 (0.10) & 13.67 & $-$3.22 & 14.08 &  12 & 9.24 &     \\
2328$+$510  & GD 406            & 14,500 ( 362) & 8.03 (0.16) & $ -$6.47 (2.09) & 0.61 (0.10) & 11.43 & $-$2.21 & 15.09 &  54 & 8.36 &     \\
2354$+$159  & PG 2354$+$159     & 24,830 (1670) & 8.15 (0.03) & $<-$4.59 (1.78) & 0.70 (0.02) & 10.67 & $-$1.34 & 15.78 & 105 & 7.58 &     \\
\enddata
\tablecomments{
(1) Variable white dwarf of the V777 Her class.
(2) Hydrogen abundance based on \hb.
}
\end{deluxetable}

\clearpage
\begin{deluxetable}{llrrrccccccc}
\tabletypesize{\scriptsize}
\tablecolumns{12}
\tablewidth{0pt}
\rotate
\tablecaption{Atmospheric Parameters of Cool, He-rich DA White Dwarfs \label{table_fits_DCA}}
\tablehead{
\colhead{WD} &
\colhead{Name} &
\colhead{$\Te$ (K)} &
\colhead{log $g$} &
\colhead{log H/He} &
\colhead{$M/$\msun} &
\colhead{$M_g$} &
\colhead{log $L/$\lsun} &
\colhead{$g$} &
\colhead{$D ({\rm pc})$} &
\colhead{log $\tau$} &
\colhead{Notes}}
\startdata
0042$+$141  & SDSS J004513.88$+$142248.1 &   6980 (197) & 8.00 & $-$2.28 (0.13) & 0.57 & 13.68 & $-$3.48 & 19.20 & 126 & 9.19 &     \\
0107$-$003  & SDSS J011012.48$-$000313.5 &   9790 (434) & 8.00 & $-$4.05 (0.23) & 0.58 & 12.34 & $-$2.88 & 19.38 & 255 & 8.82 &     \\
0236$-$094  & SDSS J023856.77$-$092653.6 & 10,940 (463) & 8.00 & $-$4.98 (0.26) & 0.58 & 12.00 & $-$2.69 & 18.43 & 193 & 8.69 &     \\
0528$+$615  & SDSS J074250.80$+$222444.7 & 10,730 (462) & 8.00 & $-$4.47 (0.20) & 0.58 & 12.05 & $-$2.72 & 19.27 & 277 & 8.71 &     \\
0739$+$225  & SDSS J074250.80$+$222444.7 &   8330 (236) & 8.00 & $-$3.44 (0.30) & 0.58 & 12.93 & $-$3.17 & 19.13 & 173 & 9.01 &     \\
0748$+$314  & SDSS J075113.24$+$313249.4 & 11,490 (550) & 8.00 & $-$4.72 (0.09) & 0.58 & 11.85 & $-$2.60 & 19.19 & 293 & 8.63 &     \\
0829$+$532  & SDSS J083317.40$+$531335.5 & 11,350 (534) & 8.00 & $-$4.76 (0.16) & 0.58 & 11.89 & $-$2.62 & 18.82 & 243 & 8.65 &     \\
0838$+$204  & SDSS J084113.94$+$203018.7 &   8980 (277) & 8.00 & $-$3.39 (0.17) & 0.58 & 12.65 & $-$3.04 & 18.66 & 159 & 8.92 &     \\
0844$+$364  & SDSS J084757.57$+$362649.1 &   9400 (303) & 8.00 & $-$3.60 (0.24) & 0.58 & 12.48 & $-$2.96 & 19.46 & 248 & 8.86 &     \\
0859$+$094  & SDSS J090150.74$+$091211.3 &   8320 (218) & 8.00 & $-$2.71 (0.20) & 0.58 & 12.94 & $-$3.17 & 18.60 & 135 & 9.01 &     \\
1023$+$142  & SDSS J102626.01$+$135745.0 &   8780 (255) & 8.00 & $-$3.44 (0.22) & 0.58 & 12.73 & $-$3.07 & 18.92 & 172 & 8.94 &     \\
1157$+$072  & SDSS J115948.51$+$070708.7 &   9940 (328) & 8.00 & $-$4.61 (0.18) & 0.58 & 12.30 & $-$2.86 & 17.55 & 112 & 8.80 &     \\
1203$+$343  & SDSS J120555.16$+$341813.4 & 11,330 (445) & 8.00 & $-$5.31 (0.15) & 0.58 & 11.89 & $-$2.63 & 18.29 & 190 & 8.65 &     \\
1246$+$021  & SDSS J124909.03$+$015559.3 &   7350 (176) & 8.00 & $-$2.38 (0.14) & 0.57 & 13.45 & $-$3.38 & 18.84 & 119 & 9.13 &     \\
1307$+$454  & SDSS J130916.90$+$452342.6 &   9750 (325) & 8.00 & $-$3.82 (0.17) & 0.58 & 12.36 & $-$2.89 & 18.82 & 195 & 8.82 &     \\
1345$+$513  & SDSS J134710.47$+$511640.8 &   9280 (263) & 8.00 & $-$3.46 (0.16) & 0.58 & 12.53 & $-$2.98 & 18.49 & 155 & 8.88 &     \\
1409$+$114  & SDSS J141209.94$+$112902.6 &   7280 (162) & 8.00 & $-$2.12 (0.15) & 0.57 & 13.49 & $-$3.40 & 18.65 & 107 & 9.14 &     \\
1412$-$009  & SDSS J141516.10$-$010912.1 &   8520 (225) & 8.00 & $-$3.25 (0.17) & 0.58 & 12.85 & $-$3.13 & 18.26 & 120 & 8.98 & 1   \\
1506$+$017  & SDSS J150856.93$+$013557.0 &   8060 (186) & 8.00 & $-$3.30 (0.15) & 0.58 & 13.07 & $-$3.22 & 17.98 &  96 & 9.04 &     \\
1519$+$397  & SDSS J152145.91$+$393128.1 &   7980 (182) & 8.00 & $-$2.95 (0.17) & 0.58 & 13.11 & $-$3.24 & 18.54 & 122 & 9.05 &     \\
1556$+$110  & SDSS J155903.81$+$105614.8 &   8530 (213) & 8.00 & $-$2.87 (0.15) & 0.58 & 12.84 & $-$3.12 & 18.52 & 136 & 8.98 & 1   \\
1558$+$077  & SDSS J160053.84$+$074803.4 &   6870 (155) & 8.00 & $-$2.14 (0.14) & 0.57 & 13.76 & $-$3.50 & 19.39 & 133 & 9.22 & 1   \\
1617$-$003  & SDSS J161948.91$+$003445.3 & 10,440 (359) & 8.00 & $-$4.23 (0.18) & 0.58 & 12.14 & $-$2.77 & 17.88 & 140 & 8.75 &     \\
1623$+$222  & SDSS J162535.21$+$221516.4 &   8620 (242) & 8.00 & $-$3.52 (0.18) & 0.58 & 12.80 & $-$3.11 & 18.75 & 154 & 8.97 & 2   \\
1625$+$305  & SDSS J162721.62$+$304320.2 &   8460 (224) & 8.00 & $-$3.60 (0.23) & 0.58 & 12.87 & $-$3.14 & 18.74 & 149 & 8.99 &     \\
1644$+$202  & SDSS J164645.22$+$200701.5 &   9370 (311) & 8.00 & $-$3.91 (0.20) & 0.58 & 12.49 & $-$2.96 & 19.37 & 237 & 8.87 &     \\
2116$+$110  & SDSS J211852.10$+$111756.5 & 11,080 (412) & 8.00 & $-$5.37 (0.19) & 0.58 & 11.96 & $-$2.67 & 16.66 &  87 & 8.68 &     \\
2134$+$112  & SDSS J213621.56$+$113726.8 & 11,390 (521) & 8.00 & $-$4.61 (0.16) & 0.58 & 11.88 & $-$2.62 & 18.29 & 191 & 8.64 &     \\
\enddata

\tablecomments{
(1) Possible unresolved DA$+$DC degenerate binary.
(2) $g$ band omitted during atmospheric parameter determination.
}
\end{deluxetable}

\clearpage
\begin{deluxetable}{cccccc}
\tablecolumns{6}
\tablewidth{0pt}
\tablecaption{Hydrogen- to Helium-Atmosphere Transition Temperatures\label{table_mixingT}}
\tablehead{
\colhead{} &
\multicolumn{2}{c}{ML2/$\alpha=0.6$} &
\colhead{} &
\multicolumn{2}{c}{ML2/$\alpha=2.0$}\\
\cline{2-3}\cline{5-6}\\
\colhead{log $\mh$/\msun} &
\colhead{$\Te$ (K)} &
\colhead{$\logh$} &
\colhead{} &
\colhead{$\Te$ (K)} &
\colhead{$\logh$}
}

\startdata
$-$15.0 & 22,100 & $-$2.8 && 31,500 & $-$0.3 \\
$-$14.5 & 21,000 & $-$2.6 && 27,950 & $-$2.7 \\
$-$14.0 & 19,350 & $-$5.1 && 22,250 & $-$3.8 \\
$-$13.5 & 16,850 & $-$6.6 && 18,000 & $-$6.6 \\
$-$13.0 & 11,250 & $-$7.5 && 12,350 & $-$7.3 \\
$-$12.5 & 11,050 & $-$7.0 && 12,100 & $-$6.9 \\
$-$12.0 & 10,850 & $-$6.7 && 11,800 & $-$6.4 \\
$-$11.5 & 10,550 & $-$6.1 && 11,450 & $-$6.0 \\
$-$11.0 & 10,300 & $-$5.6 && 11,150 & $-$5.5 \\
$-$10.5 & 10,050 & $-$5.1 && 10,800 & $-$5.0 \\
$-$10.0 &   9700 & $-$4.6 && 10,400 & $-$4.6 \\
 $-$9.5 &   9250 & $-$4.0 &&   9900 & $-$4.0 \\
 $-$9.0 &   8650 & $-$3.3 &&   9100 & $-$3.4 \\
 $-$8.5 &   7850 & $-$2.6 &&   8150 & $-$2.6 \\
 $-$8.0 &   6850 & $-$1.9 &&   7000 & $-$1.9 \\
\enddata


\end{deluxetable}

\section{Online Material}

\begin{figure}[bp]
\centering
\includegraphics[width=0.8\linewidth]{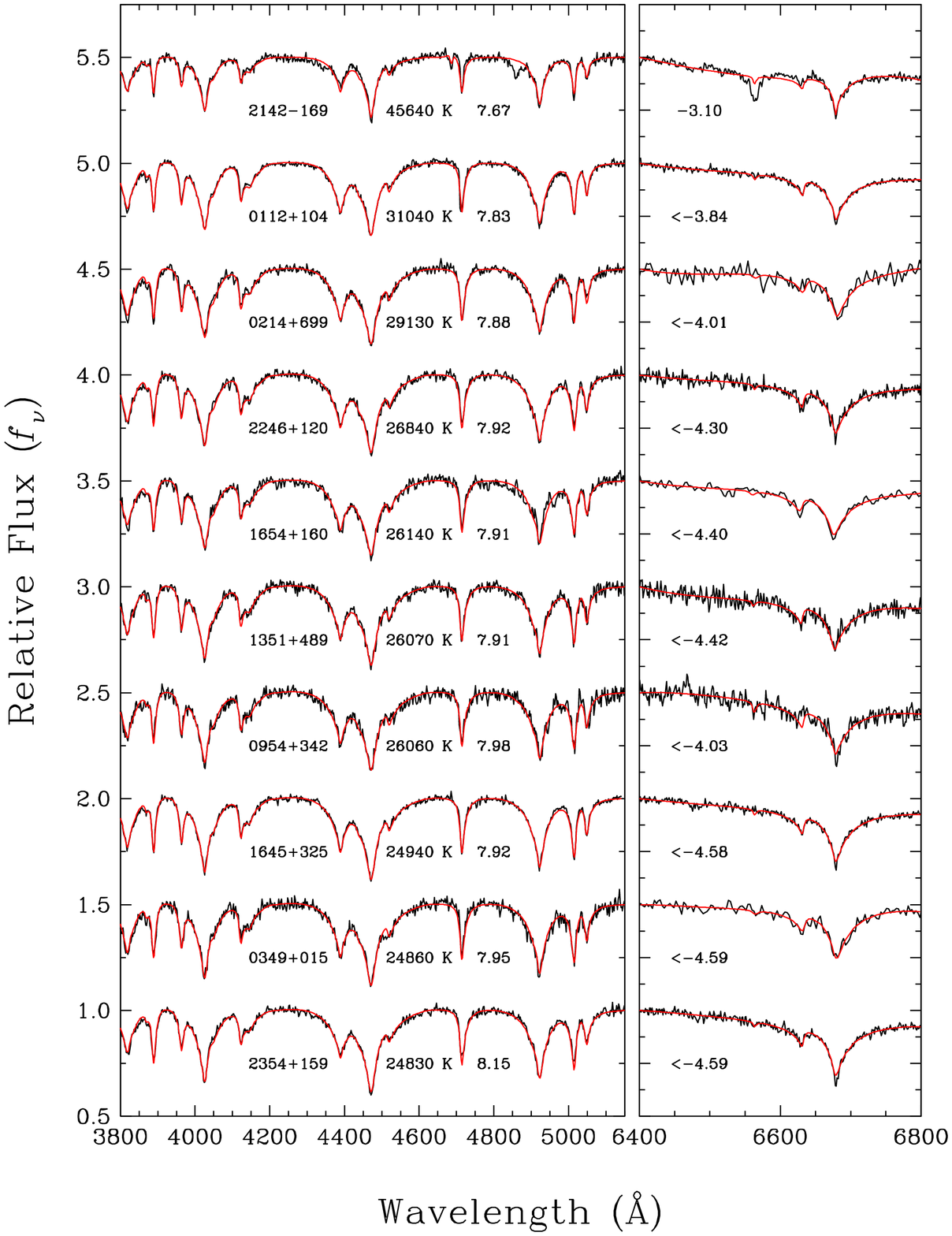}
\caption{Spectroscopic fits for all DB and DBA stars in our sample, in
  order of decreasing effective temperature. The atmospheric
  parameters ($\Te$, $\logg$, $\logh$) of each object are given in the
  figure. The region near \ha\ (right panel) is used to measure, or to
  constrain, the hydrogen abundance. In the case of DB stars, these
  spectra only provide \textit{upper limits} on the hydrogen-to-helium
  abundance ratio.}
\end{figure}

\begin{figure}[bp]
\figurenum{18}
\centering
\includegraphics[width=0.8\linewidth]{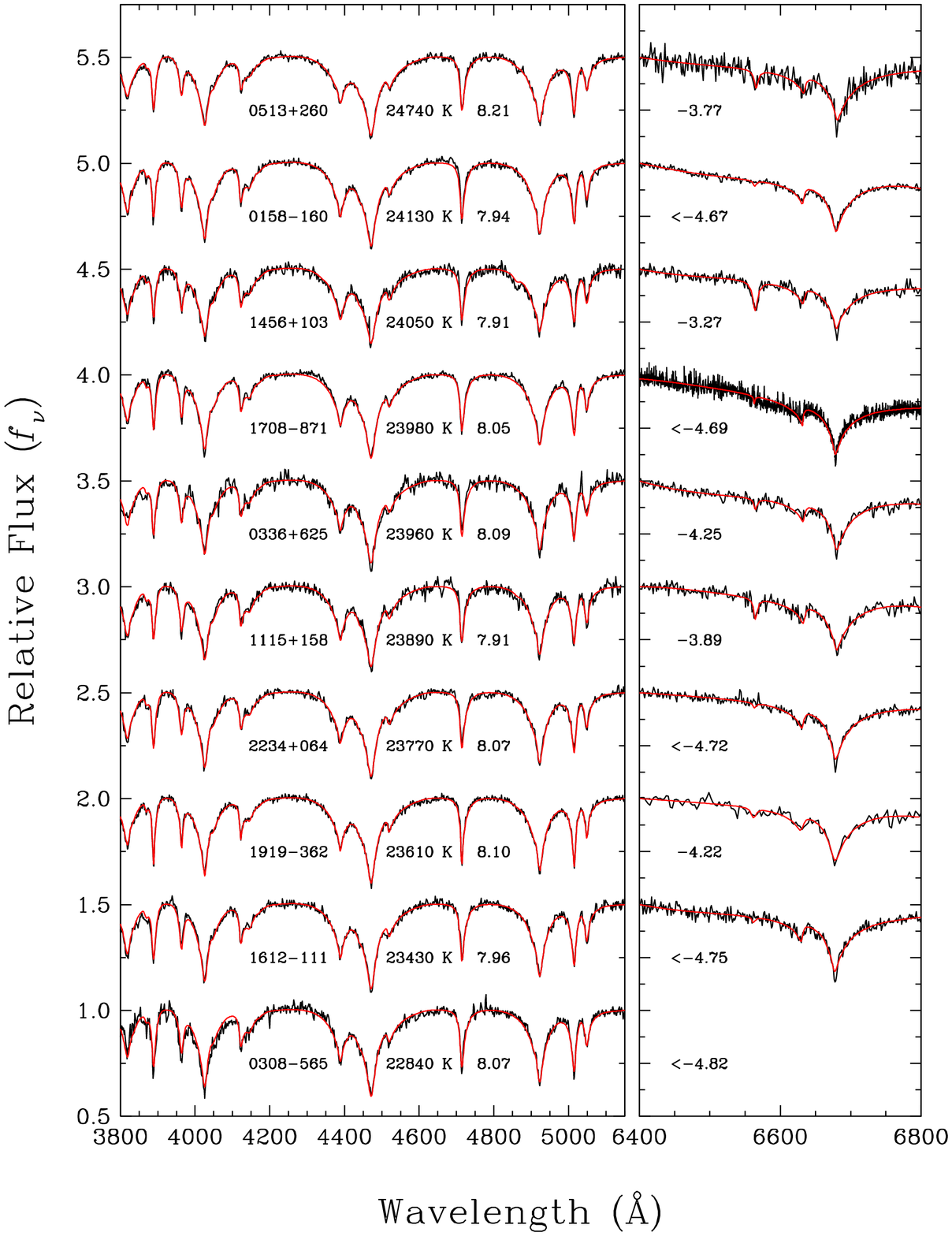}
\caption{(Continued)}
\end{figure}

\begin{figure}[bp]
\figurenum{18}
\centering
\includegraphics[width=0.8\linewidth]{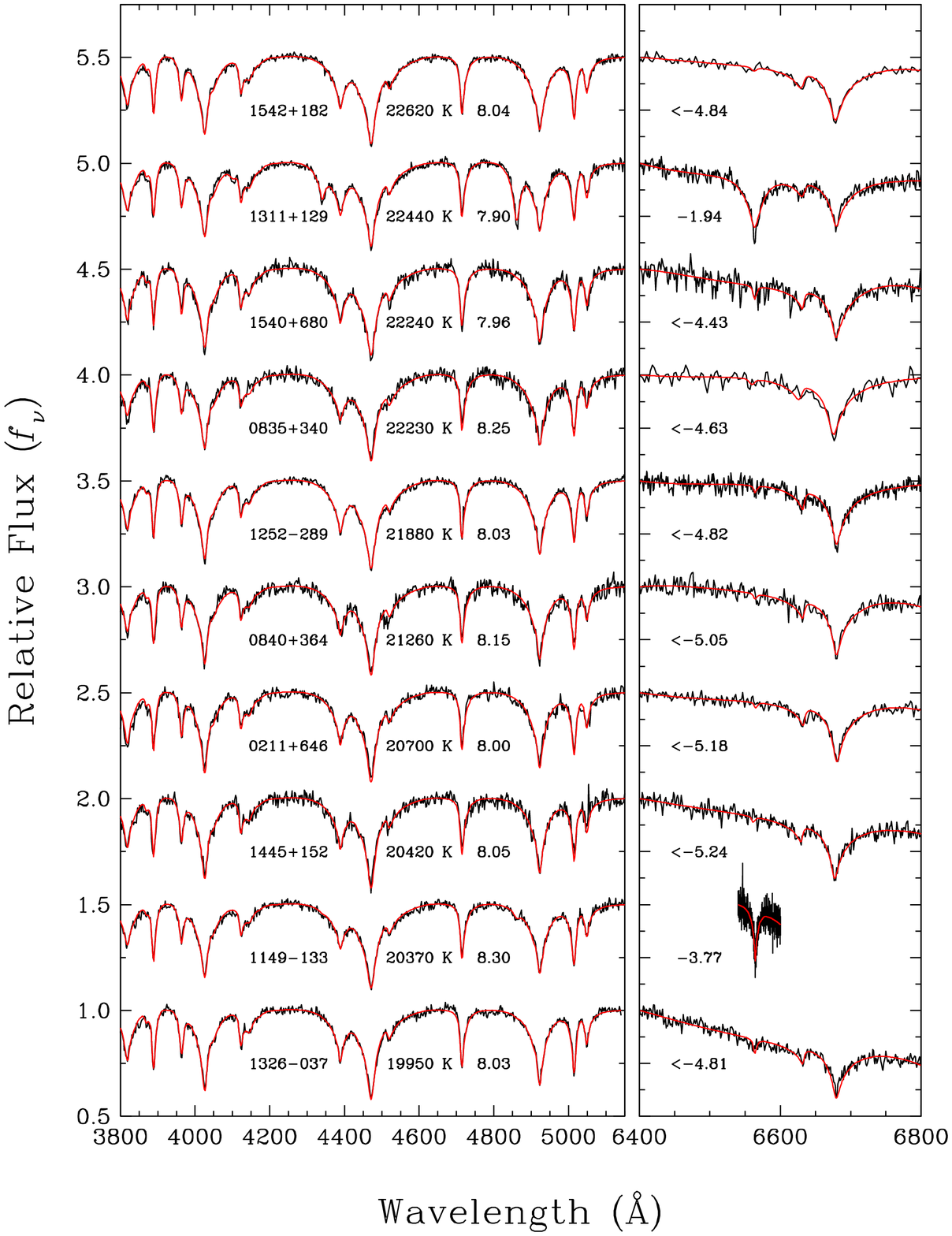}
\caption{(Continued)}
\end{figure}

\begin{figure}[bp]
\figurenum{18}
\centering
\includegraphics[width=0.8\linewidth]{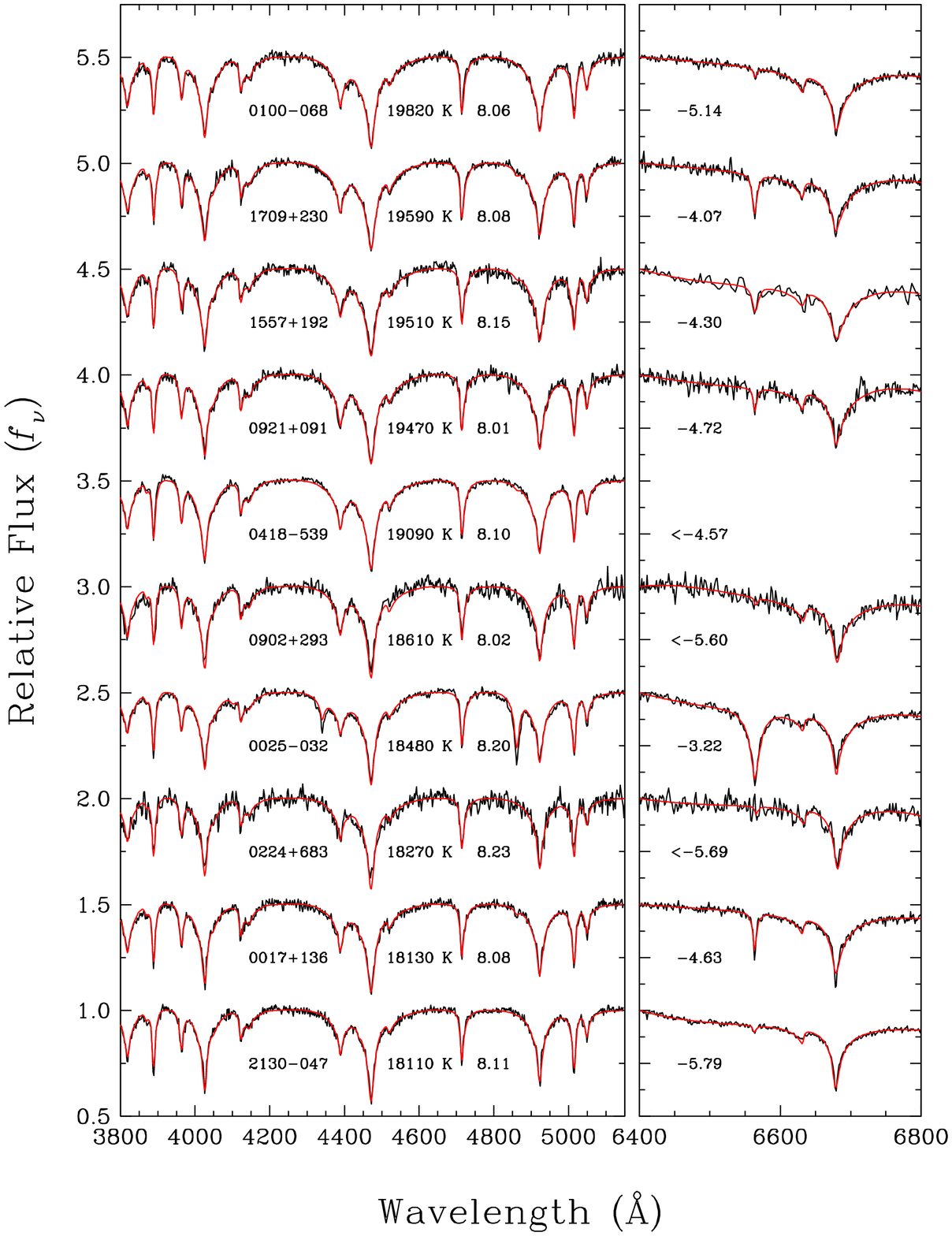}
\caption{(Continued)}
\end{figure}

\begin{figure}[bp]
\figurenum{18}
\centering
\includegraphics[width=0.8\linewidth]{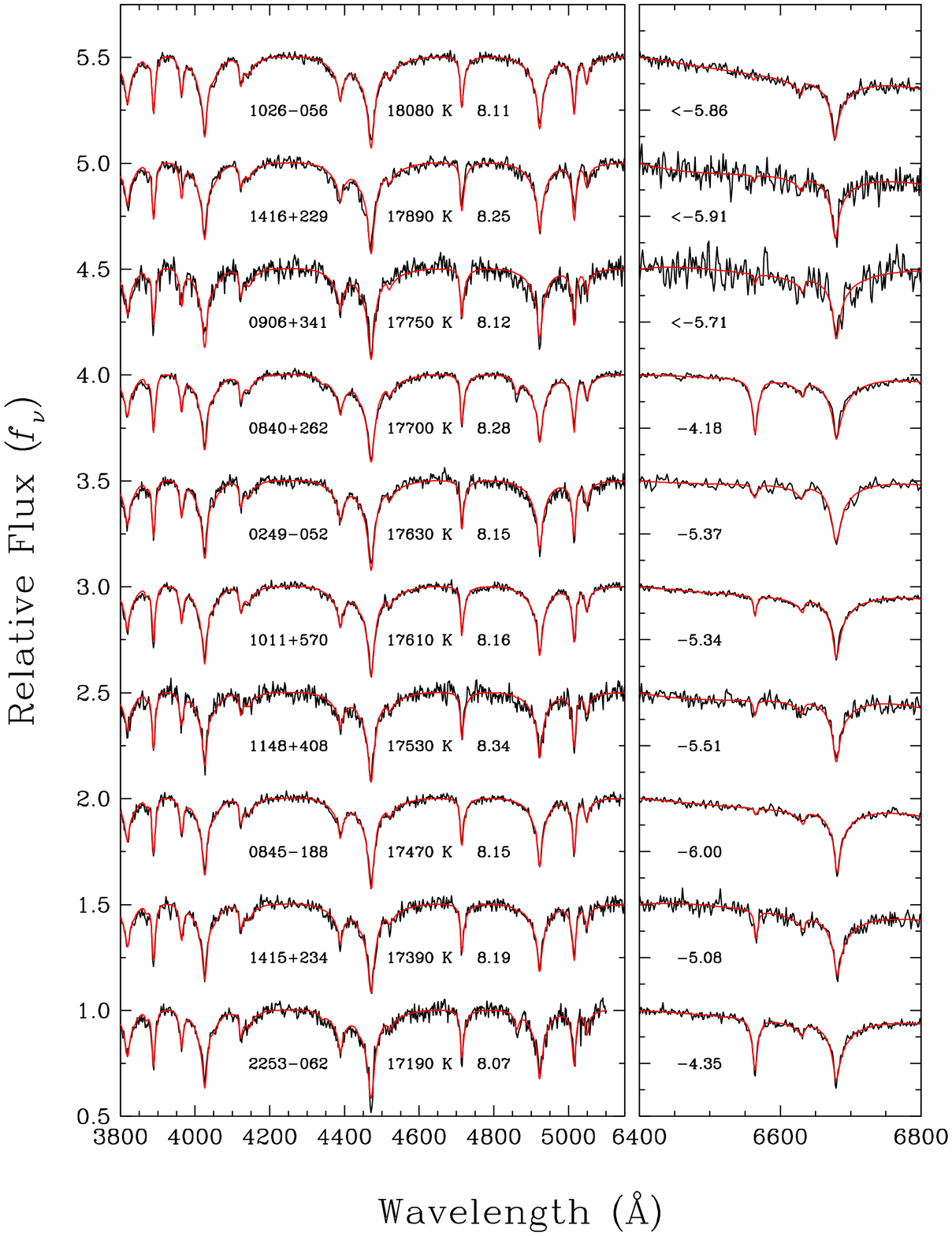}
\caption{(Continued)}
\end{figure}

\begin{figure}[bp]
\figurenum{18}
\centering
\includegraphics[width=0.8\linewidth]{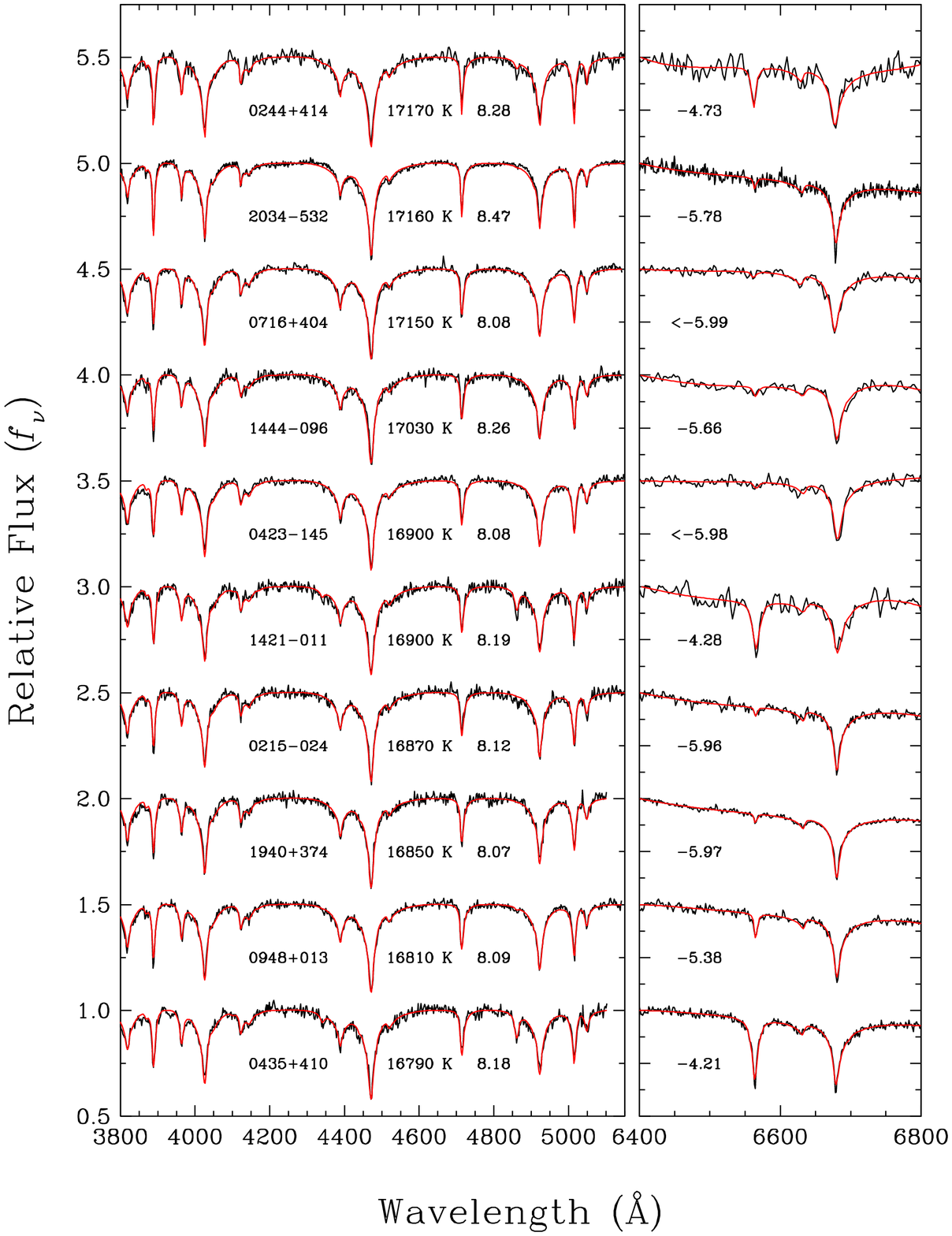}
\caption{(Continued)}
\end{figure}

\begin{figure}[bp]
\figurenum{18}
\centering
\includegraphics[width=0.8\linewidth]{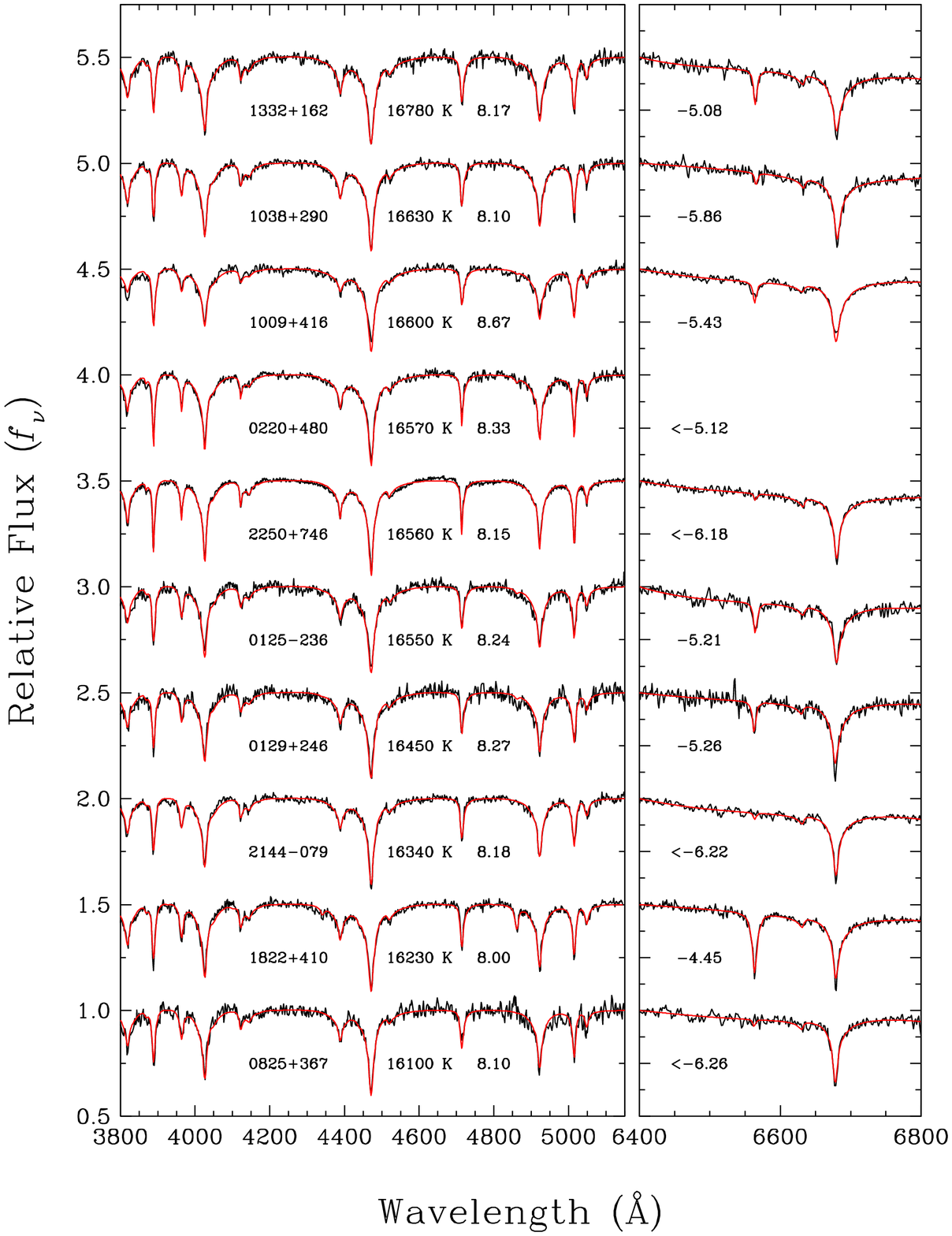}
\caption{(Continued)}
\end{figure}

\begin{figure}[bp]
\figurenum{18}
\centering
\includegraphics[width=0.8\linewidth]{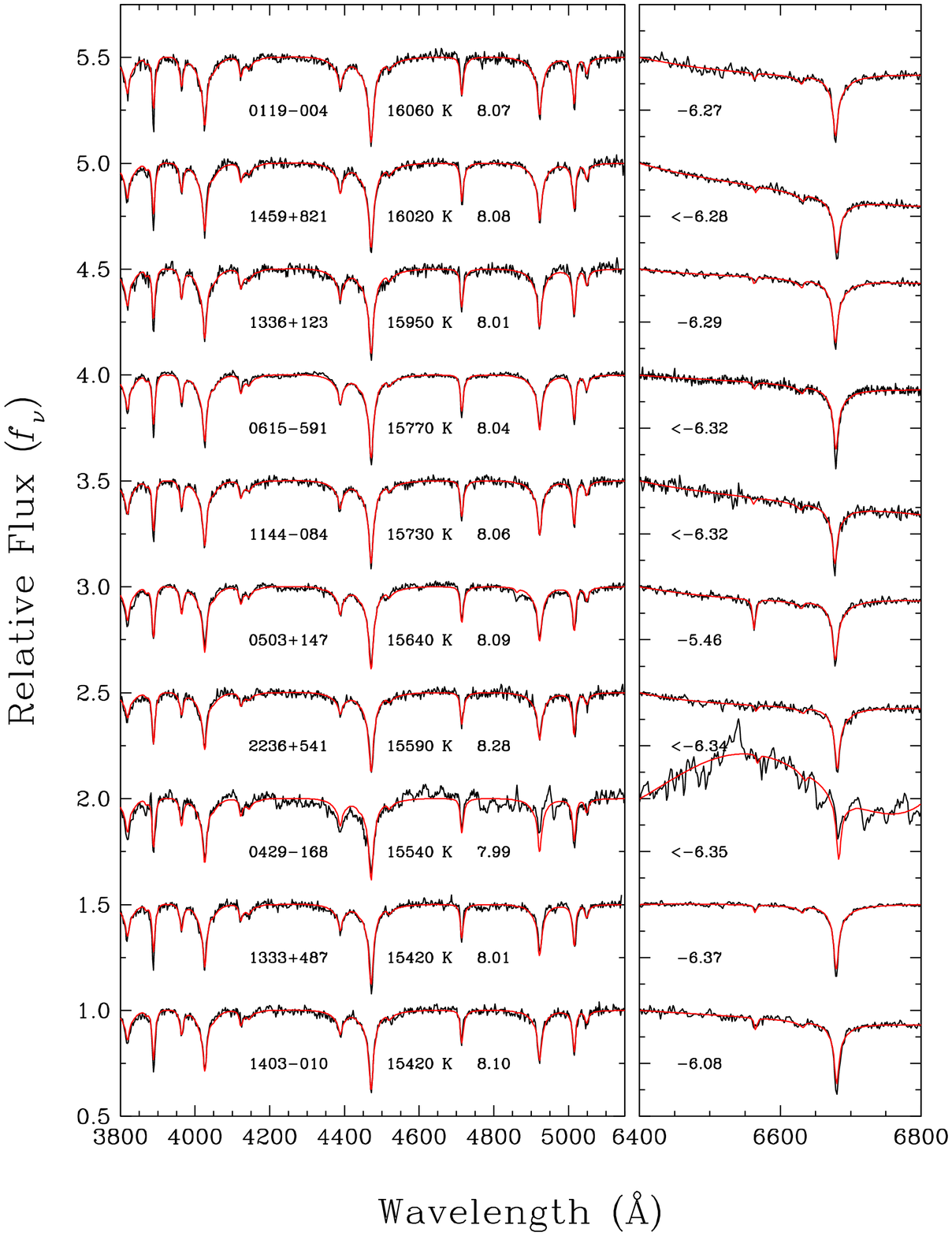}
\caption{(Continued)}
\end{figure}

\begin{figure}[bp]
\figurenum{18}
\centering
\includegraphics[width=0.8\linewidth]{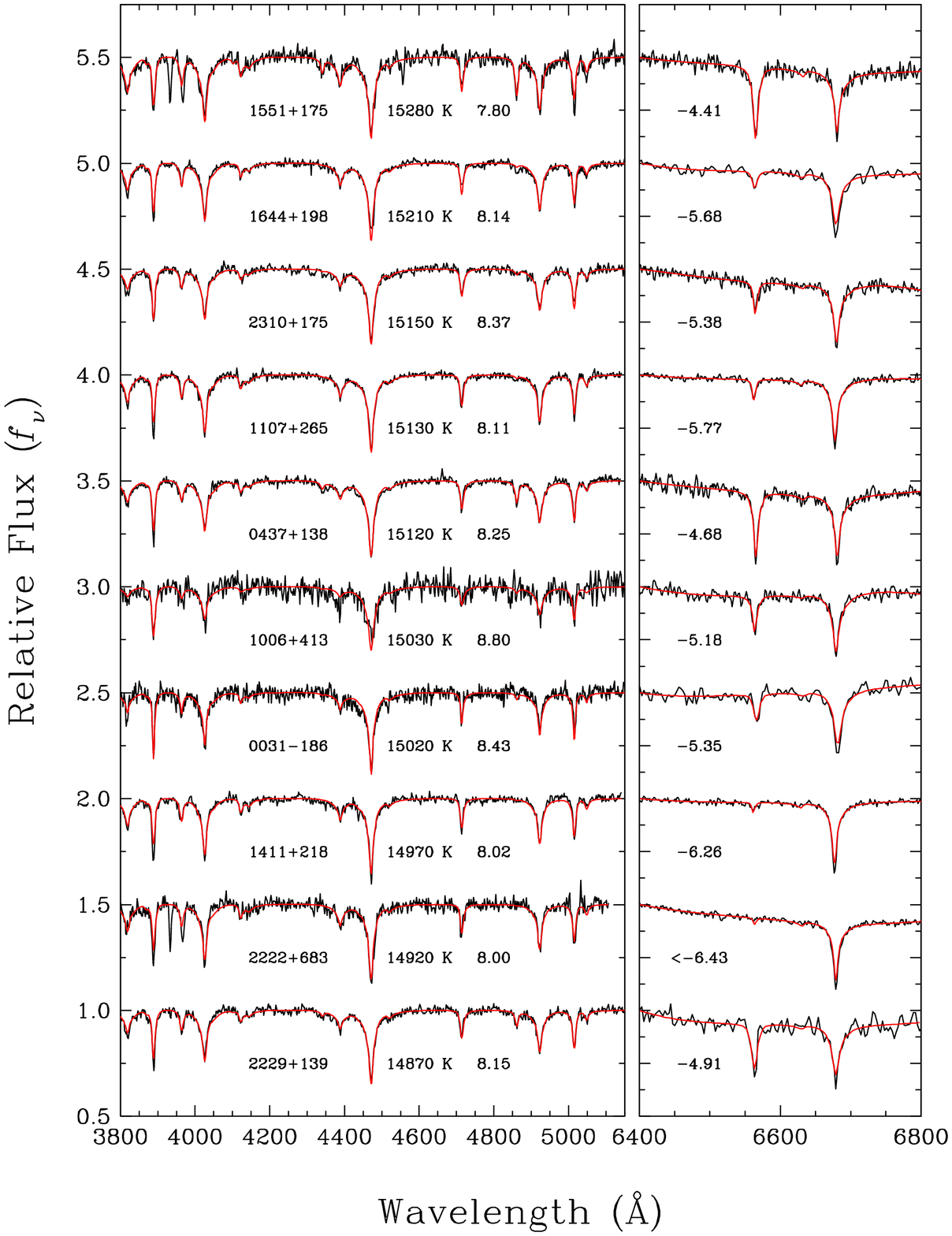}
\caption{(Continued)}
\end{figure}

\begin{figure}[bp]
\figurenum{18}
\centering
\includegraphics[width=0.8\linewidth]{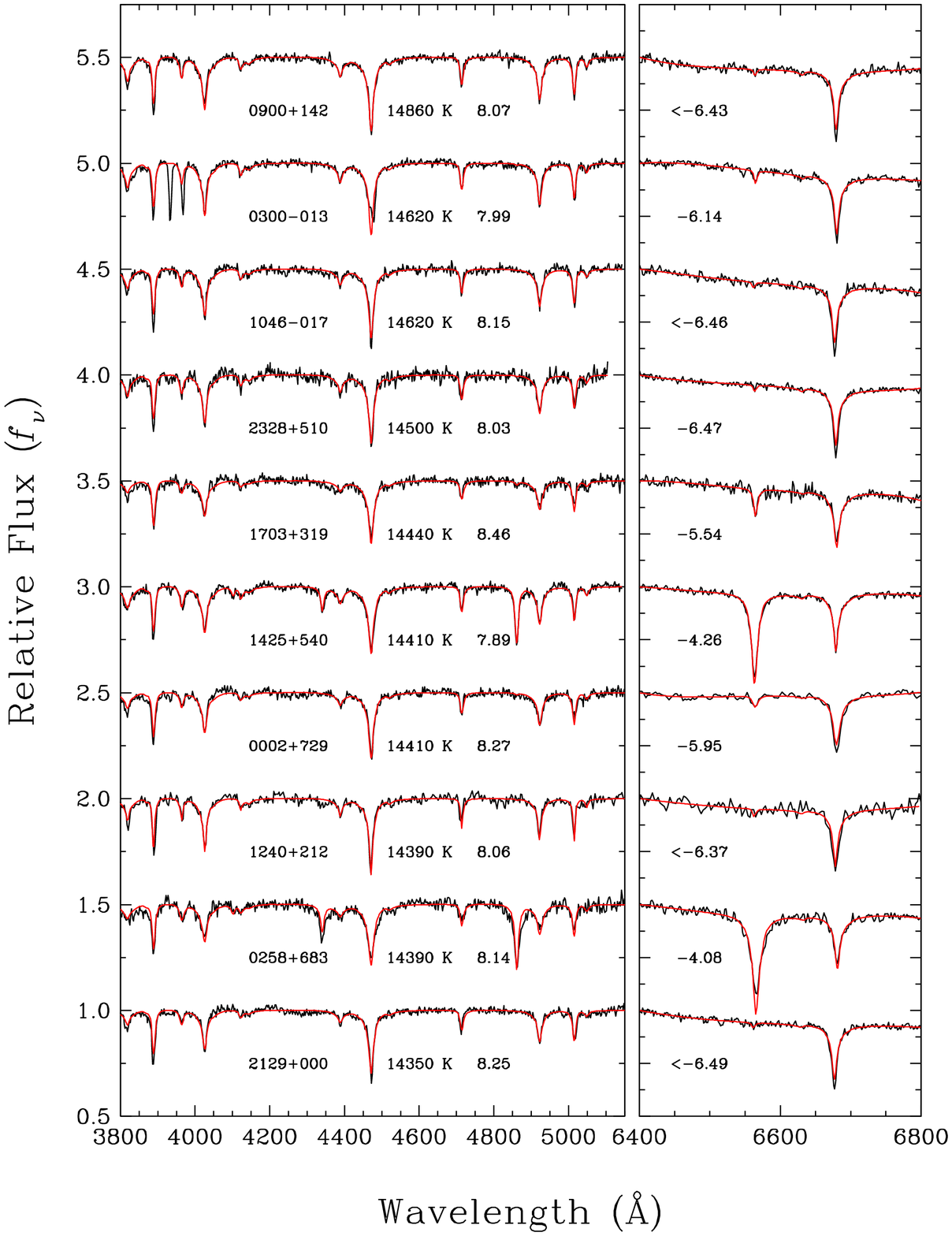}
\caption{(Continued)}
\end{figure}

\begin{figure}[bp]
\figurenum{18}
\centering
\includegraphics[width=0.8\linewidth]{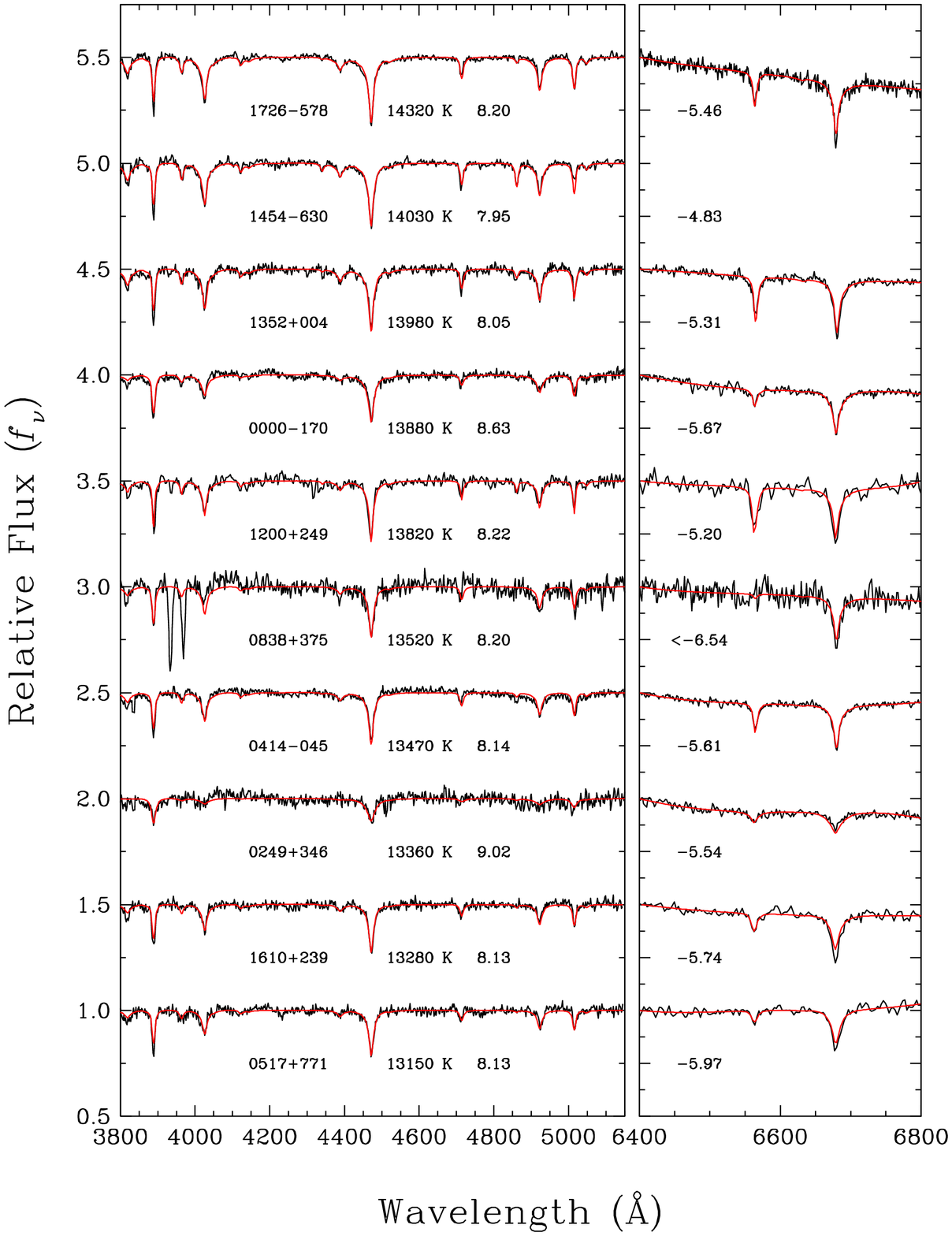}
\caption{(Continued)}
\end{figure}

\begin{figure}[bp]
\figurenum{18}
\centering
\includegraphics[width=0.8\linewidth]{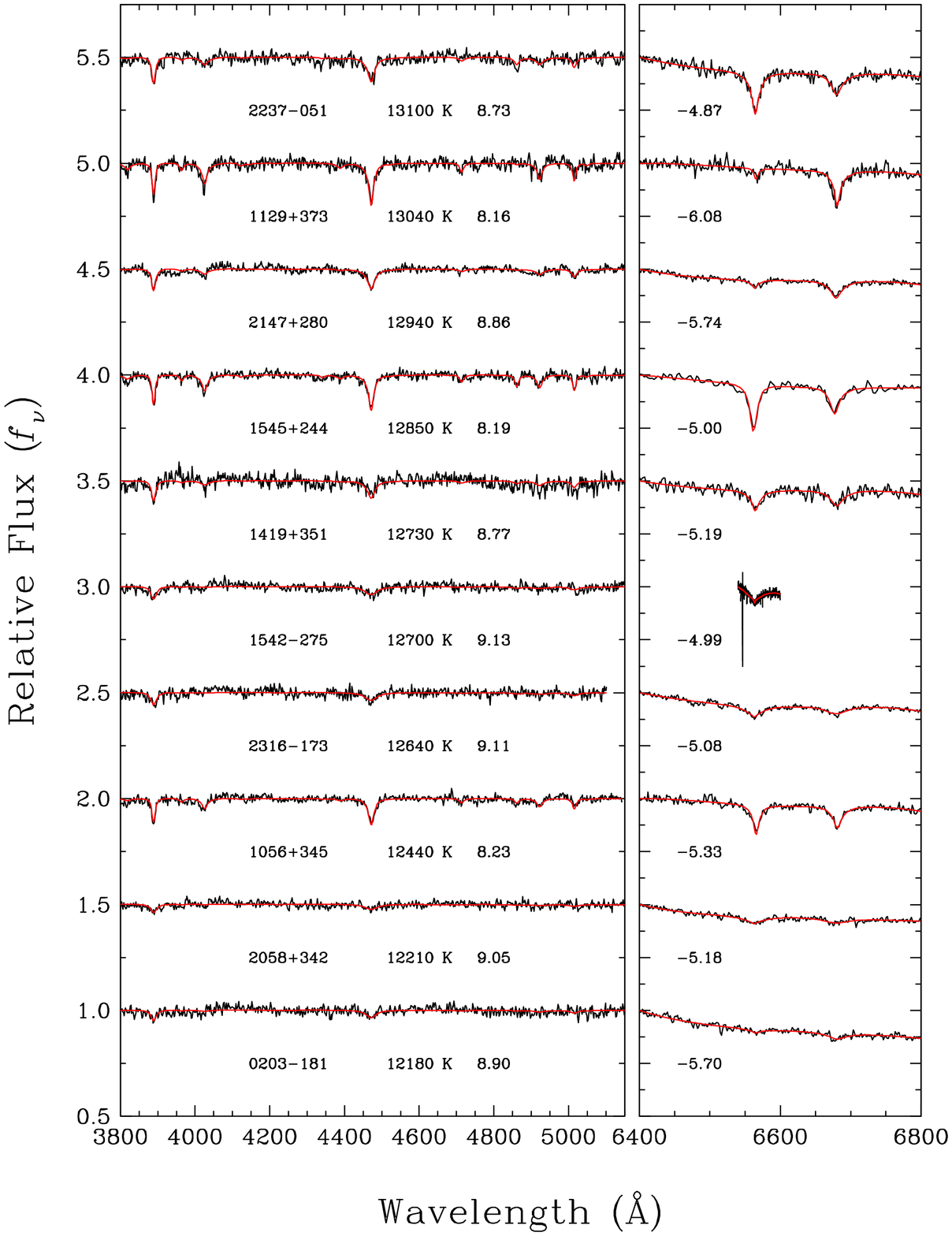}
\caption{(Continued)}
\end{figure}

\end{document}